\documentclass[journal]{vgtc}                     % final (journal style)}
\onlineid{1535}

\vgtccategory{IEEE VIS Full Paper}

\vgtcpapertype{Data Transformations}

\title{VolSegGS: Segmentation and Tracking in Dynamic Volumetric Scenes via Deformable 3D Gaussians}

\author{%
  \authororcid{Siyuan Yao}{0000-0002-4093-193X} and 
  \authororcid{Chaoli Wang}{0000-0002-0859-3619}
}

\authorfooter{
  \item The authors are with %the University of Notre Dame. 
the Department of Computer Science and Engineering, University of Notre Dame, Notre Dame, IN 46556, USA.\\
E-mail: \{syao2, chaoli.wang\}@nd.edu.
}

\abstract{Visualization of large-scale time-dependent simulation data is crucial for domain scientists to analyze complex phenomena, but it demands significant I/O bandwidth, storage, and computational resources. To enable effective visualization on local, low-end machines, recent advances in view synthesis techniques, such as neural radiance fields, utilize neural networks to generate novel visualizations for volumetric scenes. However, these methods focus on reconstruction quality rather than facilitating interactive visualization exploration, such as feature extraction and tracking. We introduce VolSegGS, a novel Gaussian splatting framework that supports interactive segmentation and tracking in dynamic volumetric scenes for \hot{exploratory visualization and analysis}. Our approach utilizes deformable 3D Gaussians to represent a dynamic volumetric scene, allowing for real-time novel view synthesis. For accurate segmentation, we leverage the view-independent colors of Gaussians for coarse-level segmentation and refine the results with an affinity field network for fine-level segmentation. Additionally, by embedding segmentation results within the Gaussians, we ensure that their deformation enables continuous tracking of segmented regions over time. We demonstrate the effectiveness of VolSegGS with several time-varying datasets and compare our solutions against state-of-the-art methods. With the ability to interact with a dynamic scene in real time and provide flexible segmentation and tracking capabilities, VolSegGS offers a powerful solution under low computational demands. This framework unlocks exciting new possibilities for time-varying volumetric data analysis and visualization.}


\teaser{
  \centering
  \includegraphics[width=0.9\linewidth]{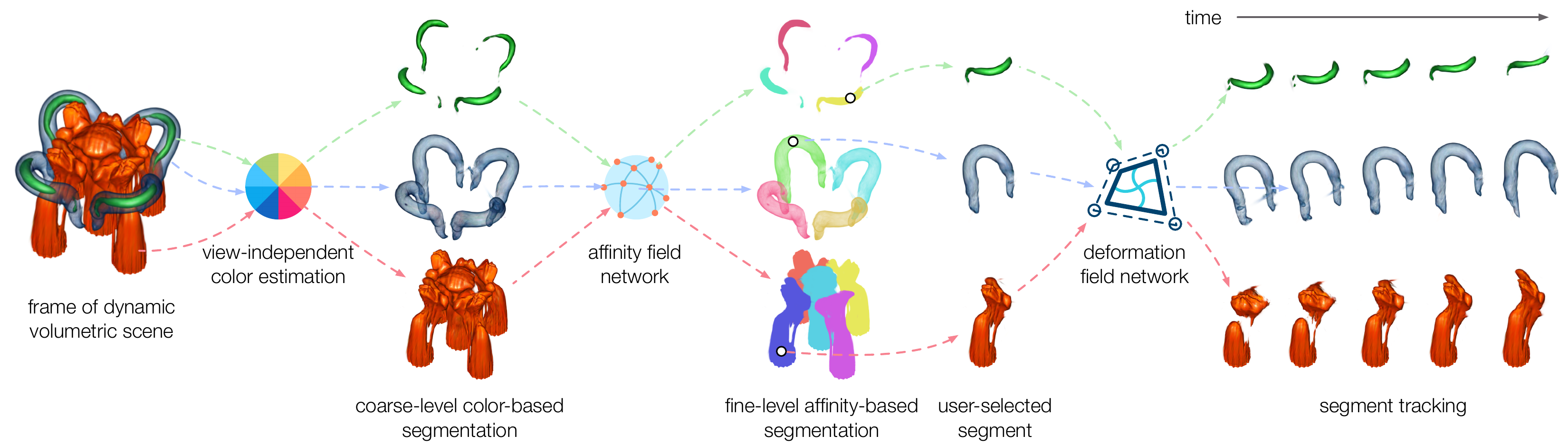}
  \vspace{-0.1in}
  \caption{Overview of VolSegGS. Deformable 3D Gaussians are learned to represent a dynamic scene. Segmentation is then performed in two stages: (1) coarse-level segmentation based on approximate view-independent colors of the Gaussians and (2) fine-level segmentation leveraging an affinity field network. Tracking over the dynamic scene is achieved using a deformation field network.}
  \label{fig:overview}
}

\graphicspath{{figs/}{figures/}{pictures/}{images/}{./}} % where to search for the images

\usepackage{tabu}                      % only used for the table example
\usepackage{booktabs}                  % only used for the table example
\usepackage{lipsum}                    % used to generate placeholder text
\usepackage{mwe}                       % used to generate placeholder figures

\usepackage{mathptmx}                  % use matching math font

\usepackage{graphicx}
\usepackage{times}
\usepackage{algorithm}
\usepackage{algorithmic}
\usepackage{amsfonts}
\usepackage{booktabs}         % only used for the table example
\usepackage{gensymb}
\usepackage{amsmath}
\usepackage{comment}
\usepackage{times} % we use Times as the main font
\usepackage{caption}
\usepackage{bm}
\usepackage{multirow}
\usepackage{color} %ref: http://en.wikibooks.org/wiki/LaTeX/Colors
\usepackage{anyfontsize}
\usepackage{soul}
\usepackage[cal=cm]{mathalfa}

\usepackage{arydshln}

\DeclareMathOperator*{\tv}{TV}

\DeclareMathOperator*{\ssim}{SSIM}
\DeclareMathOperator*{\dssim}{DSSIM}
\DeclareMathOperator*{\feat}{feat}

\DeclareMathOperator*{\ltwo}{L2}

\newcommand{\hot}[1]{{\color{black} #1}}

\newenvironment{myitemize}{
\begin{itemize}
 \setlength{\itemsep}{1pt}
 \setlength{\parskip}{0pt}
 \setlength{\parsep}{0pt}}{\end{itemize}
}

\begin{document}

\firstsection{Introduction}
\maketitle

% {\em direct volume rendering} (DVR)\\
% {\em transfer function} (TF)\\
% {\em neural radiance field} (NeRF)\\
% {\em Gaussian splatting} (GS)

Domain scientists across scientific and engineering disciplines often model complex phenomena through large-scale simulations. 
They typically run these simulations on high-performance computing resources to analyze time-dependent processes. 
As a result, these simulations generate vast amounts of high-resolution raw volumetric data over hundreds or even thousands of timesteps.
Standard {\em direct volume rendering} (DVR) techniques require access to these volumetric data during rendering. 
The large data size demands significant I/O bandwidth, storage, and computational power for visualization and analysis. 
This poses considerable challenges, making it difficult for domain scientists to perform these tasks on standard desktop computers.

In DL4SciVis~\cite{Wang-TVCG}, recent advances in {\em visualization generation}~\cite{Berger-TFGAN-TVCG19, Hong-DNN-VolVis-PVIS19, He-InSituNet-TVCG20, Han-CoordNet-TVCG} aim to tackle these challenges by enabling {\em novel view synthesis} (NVS) of volumetric scenes using multi-view 2D rendering images instead of 3D volumetric data. 
These methods go beyond {\em data generation}~\cite{Han-VIS20, Gu-CGA21, Han-CG22, Han-VI22, Yao-CG23, Gu-PVIS22, Tang-CG24} and {\em neural compression}~\cite{Tang-PVIS24, Gu-CG23, YF-Lu-VISSP24, Yang-PVISVN25, Son-VISSP25}. 
They significantly reduce data transfer and storage requirements by ensuring the solutions remain independent of the underlying data resolution. 
For instance, InSituNet~\cite{He-InSituNet-TVCG20} and CoordNet~\cite{Han-CoordNet-TVCG} synthesize high-quality visualization images by training deep neural networks on 2D rendering images. 
However, using generative and global networks leads to slow training, making these solutions less desirable for interactive applications. 
Furthermore, these works rely on interpolating 2D images for synthesizing novel views without incorporating 3D awareness, yielding subpar-quality synthesized visualizations. 

In contrast, a more recent work of ViSNeRF~\cite{Yao-PVIS25} utilizes {\em neural radiance fields} (NeRF)~\cite{Mildenhall-NeRF-ECCV20} to represent dynamic volumetric scenes. 
Leveraging the factorization techniques from TensoRF~\cite{Chen-TensoRF-ECCV22}, ViSNeRF requires a small number of training images, achieves fast training, and produces synthesized results with superior quality. 
Nevertheless, ViSNeRF still falls short in rendering speeds due to the required DVR computations inherent in NeRF.

Another key limitation of these visualization generation techniques is that, in the absence of the original volumetric data and the corresponding {\em transfer function} (TF), they do not support altering the appearance of visualizations at runtime. 
This restriction hampers the flexibility to adjust visualizations and examine specific regions in detail. 
StyleRF-VolVis~\cite{Tang-VIS24} addresses this limitation by enabling color-based content segmentation of a volumetric scene learned by NeRF to support downstream targeted appearance editing. 
Nevertheless, this solution is limited to static 3D scenes, lacks real-time rendering for certain style edits, and restricts segmentation to color-based regions without refinement for precise segmentation.

To enhance the performance and flexibility in visualization generation and provide the previously unavailable capability of segment extraction and tracking, we propose VolSegGS, {\bf Vol}umetric scene {\bf Seg}mentation based on {\bf G}aussian {\bf S}platting (GS). 
As an alternative solution for NeRF, the revival of direct projection of Gaussian splats for volume visualization through the tremendous success of 3DGS~\cite{Kerbl-TOG23} points out a promising direction for real-time NVS. 
Our VolSegGS is a novel framework that utilizes {\em deformable} 3D Gaussians for {\em dynamic} volumetric scene representation, offering new 3D segmentation and tracking capabilities. 
The primary objective is to retain the dynamic scene representation of ViSNeRF while achieving real-time rendering. 
Specifically, instead of using ViSNeRF to store the densities and colors of a dynamic scene, we leverage its architecture to create an efficient {\em deformation field network} to capture the deformation of 3D Gaussians. 
This allows us to represent the dynamic volumetric scene while maintaining interactive rendering through the rasterization of 3D Gaussians, avoiding the computationally intensive DVR process embedded in NeRF.

Furthermore, we propose a {\em two-level segmentation} strategy for flexible segmentation of the 3D volumetric scene. 
First, the {\em coarse-level} segmentation is based on the colors of 3D Gaussians. 
Second, the {\em fine-level} segmentation utilizes an {\em affinity field network} trained with 2D masks generated by the {\em segment anything model} (SAM)~\cite{Kirillov-SAM-ICCV23}. 
Users can select a segment of interest at an appropriate level and continuously track it throughout the dynamic scene. 
This is possible because our segmentation is performed directly on the Gaussians. 
Therefore, the segmented regions naturally follow temporal deformations of 3D Gaussians, ensuring consistent tracking over time. 

Our VolSegGS framework provides a unified solution for segmentation, tracking, and rendering, empowering the exploration of a dynamic volumetric scene with greater efficiency and flexibility. 
To summarize, our contributions are as follows: 
\begin{myitemize} 
\vspace{-0.025in}
\item VolSegGS employs deformable 3D Gaussians to represent dynamic volumetric scenes learned from multi-view 2D volume rendering images. This eliminates the need for the original 3D volume and enables real-time \hot{exploratory visualization}. %dynamic visualization exploration.
\item We propose a two-level segmentation strategy for flexible volumetric scene segmentation and utilize the deformation of 3D Gaussians to continuously track arbitrary segments in real time.
\item We evaluate VolSegGS against state-of-the-art methods for dynamic scene representation and 3D segmentation, demonstrating its effectiveness in real-time NVS and 3D segmentation across multiple time-varying datasets. 
\item \hot{We showcase the real-time segment tracking capabilities of the complete VolSegGS framework across various dynamic scenarios, supported by quantitative evaluations that highlight its accuracy and consistency.}
\vspace{-0.05in}
\end{myitemize}

\vspace{-0.05in}
\section{Related work}

{\bf Visualization generation.}
Visualizing large-scale, time-dependent simulation data demands significant I/O bandwidth, storage, and computational resources. 
To mitigate these challenges, deep learning techniques have been developed to generate visualizations directly via neural networks. 
These methods facilitate TF optimization~\cite{Berger-TFGAN-TVCG19}, rendering effects adjustment~\cite{Hong-DNN-VolVis-PVIS19}, image super-resolution~\cite{Weiss-SRNet-TVCG21, Weiss-LearningAdaptive-TVCG22, Bauer-FoVolNet-TVCG23}, NVS~\cite{Han-CoordNet-TVCG}, and parameter space exploration~\cite{He-InSituNet-TVCG20, Shi-VDL-TVCG22}, all without requiring access to the original volumetric data.
Recent advances have introduced neural representations of volumetric data and 3D scenes via {\em scene representation networks} (SRNs)~\cite{Weiss-CGF22, Wurster-TVCG24, Wu-TVCG24}, enabling high-fidelity visualization with strong 3D consistency while greatly reducing storage and I/O costs.

Building upon the SRN framework, VolSegGS leverages deformable 3D Gaussians to represent dynamic volumetric scenes. 
It is computationally efficient compared to existing approaches and enables real-time NVS, facilitating interactive and dynamic visualization exploration.

{\bf Novel view synthesis.}
For NVS, NeRF~\cite{Mildenhall-NeRF-ECCV20} has gained significant attention for its ability to reconstruct high-fidelity 3D scenes from 2D images.
NeRF represents a 3D scene as a continuous function, utilizing a {\em multi-layer perceptron} (MLP) to map spatial coordinates to density and color values.
However, NeRF suffers from slow training and rendering, primarily due to its implicit representation and the computational cost of DVR.
In response, a series of follow-up works~\cite{Yu-PlenOctrees-ICCV21, Fridovich-Keil-Plenoxels-CVPR22, Chen-TensoRF-ECCV22, Thomas-InstantNGP} have introduced explicit feature grids, either alone or combined with lightweight MLPs, to enhance 3D scene representation efficiency.
While these approaches have reduced training times to minutes, they still fail to achieve truly interactive rendering. 
A more recent advance, 3DGS~\cite{Kerbl-TOG23}, represents 3D scenes using 3D Gaussians and replaces DVR with efficient rasterization, enabling real-time rendering without relying on neural networks.
Although these methods have been limited to static scenes, numerous studies~\cite{Park-Nerfies-ICCV21, Pumarola-D-NeRF-CVPR21, Wu-CVPR24, Yang-CVPR24, Yang-ICLR24} have explored deformable radiance fields to model temporal changes in dynamic scenes.

In volume visualization, researchers have leveraged NeRF and GS for high-fidelity visualization.
For instance, ViSNeRF~\cite{Yao-PVIS25} employs a multi-dimensional NeRF to facilitate flexible visualization synthesis and parameter exploration. 
ReVolVE~\cite{Yao-CG25} reconstructs volumes from multi-view training images for visualization enhancement. 
StyleRF-VolVis~\cite{Tang-VIS24} applies NeRF for expressive volumetric stylization, and iVR-GS~\cite{Tang-PVIS25} and TexGS-VolVis~\cite{Tang-VIS25} utilize editable GS for inverse volume rendering and stylization. 

VolSegGS adopts deformable 3D Gaussians to represent dynamic volumetric scenes.
Unlike ViSNeRF, which also handles dynamic scenes, VolSegGS achieves real-time rendering while enabling temporal tracking of 3D segments via Gaussian deformation.
Additionally, incorporating ViSNeRF's hybrid representation for the deformation field, VolSegGS improves reconstruction quality and training efficiency over prior deformable radiance field methods.

{\bf 3D segmentation.}
In conventional volume visualization using DVR, TFs map voxel intensities to color and opacity, effectively serving as a basic form of 3D segmentation.
Building on this concept, traditional methods~\cite{Huang-RGVis-PG03, Tzeng-HiDimCla-TVCG05, Ip-HistSeg-TVCG12, Soundararajan-LPTF-CGF15, Ma-FeatCla-TVCG18, Quan-H3DCSC-TVCG18} extract high-dimensional features from volume data to enable more advanced classification using multi-dimensional TFs.
However, handling high-dimensional features for large-scale volume data is impractical on local machines, and manually designing multi-dimensional TFs is both complex and time-consuming.
Another research direction~\cite{Cicek-3DUNet-MICCAI16, Toubal-AIPR20} applies deep learning for semantic segmentation of 3D volume data.
While these methods produce expert-quality results, their reliance on manual annotations makes them costly in both time and human effort.

A more recent trend leverages 2D-based foundation segmentation models, performing slice-by-slice segmentation on volume data.
For example, MedSAM~\cite{Ma-MedSAM-NC24} applies the {\em segment anything model} (SAM)~\cite{Kirillov-SAM-ICCV23} to segment medical images from CT and MRI scans.
Although foundation models reduce training costs, slice-by-slice processing remains inherently inefficient.
With the emergence of NeRF and 3DGS, researchers have explored more diverse and efficient approaches for integrating 2D foundation models into 3D segmentation.
These approaches include: 
transforming multi-view 2D masks into 3D masks~\cite{Cen-NIPS23, Chen-arXiv23, Hu-arXiv24}, 
distilling semantic knowledge from foundation models into 3D representations~\cite{Kobayashi-NIPS22, Kerr-LERF-ICCV23, Goel-CVPR23}, and training affinity fields with 2D mask supervision from foundation models~\cite{Cen-arXiv23, Kim-CVPR24, Ye-ECCV24, Choi-ECCV24}.

VolSegGS introduces a two-level strategy for flexible 3D segmentation of volumetric scenes: 
(1) coarse-level segmentation relies on the color attributes of 3D Gaussians, aligning with TF-based classification; 
(2) fine-level segmentation employs an affinity field network trained using 2D masks generated by SAM, allowing foundation models to be efficiently integrated into 3D segmentation. 
This design ensures compatibility with traditional TF-based segmentation methods while leveraging the strengths of modern foundation models.
The affinity field enables multi-scale scene decomposition with a single training pass, making it an efficient alternative to traditional segmentation approaches.
Moreover, semantic priors learned from real-world data may not generalize well to simulated datasets, making knowledge distillation approaches less effective for scientific visualization, further highlighting the advantages of our affinity field-based approach.

{\bf Segment tracking.}
Traditional approaches track segments of interest in time-varying volumetric data by explicitly comparing geometric features across consecutive timesteps.
Early methods~\cite{Silver-TVCG97, Ji-VIS03, Muelder-PVIS09, Dutta-TVCG16, Schnorr-TVCG20} use algorithms that match overlapping regions with similar characteristics between adjacent timesteps.
While these methods achieve accurate tracking, they require each target segment to be processed separately across all timesteps, leading to significant inefficiencies in practical applications.
Another class of approaches~\cite{Widanagamaachchi-LDAV12, Saikia-CGF17} leverages merge trees to capture the hierarchical topological structure of volume data.
By comparing merge trees and constructing connection graphs, these methods track regions defined by the merge trees.
While this approach enables global tracking, computing merge trees for every timestep remains computationally expensive.

Unlike previous methods, VolSegGS models the deformation of 3D Gaussians to capture temporal variations in dynamic volumetric scenes.
This allows for real-time visualization exploration and segment tracking at any point in time, including unseen intermediate timesteps.

\begin{comment}
\begin{figure*}[ht]
%\vspace{-0.1in}
  \centering
  \includegraphics[width=0.9\linewidth]{figs/overview.pdf}
  \vspace{-0.1in}
  \caption{Overview of VolSegGS. The process begins by learning deformable 3D Gaussians to represent a dynamic scene. Segmentation is then performed in two stages: (1) coarse-level segmentation based on approximate view-independent colors of the Gaussians and (2) fine-level segmentation leveraging an affinity field network. Consistent tracking throughout the dynamic scene is achieved using a deformation field network.}
  \label{fig:overview}
\end{figure*}
\end{comment}

\vspace{-0.05in}
\section{VolSegGS}

While methods like InSituNet~\cite{He-InSituNet-TVCG20} and ViSNeRF~\cite{Yao-PVIS25} have successfully enabled the exploration of dynamic visualization scenes without requiring the original volume data, VolSegGS extends this capability by integrating robust 3D segmentation techniques and enabling real-time tracking and exploration of segmented regions across a dynamic scene. 

As shown in Figure~\ref{fig:overview}, VolSegGS begins by optimizing deformable 3D Gaussians (Sections~\ref{subsec:3dgs}~and~\ref{subsec:4dgs}) to represent the dynamic scene. 
Users can then select a scene frame at any desired timestep for 3D segmentation, which is applied to the corresponding deformed Gaussians.
The segmentation process consists of two key stages:
(1) coarse-level color-based segmentation (Section~\ref{subsec:color-seg}), where Gaussians are grouped based on their approximate view-independent colors; and 
(2) fine-level affinity-based segmentation (Section~\ref{subsec:feat-seg}), which refines the segmentation by learning affinity features from 2D masks generated by SAM.
Users choose a scene frame to pick a segment of interest for tracking. 
Since segmentation is performed directly on the Gaussians, the segmented regions naturally follow Gaussian deformations, ensuring consistent tracking throughout the dynamic scene.

\vspace{-0.05in}
\subsection{3D Gaussian Splatting}
\label{subsec:3dgs}

3DGS~\cite{Kerbl-TOG23} is a highly efficient alternative to NeRF~\cite{Mildenhall-NeRF-ECCV20} for representing 3D scenes.
Unlike NeRF, which relies on neural network-based implicit representations, 3DGS employs a rasterization-based approach with a fully explicit representation, enabling real-time rendering.
The Gaussian function $G({\mathbf x})$ at a spatial position ${\mathbf x}$ is defined as
\vspace{-0.05in}
\begin{equation}
      G({\mathbf x}) = \exp\left(-\frac{1}{2}({\mathbf x}-{\bm \mu})^T\Sigma^{-1}({\mathbf x}-{\bm \mu})\right),
      \label{eqn:3d-gaussian}
    \vspace{-0.05in}
\end{equation}
where ${\bm \mu}$ represents the spatial mean, and $\Sigma$ is the covariance matrix, which encodes the anisotropic shape of the Gaussian.
The covariance matrix $\Sigma$ can be decomposed as
\vspace{-0.05in}
\begin{equation}
      \Sigma = {\mathbf R}{\mathbf S}{\mathbf S}^T{\mathbf R}^T, 
      \label{eqn:covariance-matrix}
    \vspace{-0.05in}
\end{equation}
where ${\mathbf R}$ is the rotation matrix and ${\mathbf S}$ is the scaling matrix. 
To facilitate separate optimization of these factors, we parameterize the scaling vector ${\mathbf s}$ and the rotation quaternion ${\mathbf r}$ as independent Gaussian attributes rather than directly optimizing the covariance matrix $\Sigma$.

During rendering, 3D Gaussians are projected onto the 2D image plane.
This requires transforming the 3D covariance matrix $\Sigma$ into a 2D covariance matrix $\Sigma'$, defined as
\vspace{-0.05in}
\begin{equation}
      \Sigma' = {\mathbf J}{\mathbf W}\Sigma {\mathbf W}^T {\mathbf J}^T, 
      \label{eqn:covariance-projection}
    \vspace{-0.05in}
\end{equation}
where ${\mathbf W}$ is the viewing transformation matrix, and ${\mathbf J}$ is the Jacobian matrix of the affine approximation of the projective transformation.

The final pixel color ${\mathbf C}$ in the 2D rendered image is determined by blending $N$ overlapping Gaussians in order, computed as
\vspace{-0.05in}
\begin{equation}
      \mathbf{C} = \sum_{i\in N}{{\mathbf c}_i}{\alpha_i} \prod^{i-1}_{j=1}{(1-\alpha_j)}, 
      \label{eqn:pixel-color}
    \vspace{-0.05in}
\end{equation}
where ${\mathbf c}_i$ is the view-dependent color of the $i$-th Gaussian, parameterized by {\em spherical harmonic} (SH) coefficients. 
The opacity term $\alpha_i$ is computed as a weighted product of the Gaussian's covariance matrix $\Sigma$ and its intrinsic opacity $o$.
\hot{
Note that in volume visualization, view-dependent color is produced during the shading process in DVR. 
In this work, we employ Blinn-Phong shading with ambient, diffuse, and specular components to enhance perceptual clarity. 
While specular reflection is inherently view-dependent, our use of headlight illumination also introduces view dependency into the diffuse component. 
In DVR volumetric scenes, the opacity attribute of Gaussians plays a critical role in representing the scene's semi-transparency, enabling the visualization of overlapping structures.
}
As a result, each 3D Gaussian at a spatial position ${\mathbf x}$ is parameterized by five learnable attributes: (${\bm \mu}$, ${\mathbf r}$, ${\mathbf s}$, ${\mathbf c}$, $o$), representing spatial mean, rotation quaternion, scaling vector, view-dependent color, and opacity. 

\vspace{-0.05in}
\subsection{Deformable 3D Gaussians for Dynamic Scene}
\vspace{-0.025in}
\label{subsec:4dgs}

Recent works~\cite{Yang-ICLR24, Luiten-3DV24, Yang-CVPR24, Wu-CVPR24} have extended 3DGS to represent dynamic scenes by introducing time-dependent modifications to Gaussian attributes. 
These modifications, often called {\em deformations}, allow 3D Gaussians to adapt over time, leading to {\em deformable 3D Gaussians}, which incorporate time-varying attributes. 
Here, we present the formulation of deformable 3D Gaussians as used in VolSegGS.

{\bf Deformable 3D Gaussian.}
We define the deformation of a canonical 3D Gaussian $G$ as $\Delta G$, i.e., $\Delta G_t = G_t - G$, where $G_t$ represents the deformed 3D Gaussian at timestep $t$. 
The deformation $\Delta G$ captures changes in mean position $\Delta {\bm \mu}$, rotation $\Delta {\mathbf r}$, scaling $\Delta {\mathbf s}$, and opacity $\Delta o$. 
In particular, geometric deformations—including $\Delta \bm \mu$, $\Delta {\mathbf r}$, and $\Delta {\mathbf s}$—model changes in the geometry of visible regions in a dynamic scene.
Meanwhile, opacity deformation $\Delta o$ captures the appearance and disappearance of scene regions over time.
The color ${\mathbf c}$ remains unchanged to ensure consistent coarse-level segment tracking. 
By predicting $\Delta G_t$, we obtain the deformed Gaussian $G_t$ with attributes ($\bm \mu_t$, ${\mathbf r}_t$, ${\mathbf s}_t$, ${\mathbf c}$, $o_t$).
These updated Gaussians $G_t$ are then used to render the scene frame at timestep $t$.

\begin{figure}[ht]
\vspace{-0.1in}
  \centering
  \includegraphics[width=0.9\columnwidth]{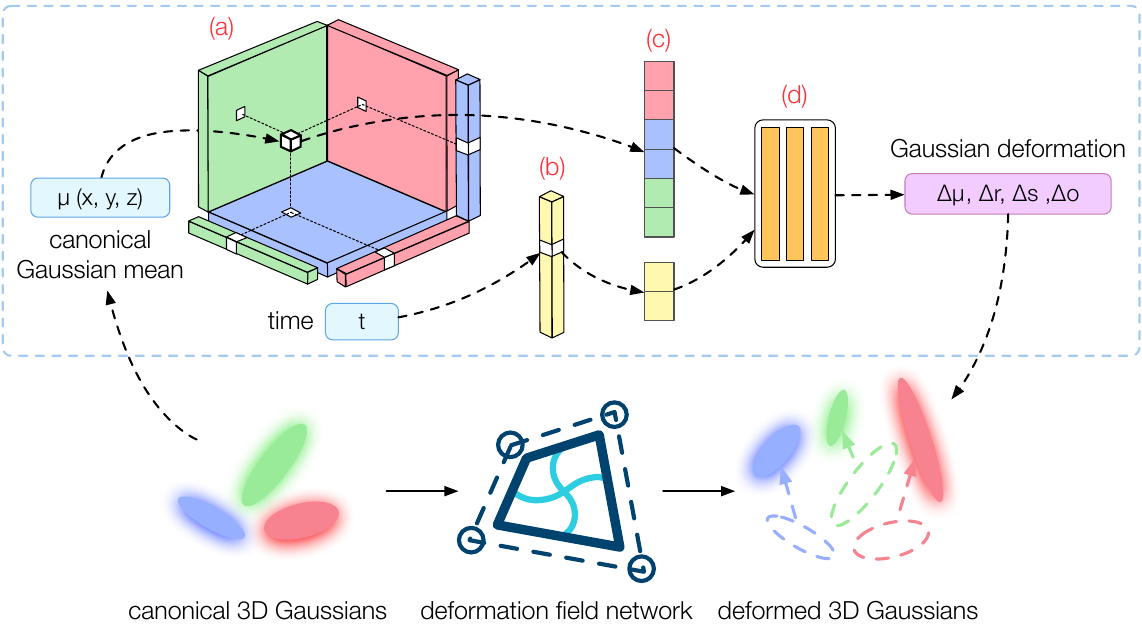}
  \vspace{-0.1in}
  \caption{The deformation field network takes the mean positions of 3D Gaussians as input and outputs their deformations. 
It features an explicit encoder structure consisting of (a) spatial feature planes and vectors, as well as (b) a temporal feature vector. 
The features are sampled from these planes and vectors, then (c) concatenated and fed into (d) a lightweight MLP decoder.}
  \label{fig:deform-net}
\end{figure}

\begin{figure}[htb]
\vspace{-0.1in}
\begin{center}
$\begin{array}{c@{\hspace{0.05in}}c@{\hspace{0.05in}}c}
    \includegraphics[width=0.3\linewidth]{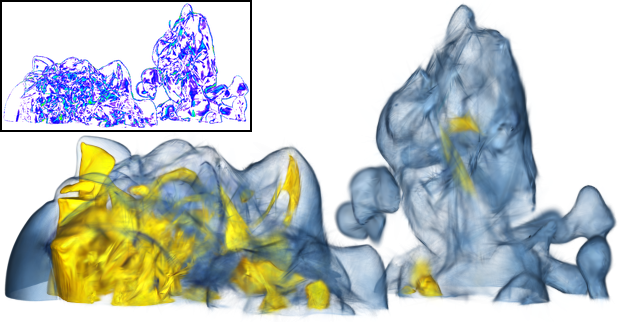}&
    \includegraphics[width=0.3\linewidth]{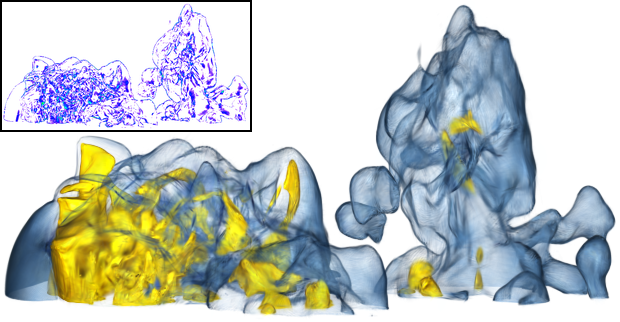}&
    \includegraphics[width=0.3\linewidth]{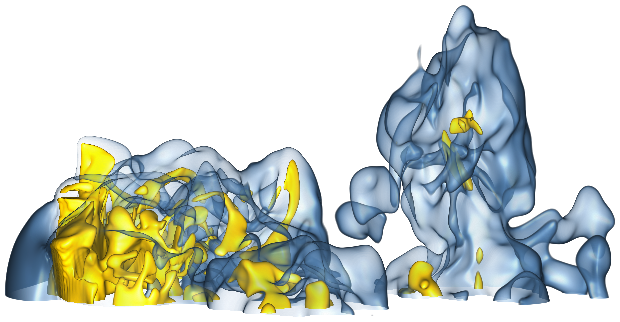}\\
    \mbox{\footnotesize (a) fully implicit} & \mbox{\footnotesize (b) hybrid} & \mbox{\footnotesize (c) GT}
\end{array}$
\end{center}
\vspace{-.25in} 
\caption{Comparison of deformation field network encoder structures via rendered images of the Tangaroa dataset. The corner images highlight the perceptible pixel-wise differences, with colors ranging from purple to green to red, corresponding to low, medium, and high differences.} 
\label{fig:deform-hybrid}
\end{figure}

{\bf Deformation field network.}
Following~\cite{Yang-CVPR24, Wu-CVPR24}, for a set of canonical 3D Gaussians $\mathcal G$, we predict the global deformation $\Delta \mathcal G$ using a {\em deformation field network} ${\mathcal F}$, such that $\Delta \mathcal G_t = {\mathcal F}(\mathcal G, t)$ at timestep $t$.
Inspired by ViSNeRF~\cite{Yao-PVIS25}, as shown in Figure~\ref{fig:deform-net}, our ${\mathcal F}$ adopts a hybrid architecture, integrating an explicit spatiotemporal structure encoder ${\mathcal H}$ and a lightweight MLP decoder ${\mathcal D}$.
In the encoder ${\mathcal H}$, the explicit 4D feature tensor ${\mathcal T^4}\in {\mathbb R}^{XYZT}$ is decomposed into three spatial feature matrices (${\mathbf M}^{XY}$, ${\mathbf M}^{XZ}$, and ${\mathbf M}^{YZ}$), three spatial feature vectors (${\mathbf v}^{X}$, ${\mathbf v}^{Y}$, and ${\mathbf v}^{Z}$), and one temporal feature vector (${\mathbf v}^{T}$).
This decomposition significantly reduces memory consumption while preserving expressiveness. 
The formulation is expressed as
\vspace{-0.05in}
\begin{align}
{\mathcal T}^4 =& {\mathcal T}^3 \circ {\mathcal T}^1\\
                =& \Bigl(\sum^{R_s}_{r=1} {\mathbf M}^{XY}_{r} \circ {\mathbf v}^Z_r + {\mathbf M}^{XZ}_{r} \circ {\mathbf v}^Y_r + {\mathbf M}^{YZ}_{r} \circ {\mathbf v}^X_r \Bigr) \circ \sum^{R_t}_{r=1} {\mathbf v}^T_r,
\end{align}
where $R_s$ and $R_t$ are the numbers of low-rank and one-rank components in spatial and temporal space, respectively. 
If the spatial resolution is $N$ (i.e., $N=X=Y=Z$), and the temporal resolution is $T$, this decomposition reduces the complexity of the deformation field from $O(N^3T)$ to $O(R_sN^2+R_sN+R_tT)$.
The encoder ${\mathcal H}$ takes the means of $\mathcal G$ as input and outputs the sampled features in spatiotemporal space.
The decoder ${\mathcal D}$ takes these encoded spatiotemporal features as input and predicts the deformation $\Delta \mathcal G$.
As shown in Figure~\ref{fig:deform-hybrid}, the hybrid structure allows VolSegGS to generate higher-quality results with sharper edges and more precise structures than the fully implicit one. 

{\bf Optimization.}
Before learning the deformation of 3D Gaussians, we train the canonical 3D Gaussians $\mathcal G$ following the 3DGS framework~\cite{Kerbl-TOG23}, without incorporating temporal information.
To optimize the Gaussians, we minimize the L2 loss between the rendered image ${\hat I}$ at a training view and the ground truth (GT) image $I$
\vspace{-0.05in}
\begin{equation}
        {\mathcal L}_{\ltwo} = \|{\hat I} - I\|_2^2. 
        \label{eqn:loss-l2}
        \vspace{-0.05in}
\end{equation}
After warming up $\mathcal G$, we jointly optimize $\mathcal G$ and ${\mathcal F}$.
Since the encoder ${\mathcal H}$ maintains an explicit representation, we introduce {\em total variation} (TV) regularization to smooth the learned features.
The TV loss for a 2D matrix $\mathbf{M}$ and a 1D vector $\mathbf{v}$ is defined as
\vspace{-0.05in}
\begin{equation}
  \begin{array}{l}
    \mathcal{L}_{\tv} = \mathcal{L}_{\tv_{1}} + \mathcal{L}_{\tv_{2}}, \vspace{0.05in} \\
    \mathcal{L}_{\tv_{1}} = \sum_{\mathbf{v}\in \mathcal{V}} \sum_{i} \|\mathbf{v}_{i}-\mathbf{v}_{i-1}\|_2^2, \\
    \mathcal{L}_{\tv_{2}} = \sum_{\mathbf{M}\in \mathcal{M}} \sum_{i,j} \Bigl( \|\mathbf{M}_{i,j}-\mathbf{M}_{i-1,j}\|_2^2 + \|\mathbf{M}_{i,j}-\mathbf{M}_{i,j-1}\|_2^2 \Bigr),
    \label{eqn:loss-tv}
\vspace{-0.05in}
  \end{array}
\end{equation}
where $\mathcal{V}$ represents the set of vectors, and $\mathcal{M}$ is the set of matrices.

In practice, we observed that using L2 loss alone introduced irregular artifacts due to overfitting.
To mitigate the artifacts and smooth the rendering results, we incorporated an additional {\em structural dissimilarity} (DSSIM) loss, defined as
\vspace{-0.05in}
\begin{equation}
  {\mathcal L}_{\dssim} = 1 - {\ssim}({\hat I}, I), 
  \label{eqn:loss-dssim}
  \vspace{-0.05in}
\end{equation}
where ${\ssim}({\hat I}, I)$ is the {\em structural similarity index measure} (SSIM) between the rendered image ${\hat I}$ and the GT image $I$.

The joint optimization of $\mathcal G$ and $\mathcal F$ is guided by the full loss function
\vspace{-0.05in}
\begin{equation}
  {\mathcal L} = {\mathcal L}_{\ltwo} + \lambda_1{\mathcal L}_{\tv} +  \lambda_2{\mathcal L}_{\dssim}, 
  \label{eqn:loss-full}
  \vspace{-0.05in}
\end{equation}
where $\lambda_1$ and $\lambda_2$ are weights that balance the contributions of TV and DSSIM losses.
In the appendix, we provide a detailed analysis of the loss functions and their impact on the performance of VolSegGS.

\vspace{-0.05in}
\subsection{Coarse-Level Color-Based Segmentation}
\label{subsec:color-seg}

In scientific visualization, DVR uses a TF to map scalar voxel values to corresponding colors and opacities, enhancing the contrast between different components in a 3D dataset. 
However, because DVR consists of overlapping semi-transparent layers, distinguishing individual components in a 2D rendering after compositing becomes challenging.
By reconstructing a 3D scene from 2D images using 3D Gaussians, we can identify distinct, coarse-level components based on the color attributes of the Gaussians. 
For simplicity, we assume that Gaussian colors remain time-invariant, ensuring consistent color-based segmentation across a dynamic scene.

\begin{figure}[htb]
%\vspace{-0.1in}
\begin{center}
$\begin{array}{c@{\hspace{0.05in}}c@{\hspace{0.05in}}c}
    \includegraphics[width=0.3\linewidth]{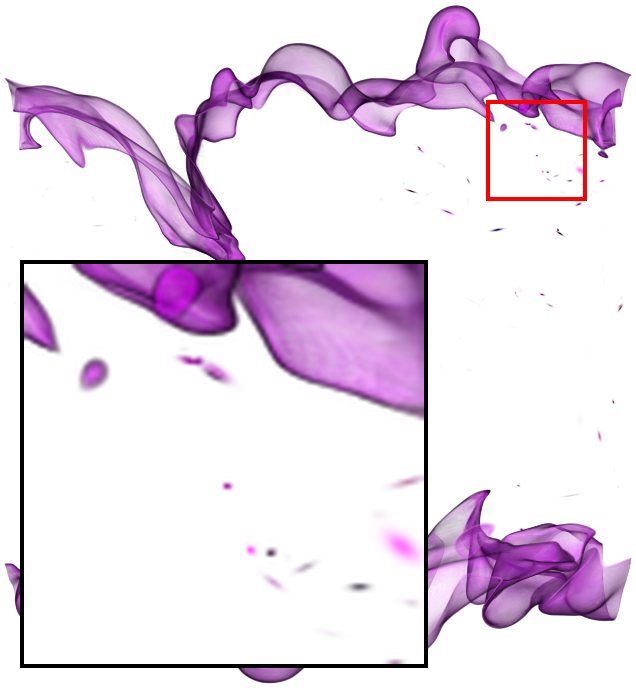}&
    \includegraphics[width=0.3\linewidth]{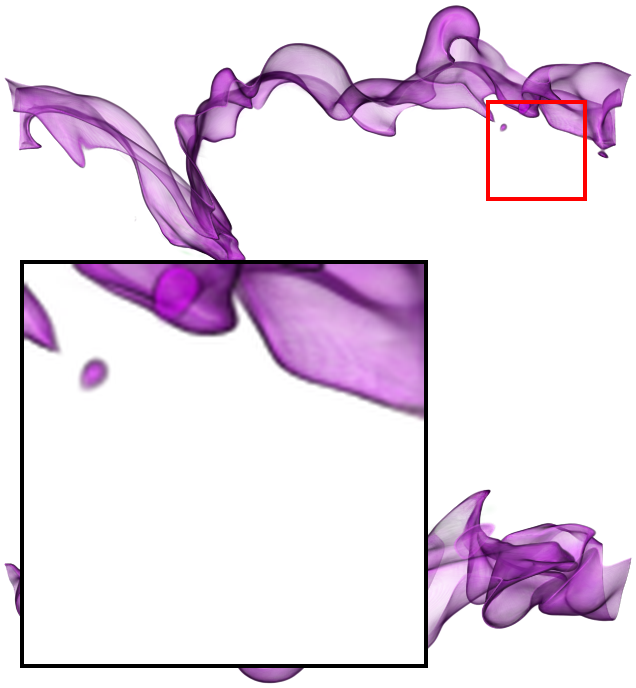}&
    \includegraphics[width=0.3\linewidth]{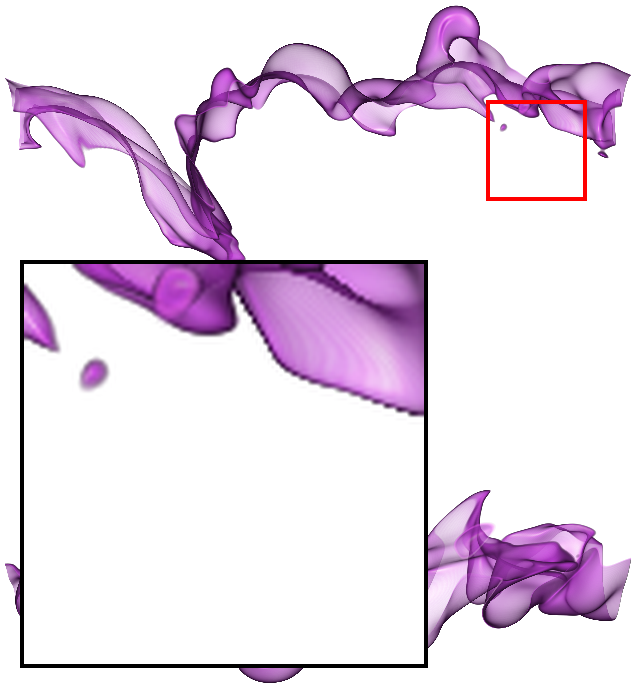}\\
    \mbox{\footnotesize (a) before removal} & \mbox{\footnotesize (b) after removal} & \mbox{\footnotesize (c) GT}
\end{array}$
\end{center}
\vspace{-.25in} 
\caption{Outlier removal using the combustion dataset shows the purple segment after coarse-level segmentation.} 
\label{fig:outlier-removal}
\end{figure}

{\bf View-independent color approximation.}
Due to DVR shading and lighting effects, a Gaussian's color ${\mathbf c}$ may vary depending on the viewing direction ${\mathbf d}$.
Thus, before clustering Gaussians, we approximate the {\em view-independent} color $\tilde{\mathbf c}$ by averaging the {\em view-dependent} colors ${\mathbf c}_{\mathbf d}$ across all sampled viewing directions ${\mathbf D}$
\vspace{-0.05in}
\begin{equation}
  \tilde{\mathbf c} \approx \frac{1}{\|{\mathbf D}\|}\sum_{{\mathbf d}\in {\mathbf D}}{{\mathbf c}_{\mathbf d}}. 
  \label{eqn:view-independent-color}
\vspace{-0.05in}
\end{equation}
We then apply a clustering algorithm, such as k-means, to group Gaussians based on their approximate view-independent colors. 
Color differences are measured using Euclidean distance in the RGB color space, and the cluster centroids serve as the representative colors for different components.
Users can select components by choosing a representative color or clicking directly on the rendered image.

{\bf Outlier removal.}
In practice, color-based segmentation can produce outliers due to the visualization's lighting effects or semi-transparent layers. 
When segments are viewed individually, these outliers become evident and can negatively affect the quality of 2D mask generation in the subsequent fine-level segmentation. 
To mitigate this issue, we employ an outlier removal technique. 
Specifically, we search for neighbors within a small-radius sphere for each Gaussian. 
The Gaussian is removed from the segmentation if the number of neighbors falls below a predefined threshold. 
Although simple, this technique effectively removes outliers and improves the overall quality of the segmentation results, as shown in Figure~\ref{fig:outlier-removal}. 

\vspace{-0.05in}
\subsection{Fine-Level Affinity-Based Segmentation}
\label{subsec:feat-seg}

Color-based segmentation identifies components in the 3D scene based on the colors of 3D Gaussians. 
However, it cannot distinguish fine-grained structures within components that share similar colors.
To address this limitation, we introduce fine-level affinity-based segmentation, which lifts 2D masks generated from SAM to a 3D affinity field, subdividing the coarse-level color-based segmentation results.

\begin{figure}[ht]
\vspace{-0.1in}
  \centering
  \includegraphics[width=0.9\columnwidth]{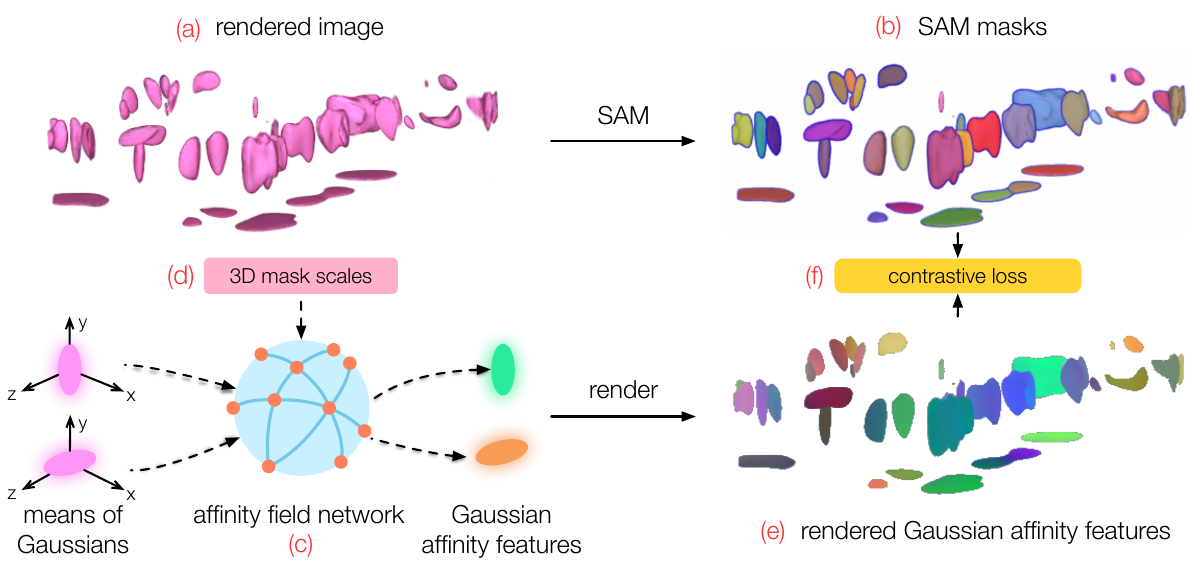}
  \vspace{-0.1in}
  \caption{Affinity field network. 
(a) A rendered view of the mantle dataset using VolSegGS.
(b) 2D masks generated by SAM.
(c) The affinity field network takes the means of the Gaussians and 3D mask scales as input to generate affinity features for the Gaussians.
(d) 3D mask scales control the segmentation granularity.
(e) Affinity features of the Gaussians rendered in the view.
(f) Contrastive loss used for optimizing the affinity field network.}
  \label{fig:affinity-net}
\end{figure}

\begin{figure}[htb]
\vspace{-0.1in}
\begin{center}
$\begin{array}{c@{\hspace{0.1in}}c@{\hspace{0.05in}}c@{\hspace{0.05in}}c}
    \includegraphics[width=0.22\linewidth]{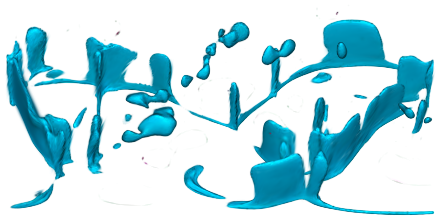}&
    \includegraphics[width=0.22\linewidth]{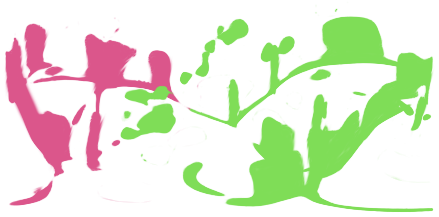}&
    \includegraphics[width=0.22\linewidth]{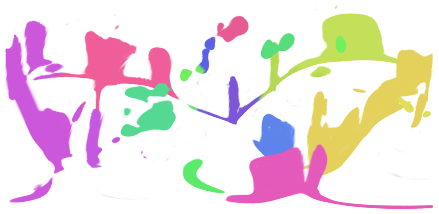}&
    \includegraphics[width=0.22\linewidth]{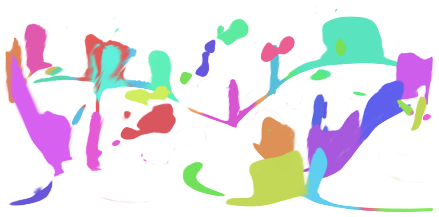}\\
    \includegraphics[width=0.22\linewidth]{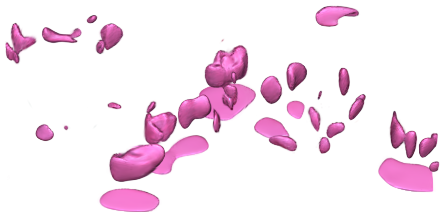}&
    \includegraphics[width=0.22\linewidth]{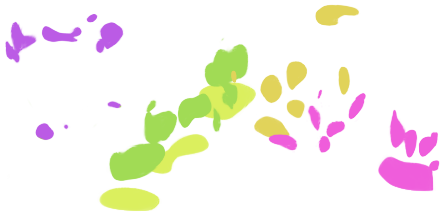}&
    \includegraphics[width=0.22\linewidth]{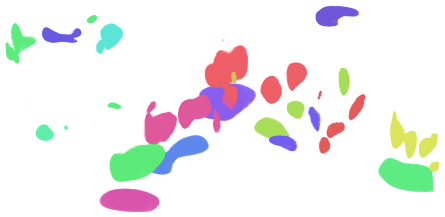}&
    \includegraphics[width=0.22\linewidth]{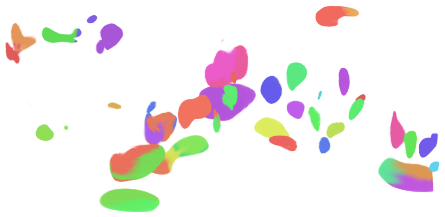}\\
    \mbox{\footnotesize (a) coarse segment} & \mbox{\footnotesize (b) large} & \mbox{\footnotesize (c) medium} & \mbox{\footnotesize (d) small}
\end{array}$
\end{center}
\vspace{-.25in} 
\caption{Multi-scale fine-level segmentation with various granularities using the mantle dataset. Top row: cyan segment. Bottom row: pink segment.} 
\label{fig:affinity-scale}
\end{figure}

{\bf Affinity field network.}
Recent works on Gaussian Grouping~\cite{Kim-CVPR24, Ye-ECCV24, Choi-ECCV24} have utilized feature fields to model the affinity between Gaussians, effectively grouping those that are closely related. 
Instead of directly optimizing the Gaussian affinity feature as an additional attribute, as shown in Figure~\ref{fig:affinity-net} (c), we employ a lightweight MLP as the {\em affinity field network}, which uses the mean of each Gaussian to query its affinity feature. 
This approach offers two key advantages: (1) the implicit MLP generates a smoother, more continuous affinity field, helping mitigate the impact of inconsistent masks across different views, and (2) the model remains compact, as its size is independent of the total number of Gaussians.
To offer hierarchical fine-level segmentation, the affinity field network takes an additional 3D mask scale input to control the granularity of segmentation, as indicated in Figure~\ref{fig:affinity-net} (d) and Figure~\ref{fig:affinity-scale}.

{\bf 2D mask generation.}
As illustrated in Figure~\ref{fig:affinity-net} (b), we utilize SAM~\cite{Kirillov-SAM-ICCV23} to generate 2D masks for training the affinity field network. 
First, users select a scene frame at timestep $t$ from the dynamic scene to perform fine-level segmentation on the deformed Gaussians $\mathcal{G}_t$. 
VolSegGS then renders each coarsely segmented region in the selected scene frame separately from multiple views. 
Finally, we use SAM to generate 2D masks from the rendered images.
We generate a grid of points for each image and use the SAM predictor to produce three candidate masks per point at different scales.
We select the mask with the highest confidence score and apply {\em non-maximum suppression} (NMS)~\cite{NMS} to refine overlapping masks.
As a result, each view produces a set of masks representing the segmented regions. 
To estimate the 3D scale for each mask, we identify the Gaussians within the mask and calculate the standard deviation of their means.
Due to the challenge of consistently registering masks across different views, we train the affinity field network on a per-view basis.

{\bf Optimization.}
To optimize the affinity field network, we first select a random training view and obtain all masks generated by SAM along with their corresponding 3D scales. 
Then, we use the affinity field network to query the affinity feature $\mathbf{f}$ of each deformed Gaussian $G_t$ at the user-selected timestep $t$, using the 3D scale of its corresponding SAM mask. 
Next, we render the affinity feature using the same blending process as described in Equation~\ref{eqn:pixel-color}, i.e.,
\vspace{-0.05in}
\begin{equation}
      {\mathbf F} = \sum_{i\in N}{{\mathbf f}_i}{\alpha_i} \prod^{i-1}_{j=1}{(1-\alpha_j)}.
      \label{eqn:pixel-feature}
    \vspace{-0.05in}
\end{equation}
As shown in Figure~\ref{fig:affinity-net} (f), we compute the {\em contrastive loss} ${\mathcal L}_{\feat}$ using the rendered affinity features of a view to optimize the affinity field network
\begin{equation}
  \vspace{-0.05in}
{\mathcal L}_{\feat} = \sum_{i \in N} \sum_{j \in N} \delta ({\mathbf F}_i, {\mathbf F}_j), \text{where}
\vspace{-0.05in}
\end{equation}

\begin{equation}
  \delta ({\mathbf F}_i, {\mathbf F}_j) =
	\begin{cases}
		1 - \langle {\mathbf F}_i, {\mathbf F}_j \rangle,	& \text{if in the same mask} \\
		\langle {\mathbf F}_i, {\mathbf F}_j \rangle,		& \text{otherwise} 
	\end{cases}
      \label{eqn:loss-feature}
    \vspace{-0.05in}
\end{equation}
where ${\mathbf F}_i$ and ${\mathbf F}_j$ are rendered affinity features.
$\langle {\mathbf F}_i, {\mathbf F}_j \rangle$ denote the cosine similarity between ${\mathbf F}_i$ and ${\mathbf F}_j$.
After training the affinity field network, VolSegGS can segment the deformed Gaussians at timestep $t$ by querying their affinity features at multiple scales, thereby producing hierarchical fine-level segmentation, as shown in Figure~\ref{fig:affinity-scale}.

\vspace{-0.2in}
\begin{table}[htb]
\caption{Datasets for visualization generation of dynamic scenes, with training images in PNG format.}
\vspace{-0.1in}
\centering
% {\scriptsize
\resizebox{\columnwidth}{!}{
\begin{tabular}{c|ccccc}
% dataset & volume & \# images  & \# sampled & total \# & image \\ 
% (scenario) & resolution & per scene  & timesteps & images & resolution\\ \hline
% five jets & 128$\times$128$\times$128 & 50 & 10 & 500 & 800$\times$800 \\ 
% vortex & 128$\times$128$\times$128 & 50 & 10 & 500 & 800$\times$800 \\
% mantle & 360$\times$201$\times$180 & 50 & 10 & 500 & 800$\times$800 \\
% Tangaroa & 300$\times$180$\times$120 & 50 & 10 & 500 & 800$\times$800 \\
 & volume & volume & \# timesteps  &  \# views per & training image \\ 
dataset & resolution & size (GB) & for training  & timestep  & size (MB) \\ \hline
five jets & 256$\times$256$\times$256$\times$100 & 6.25 & 10 & 40  & 79.35 \\ 
Tangaroa & 600$\times$360$\times$240$\times$100 & 19.31 & 20 & 30  & 145.22 \\
mantle & 720$\times$402$\times$360$\times$100 & 38.82 & 20 & 20  & 43.03 \\
vortex & 512$\times$512$\times$512$\times$100 & 50 & 30 & 20  & 169.66 \\
combustion & 960$\times$1440$\times$240$\times$100 & 123.59 & 30 & 30  & 430.69 \\
\end{tabular}
}
\label{tab:nvs-dataset}
\end{table}

\vspace{-0.1in}
\section{Results and Discussion}

This section presents qualitative and quantitative results for each stage of our framework, including visualization generation, 3D segmentation, and segment tracking. %We also compare VolSegGS with state-of-the-art baseline methods and existing segmentation and tracking techniques in volume visualization, highlighting its advantages and limitations. Additionally, we provide an ablation study and hyperparameter analysis in the appendix.

\vspace{-0.05in}
\subsection{Visualization Generation for Dynamic Scenes}
\label{sec:exp-nvs}

Since effective segmentation relies on high-quality images, we assess the visualization generation quality of VolSegGS for dynamic scenes.

{\bf Datasets.}
As shown in Table~\ref{tab:nvs-dataset}, VolSegGS is evaluated using dynamic visualization scenes from five time-varying volumetric datasets. 
We uniformly sample 100 consecutive intermediate timesteps from each dataset and render the scene frames for each timestep at a fixed image resolution of 800$\times$800. 
While all 100 timesteps are used to generate inference images for evaluating the quality of VolSegGS and baseline methods, a subset of timesteps is evenly sampled for generating training images. 
The number of sampled timesteps for training is adjusted based on the variation in speed of the dynamic volumetric scene. 
To ensure an even distribution of training views around the volume data, camera positions are determined using the spherical Fibonacci point set~\cite{Marques-CGF13}, which uses a Fibonacci spiral to create a nearly uniform arrangement without clustering. The test set is rendered using a spherical camera system with 181 views, arranged along a spiral path with increasing elevation and azimuth angles, capturing the changing scene frames across the 100 timesteps.

\begin{figure*}[htb]
%\vspace{-0.1in}
\begin{center}
$\begin{array}{c@{\hspace{0.05in}}c@{\hspace{0.05in}}c@{\hspace{0.05in}}c@{\hspace{0.05in}}c@{\hspace{0.05in}}c@{\hspace{0.05in}}c}
    \includegraphics[width=0.13\linewidth]{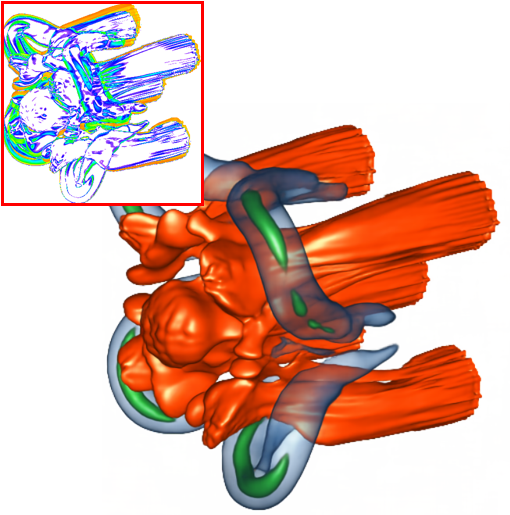}&
    \includegraphics[width=0.13\linewidth]{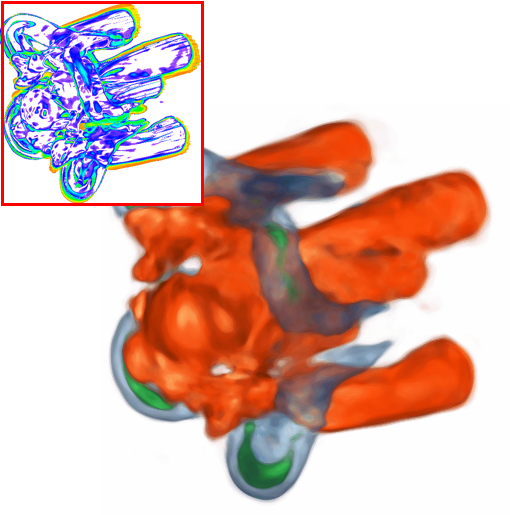}&
    \includegraphics[width=0.13\linewidth]{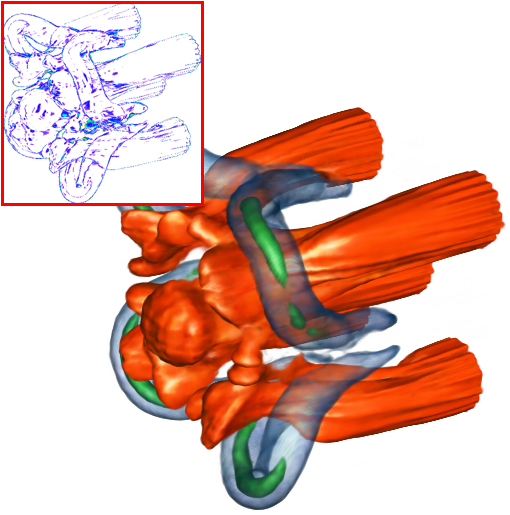}&
    \includegraphics[width=0.13\linewidth]{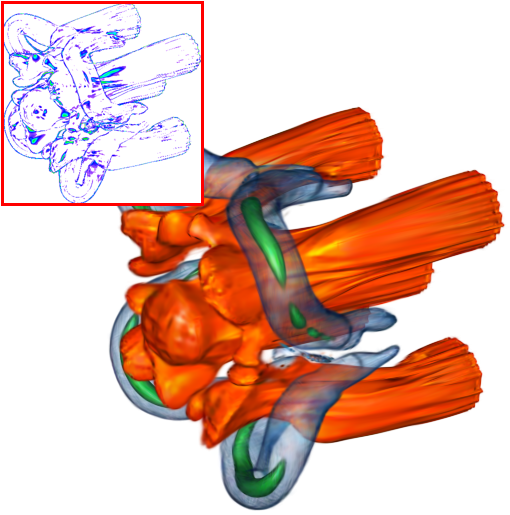}&
    \includegraphics[width=0.13\linewidth]{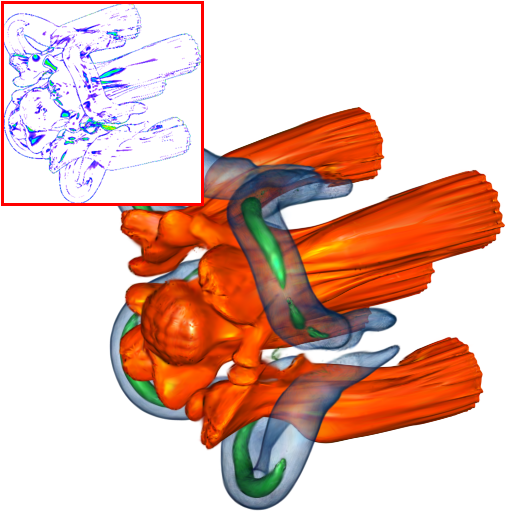}&
    \includegraphics[width=0.13\linewidth]{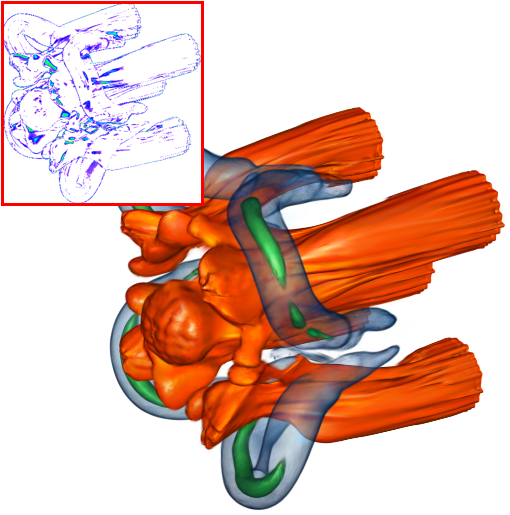}&
    \includegraphics[width=0.13\linewidth]{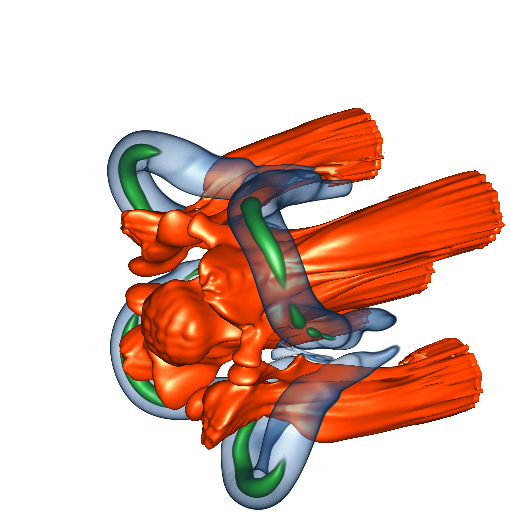}\\
    \includegraphics[width=0.13\linewidth]{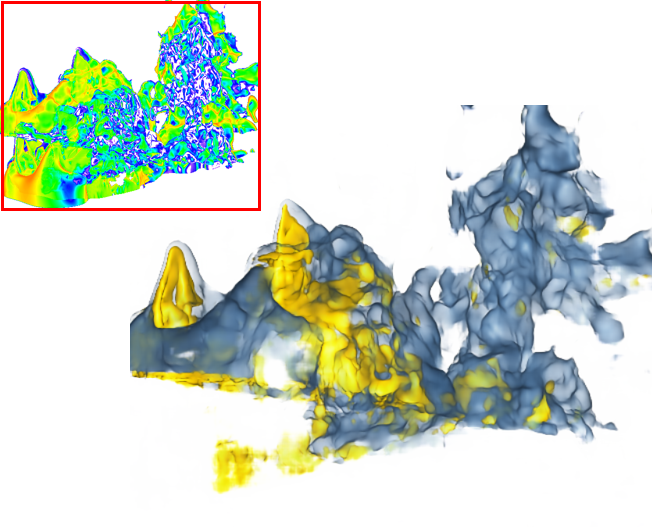}&
    \includegraphics[width=0.13\linewidth]{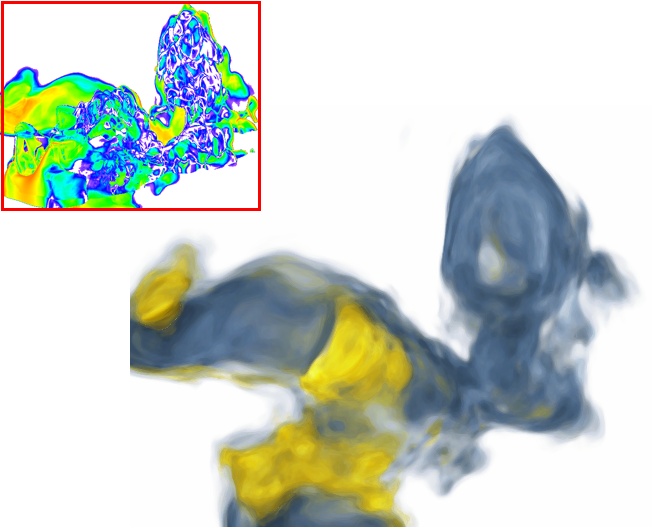}&
    \includegraphics[width=0.13\linewidth]{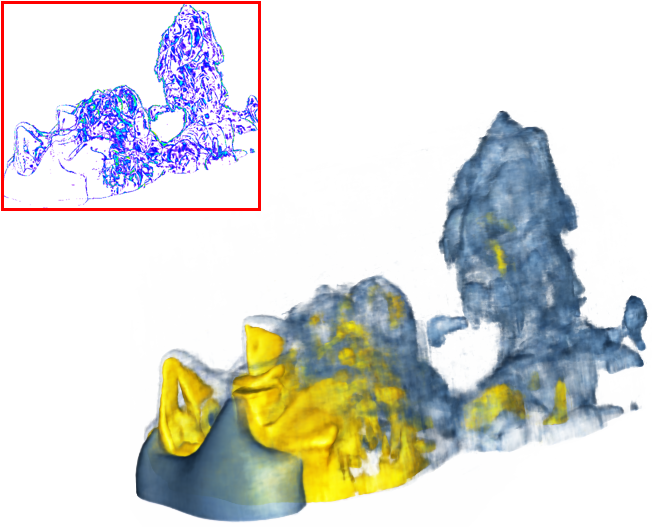}&
    \includegraphics[width=0.13\linewidth]{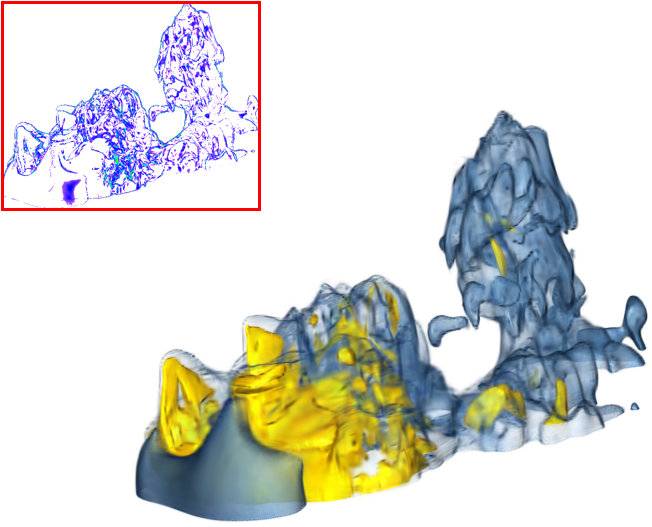}&
    \includegraphics[width=0.13\linewidth]{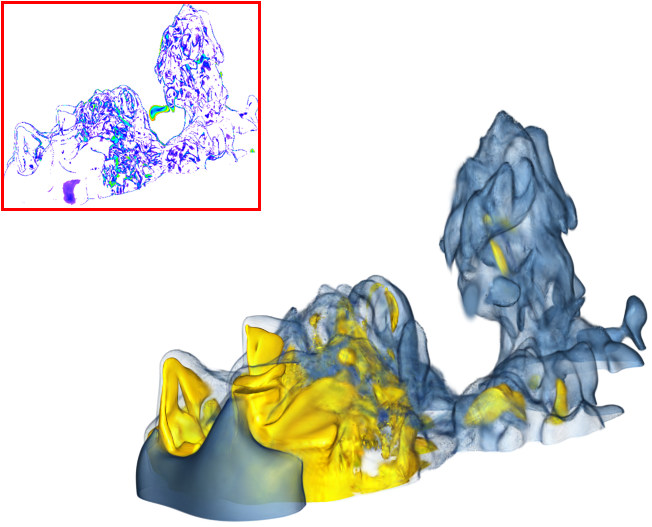}&
    \includegraphics[width=0.13\linewidth]{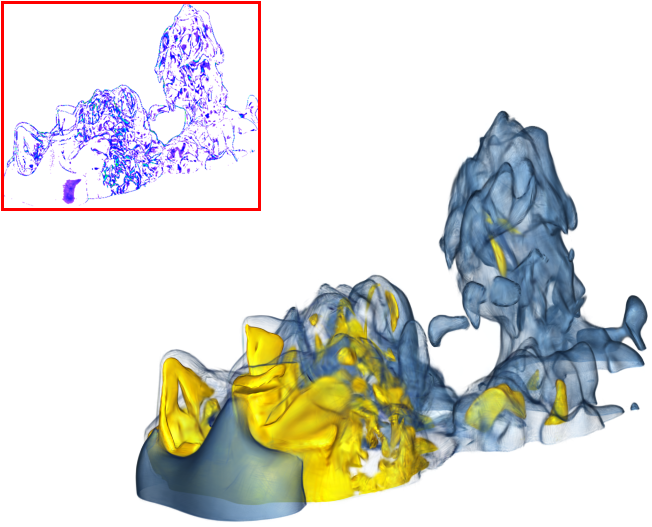}&
    \includegraphics[width=0.13\linewidth]{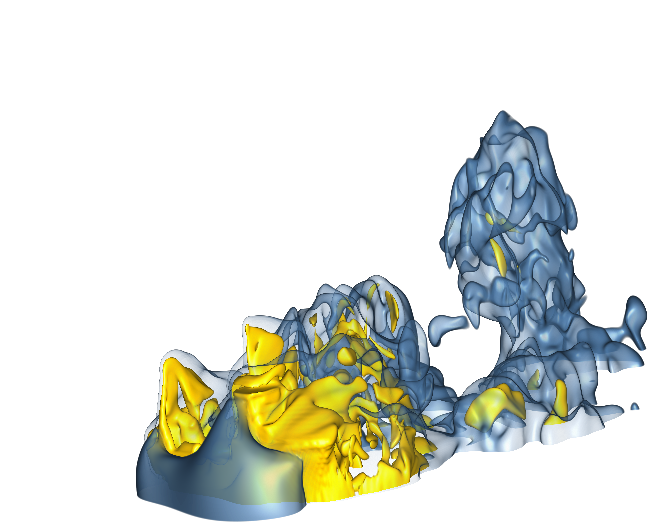}\\
    \includegraphics[width=0.13\linewidth]{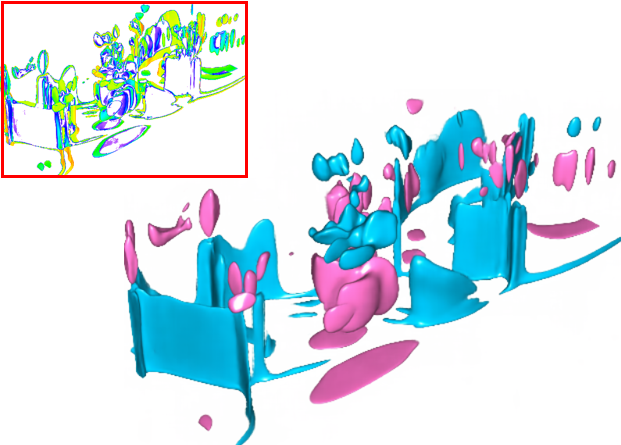}&
    \includegraphics[width=0.13\linewidth]{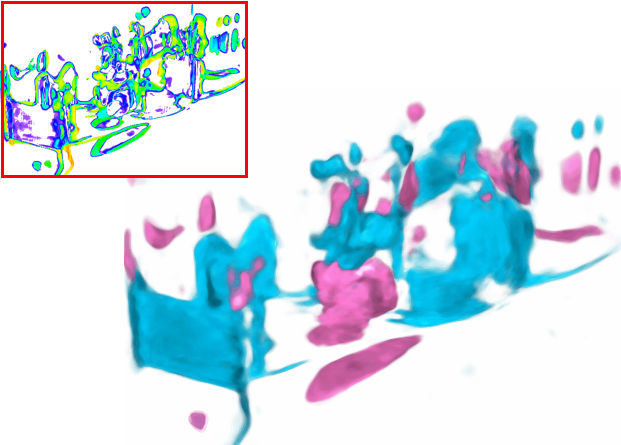}&
    \includegraphics[width=0.13\linewidth]{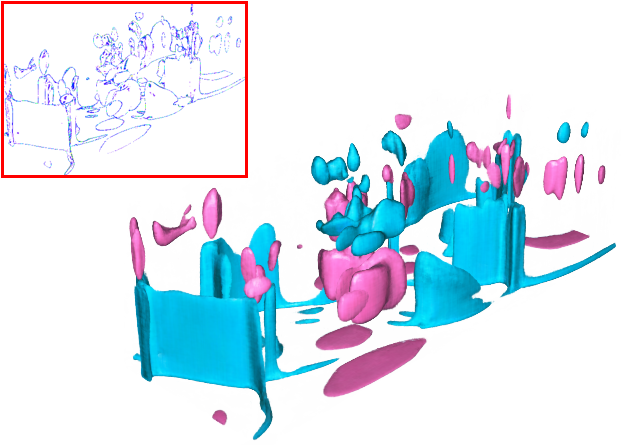}&
    \includegraphics[width=0.13\linewidth]{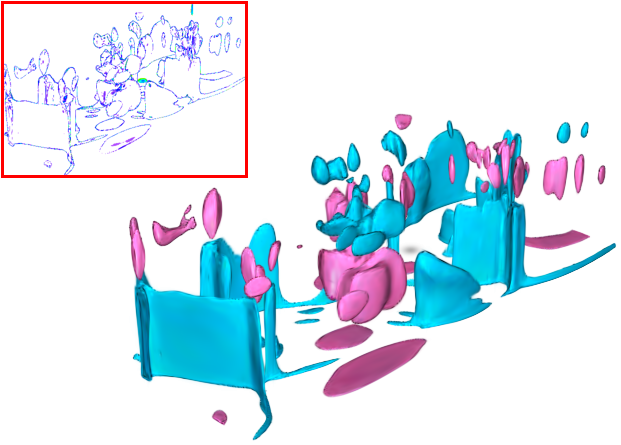}&
    \includegraphics[width=0.13\linewidth]{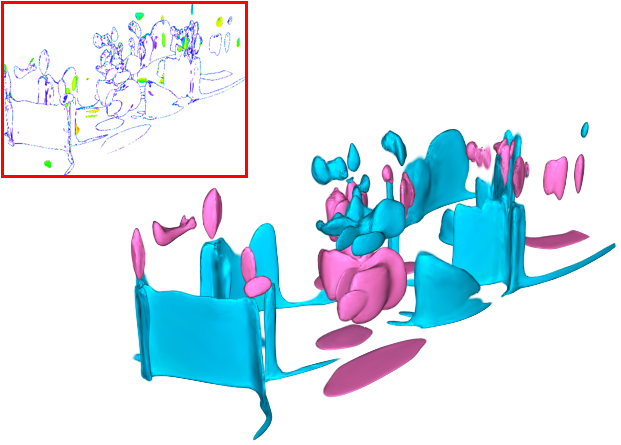}&
    \includegraphics[width=0.13\linewidth]{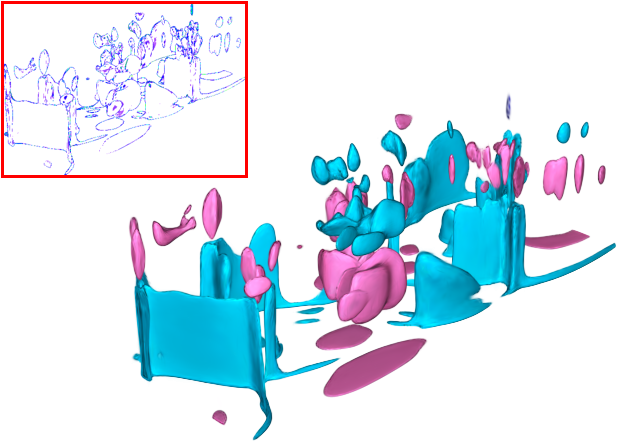}&
    \includegraphics[width=0.13\linewidth]{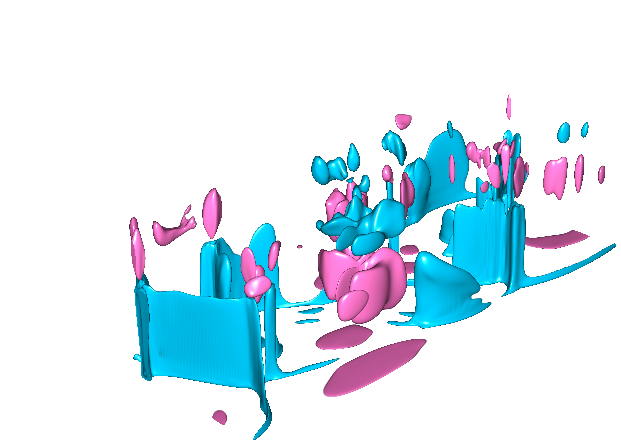}\\
    \includegraphics[width=0.13\linewidth]{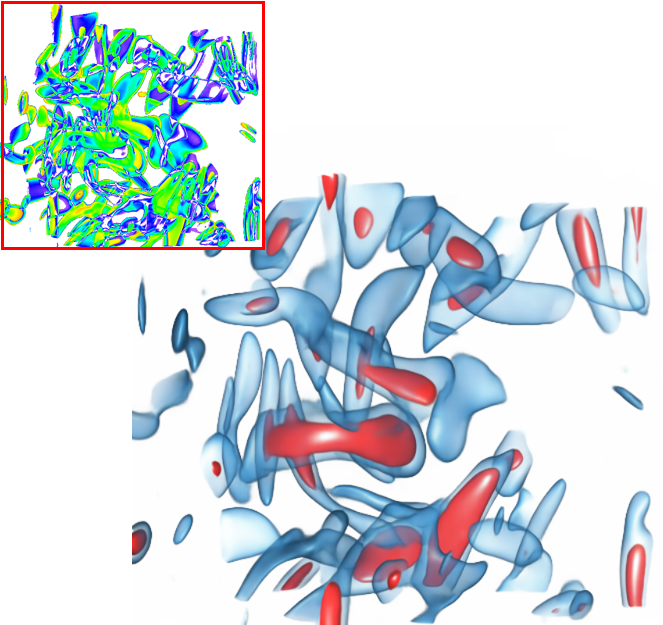}&
    \includegraphics[width=0.13\linewidth]{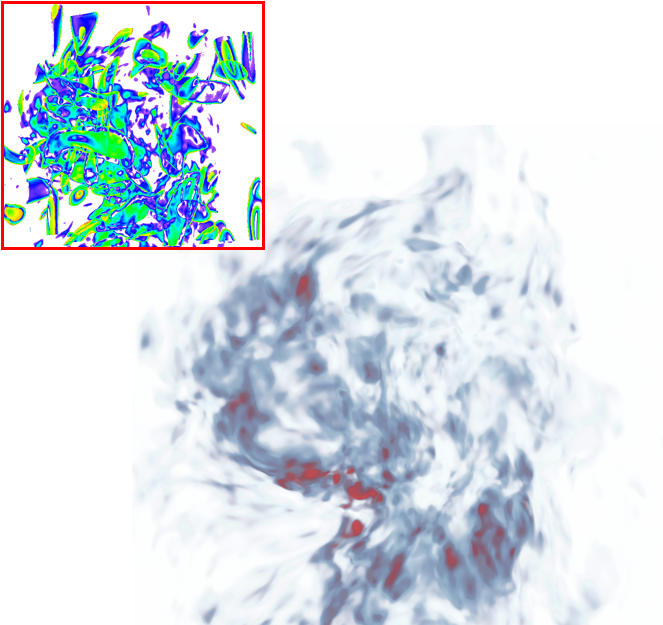}&
    \includegraphics[width=0.13\linewidth]{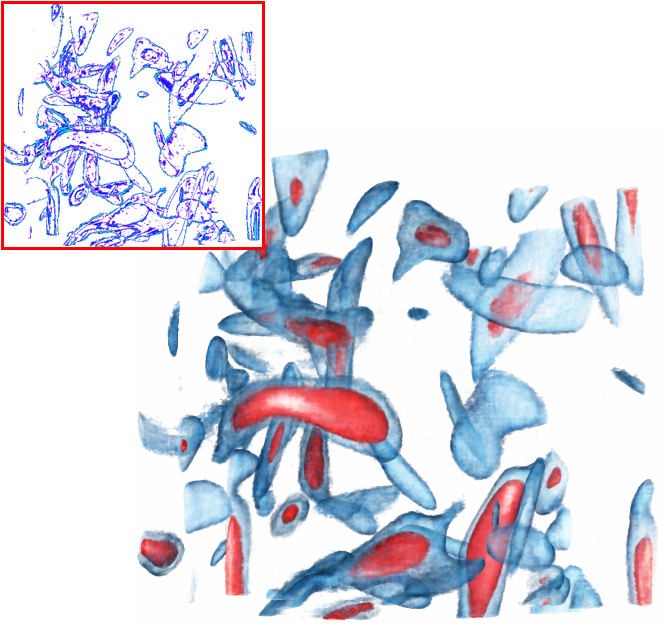}&
    \includegraphics[width=0.13\linewidth]{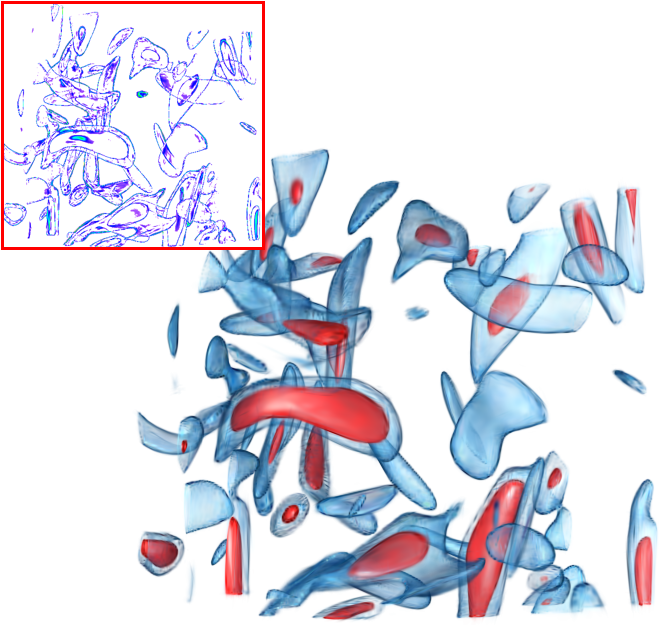}&
    \includegraphics[width=0.13\linewidth]{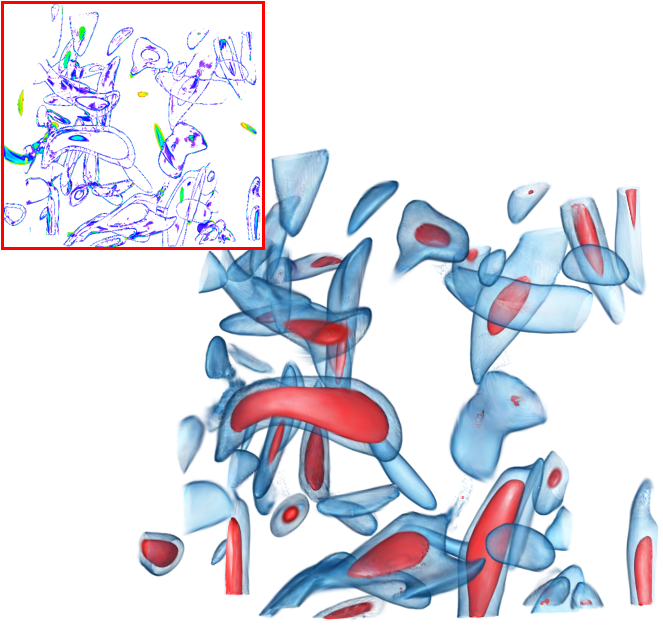}&
    \includegraphics[width=0.13\linewidth]{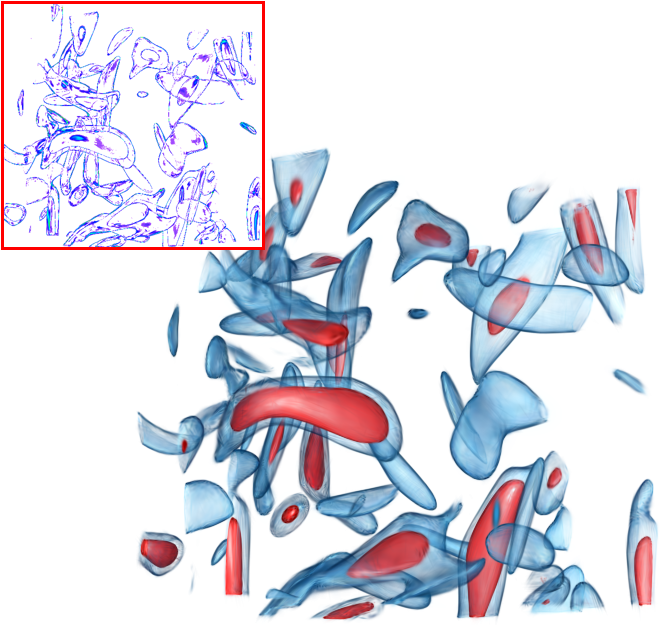}&
    \includegraphics[width=0.13\linewidth]{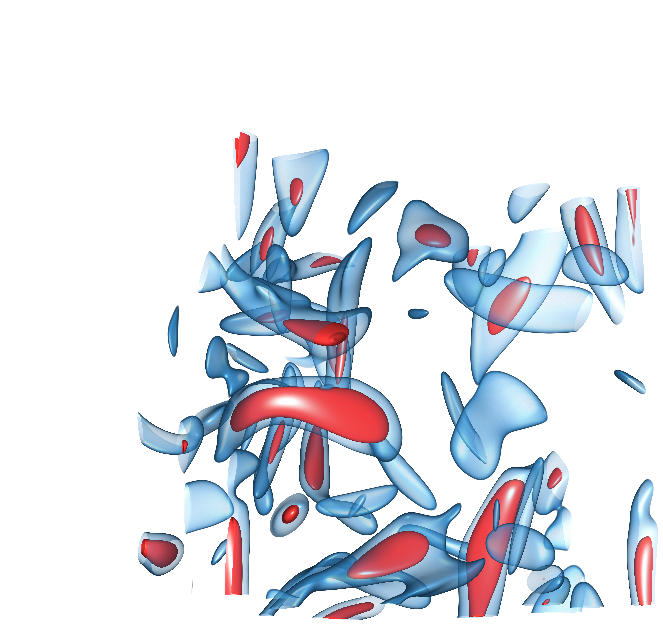}\\
    \includegraphics[width=0.13\linewidth]{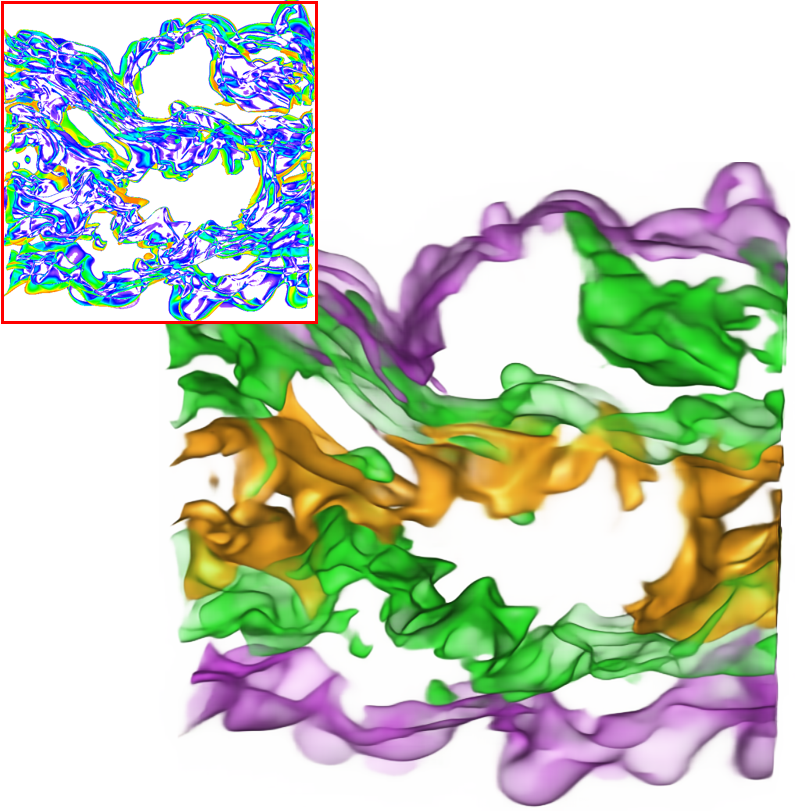}&
    \includegraphics[width=0.13\linewidth]{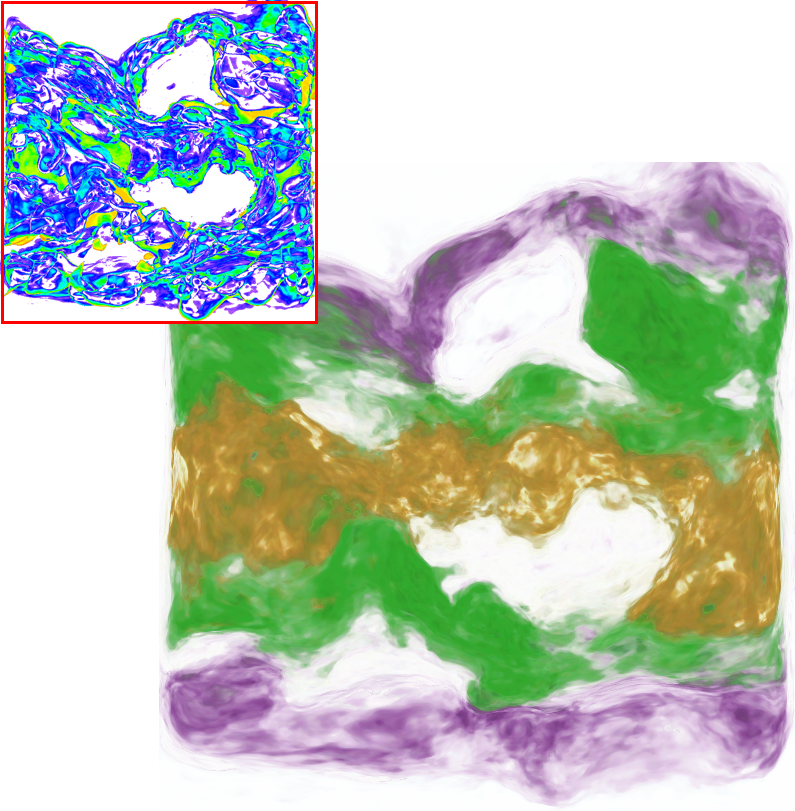}&
    \includegraphics[width=0.13\linewidth]{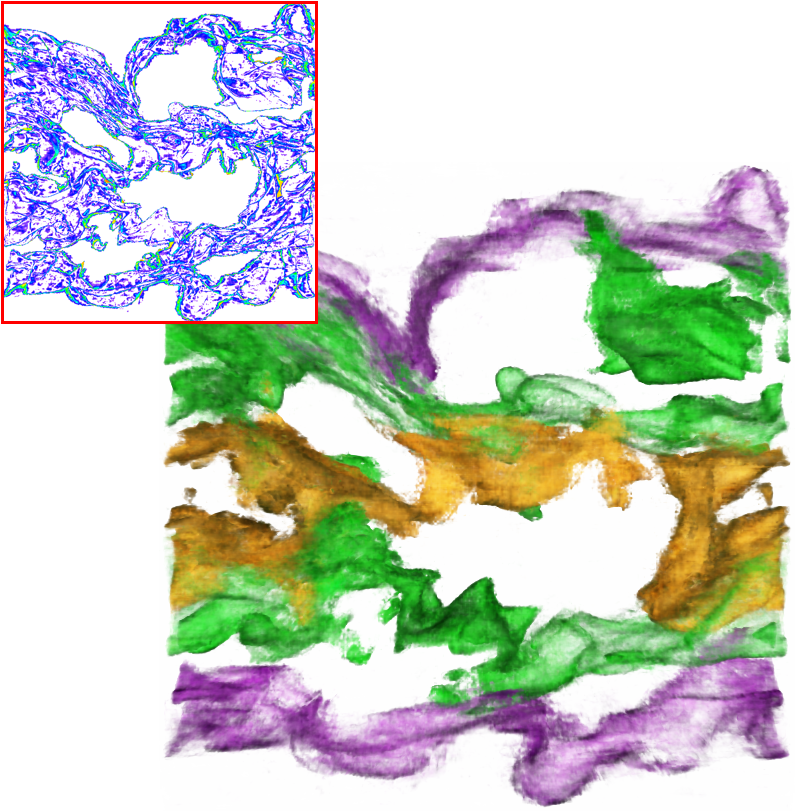}&
    \includegraphics[width=0.13\linewidth]{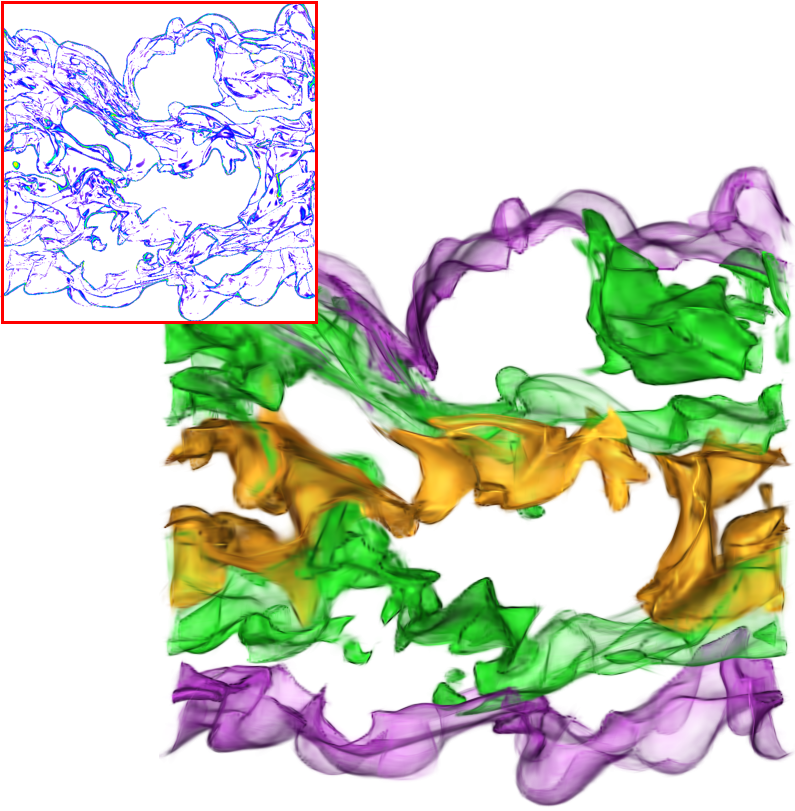}&
    \includegraphics[width=0.13\linewidth]{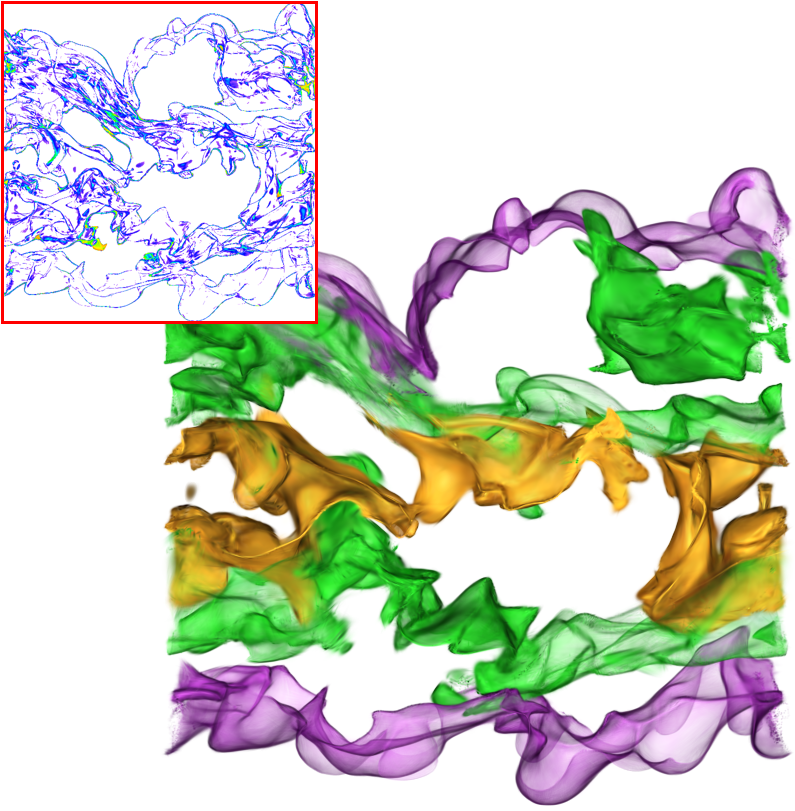}&
    \includegraphics[width=0.13\linewidth]{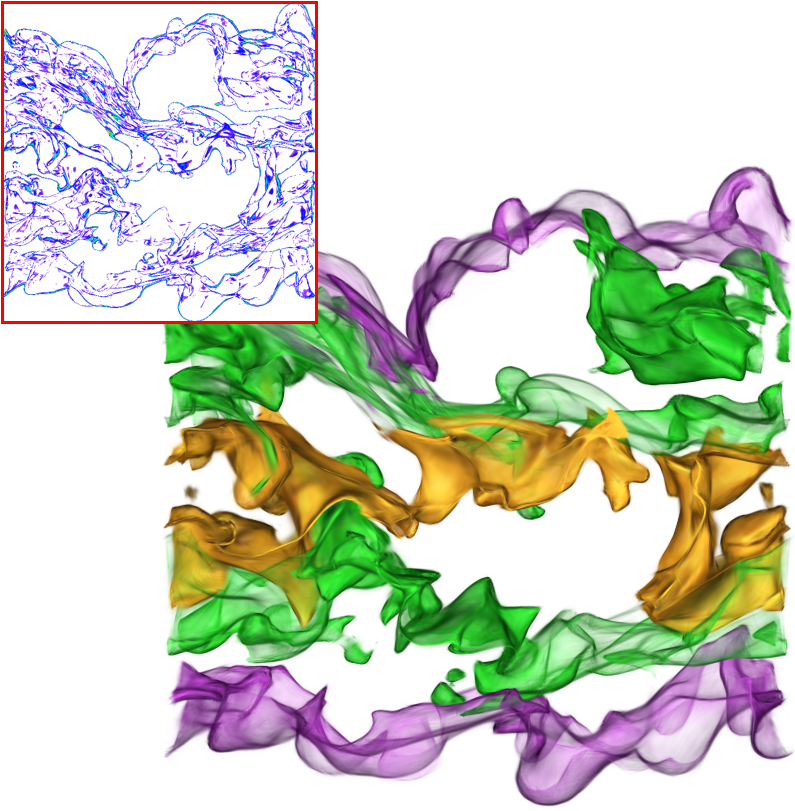}&
    \includegraphics[width=0.13\linewidth]{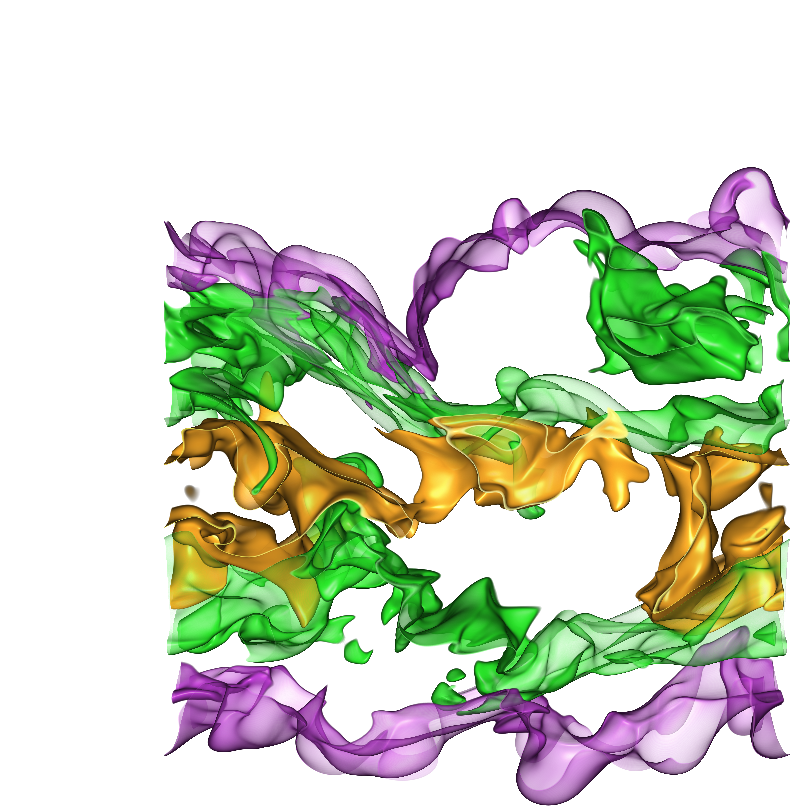}\\
    \mbox{\footnotesize (a) InSituNet} & \mbox{\footnotesize (b) CoordNet} & \mbox{\footnotesize (c) ViSNeRF} & \mbox{\footnotesize (d) 4DGS} & \mbox{\footnotesize (e) D3DGS} & \mbox{\footnotesize (f) VolSegGS} & \mbox{\footnotesize (g) GT}
\end{array}$
\end{center}
\vspace{-.25in} 
\caption{Visualization generation. Top to bottom: a selected timestep of five jets, Tangaroa, mantle, vortex, and combustion.} 
\label{fig:comp-nvs}
\end{figure*}

{\bf Baselines.}
To evaluate the performance of VolSegGS, we compare it with three methods from scientific visualization and two deformable GS techniques:
\begin{myitemize} 
\vspace{-0.05in} 
\item InSituNet~\cite{He-InSituNet-TVCG20}: A surrogate model using a GAN for parameter-space exploration of ensemble simulations. We modify InSituNet by adding upscaling convolutional blocks to output 1024$\times$1024 images, which are then resized to 800$\times$800 for evaluation. 
\item CoordNet~\cite{Han-CoordNet-TVCG}: A coordinate-based implicit neural representation model designed for multiple visualization tasks, including NVS and temporal interpolation. 
\item ViSNeRF~\cite{Yao-PVIS25}: A NeRF-based model utilizing efficient multi-dimensional factorization for dynamic visualization synthesis. 
\item 4DGS~\cite{Yang-ICLR24}: A deformable GS method that employs explicit time-conditioned 3D Gaussians and time-varying, view-dependent colors, supported by 4D SH. 
\item Deformable 3D Gaussians (D3DGS)~\cite{Yang-CVPR24}: A deformable GS method using an implicit deformation network to control the deformation of mean, rotation, and scale of Gaussians, with fixed color and opacity. 
\vspace{-0.05in} 
\end{myitemize}

{\bf Training.}
All methods are trained and evaluated on a machine with an NVIDIA A40 GPU featuring 48 GB of video memory.
% All the training parameters of VolSegGS and baselines are shown in table~\ref{tab:training}.
VolSegGS follows a two-stage training process: 
(1) warming up canonical 3D Gaussians using the training images for 3,000 iterations, and 
(2) jointly training the canonical 3D Gaussians and the deformation field network for 20,000 iterations. 
The training batch size is set to one image. 
The learning rate for Gaussian attributes is consistent with 3DGS~\cite{Kerbl-TOG23}, while the initial learning rate for the deformation field network's encoder is set to $1e^{-3}$, and for the decoder, it is set to $1e^{-4}$.
For the full loss function, we set $\lambda_1=1e^{-4}$ for ${\mathcal L}_{\tv}$. 
${\mathcal L}_{\dssim}$ is introduced only during the final 5,000 iterations of joint training, with $\lambda_2=0.2$.

Training parameters are set for the baselines following the reported configurations. 
InSituNet is trained for 125,000 iterations with a batch size of four images.
CoordNet is trained for 300 epochs with a batch size of 32,000 pixels.
ViSNeRF is trained for 90,000 iterations with a batch size of 4,096 pixels.
Both 4DGS and D3DGS are trained for 30,000 iterations with a batch size of one image.
We maintain approximately 150,000 Gaussians for 4DGS, D3DGS, and VolSegGS to ensure a fair comparison among GS methods.

% \hot{To ensure a fair comparison among GS methods, we maintain 150,000 Gaussians for each sampled timestep during training, regardless of whether a method is 3D or 4D.
% For 4DGS, the total number of Gaussians is calculated as 150,000 multiplied by the number of sampled timesteps for training since Gaussians are not shared across timesteps, resulting in a significantly larger model size. 
% In contrast, for VolSegGS and D3DGS, the total number of Gaussians remains fixed at 150,000, as the same Gaussians are shared across all sampled timesteps.} 

% \begin{table}[htb]
% \caption{Numbers of training images sampled and their resolutions. }
% \vspace{-0.1in}
% \centering
% % {\scriptsize
% \resizebox{\columnwidth}{!}{
% \begin{tabular}{c|cccccc}
% method & InSituNet & CoordNet & ViSNeRF & 4DGS & D3DGS & VolSegGS \\ \hline
% \# iterations &  &  & 90,000 & 30,000 & 30,000 & 3,000 + 20,000 \\ 
% batch size &  &  & 4,096 pixels & 1 image & 1 image & 1 image \\ 
% initial learning rate &  &  &  & 1 image & 1 &  \\ 
% \end{tabular}
% }
% \label{tab:training}
% \end{table}

%\vspace{-0.1in}
\begin{table}[htb]
    \caption{Visualization generation for dynamic scenes: average PSNR (dB), SSIM, LPIPS, and rendering framerate (FPS) across all 181 synthesized views, training time (TT, in minutes), and model size (MS, in MB). The best ones are highlighted in bold.}
    \vspace{-0.1in}
    \centering
    %{\scriptsize
    {\fontsize{5.75pt}{5.75pt}\selectfont
    %\resizebox{\columnwidth}{!}{
    \begin{tabular}{c|c|cccccc}

        dataset         & method    & PSNR$\uparrow$ & SSIM$\uparrow$ & LPIPS$\downarrow$ & FPS$\uparrow$ & TT$\downarrow$ & MS$\downarrow$ \\ \hline
                        & InSituNet & 14.93 & 0.829 & 0.167 & 5.89 & 1621.45 & 318.85 \\
                        & CoordNet  & 16.65 & 0.844 & 0.176 & 0.52 & 2065.83 & \bf 5.71 \\
        five jets       & ViSNeRF   & \bf 27.36 & 0.955 & 0.035 & 0.28 & 53.98 & 12.32 \\
                        & 4DGS      & 26.14 & 0.955 & 0.044 & \bf 214.13 & 68.87 & 96.41 \\
                        & D3DGS     & 25.45 & 0.948 & 0.037 & 37.20 & 42.82 & 37.01 \\
                        & VolSegGS  & 27.16 & \bf 0.965 & \bf 0.030 & 88.81 & \bf 16.07 & 41.62 \\ \hline
                        & InSituNet & 15.57 & 0.810 & 0.217 & 5.95 & 1591.25 & 318.85 \\
                        & CoordNet  & 16.87 & 0.828 & 0.234  & 0.49 & 3105.13 & \bf 5.71 \\
        Tangaroa        & ViSNeRF   & 25.21 & 0.906 & 0.092 & 0.31 & 60.62 & 12.45 \\
                        & 4DGS      & 27.33 & 0.945 & 0.059 & \bf 213.62 & 69.38 & 96.96 \\
                        & D3DGS     & 26.49 & 0.940 & 0.055 & 36.11 & 43.85 & 37.63 \\
                        & VolSegGS  & \bf 28.15 & \bf 0.953 & \bf 0.042 & 88.65 & \bf 16.80 & 43.48 \\ \hline
                        & InSituNet & 15.11 & 0.860 & 0.204 & 5.86 & 1632.91 & 318.85 \\
                        & CoordNet  & 15.75 & 0.861 & 0.229 & 0.51 & 2070.31 & \bf 5.71 \\
        mantle          & ViSNeRF   & \bf 29.36 & 0.975 & \bf 0.021 & 0.29 & 68.16 & 12.99 \\
                        & 4DGS      & 28.56 & 0.975 & 0.034 & \bf 213.67 & 69.67 & 97.89 \\
                        & D3DGS     & 26.69 & 0.972 & 0.046 & 36.14 & 43.35 & 37.54 \\
                        & VolSegGS  & 29.02 & \bf 0.979 & 0.030 & 89.29 & \bf 15.98 & 43.21 \\ \hline
                        & InSituNet & 14.76 & 0.748 & 0.322 & 5.78 & 1593.98 & 318.85 \\
                        & CoordNet  & 15.56 & 0.764 & 0.375 & 0.52  & 3139.56 & \bf 5.71 \\
        vortex          & ViSNeRF   & 24.94 & 0.919 & 0.096 & 0.29 & 64.68 & 12.39 \\
                        & 4DGS      & 25.82 & 0.940 & 0.081 & \bf 209.21 & 73.57 & 105.96 \\
                        & D3DGS     & 23.64 & 0.932 & 0.119 & 36.40 & 42.86 & 37.51 \\
                        & VolSegGS  & \bf 26.37 & \bf 0.947 & \bf 0.074 & 89.21 & \bf 15.90 & 43.28 \\ \hline
                        & InSituNet & 13.85 & 0.680 & 0.361 & 5.29 & 1604.72 & 318.85 \\
                        & CoordNet  & 14.72 & 0.696 & 0.370 & 0.50  & 4655.92 & \bf 5.71 \\
        combustion      & ViSNeRF   & 22.45 & 0.800 & 0.202 & 0.28 & 69.33 & 14.15 \\
                        & 4DGS      & 25.61 & 0.894 & 0.112 & \bf 213.48 & 78.92 & 105.85 \\
                        & D3DGS     & 24.55 & 0.882 & 0.107 & 36.57 & 42.72 & 37.48 \\
                        & VolSegGS  & \bf 25.76 & \bf 0.897 & \bf 0.092 & 88.35 & \bf 16.92 & 43.37 \\
  \end{tabular}
    }
    \label{tab:metrics-nvs}
\end{table}

{\bf Qualitative results.}
Figure~\ref{fig:comp-nvs} shows that 2D-based methods (InSituNet and CoordNet) produce the least satisfactory results. 
Specifically, InSituNet generates high-quality images but often deviates from the requested view or timestep. 
This issue arises because InSituNet relies on the closest available view, which may differ significantly from the requested parameters. 
As the available views are limited compared to the demand, the nearest matching view frequently remains substantially different from the requested one. 
On the other hand, the images generated by CoordNet appear blurry due to insufficient views for accurate interpolation within its implicit neural representation. 
As a result, both 2D-based methods require considerably more training images to effectively capture variations across views and timesteps.

Among the 3D-aware methods, ViSNeRF produces relatively blurry and noisy results for the Tangaroa, vortex, and combustion datasets, while 4DGS, D3DGS, and VolSegGS deliver sharper, more detailed results. 
This difference is mainly due to the rapid and simultaneous changes in scene content across these datasets. 
While 4DGS, D3DGS, and VolSegGS effectively capture these changes by modeling scene deformation, ViSNeRF struggles to maintain geometric consistency due to its reliance on interpolation between static scene frames.
Another notable observation is that ViSNeRF generally preserves specular highlights better than 4DGS, D3DGS, and VolSegGS, especially for the five jets and vortex datasets. 
This advantage arises because ViSNeRF uses a view-direction-conditioned MLP decoder to learn view-dependent colors, whereas 4DGS, D3DGS, and VolSegGS rely solely on SH.

Among the GS methods, which all model Gaussian deformation, the generation quality is generally similar across all datasets. 
However, due to the use of an implicit MLP in the deformation network, D3DGS and VolSegGS produce smoother synthesized images than 4DGS, which uses explicit 4D Gaussians. 
This difference is particularly noticeable in the transparent regions of the Tangaroa, vortex, and combustion datasets.
Additionally, D3DGS leads to more missing parts in the synthesized images compared to 4DGS and VolSegGS, particularly for the Tangaroa, mantle, and vortex datasets. 
Upon investigation, we attribute this issue to using L1 and DSSIM losses in D3DGS. 
Specifically, these losses provide insufficient gradient signals for capturing subtle structural details, preventing the deformation network from accurately positioning and representing Gaussians in these regions, resulting in missing content in the synthesized images. 
In contrast, VolSegGS employs L2 loss (which more effectively penalizes large errors in small regions) and postpones the application of DSSIM loss until the final 5,000 iterations, ensuring better structural preservation.

%Upon investigation, we attribute this issue to the combined use of L1 and DSSIM loss in D3DGS. Specifically, L1 and DSSIM losses provide insufficient gradients for subtle structural details, preventing the deformation network from accurately positioning and representing Gaussians in these regions, which leads to missing content in the synthesized images. VolSegGS, on the other hand, uses L2 loss to more effectively penalize large errors in small areas and delays using DSSIM loss until the final 5,000 iterations.

{\bf Quantitative results.}
We evaluate the quality of the synthesized images compared to the GT images using three metrics: {\em peak signal-to-noise ratio} (PSNR), {\em structural similarity index} (SSIM), and {\em learned perceptual image patch similarity} (LPIPS)~\cite{Zhang-CVPR18}. 
Additionally, we report the rendering framerate, training time, and model size to assess method efficiency. 
Quantitative results are presented in Table~\ref{tab:metrics-nvs}. 
Overall, 3D-aware methods (ViSNeRF, 4DGS, D3DGS, and VolSegGS) consistently outperform 2D-based ones (InSituNet and CoordNet) across all datasets. 
Among the 3D-aware methods, ViSNeRF achieves the highest PSNR for the five jets and mantle datasets but falls behind the GS methods (4DGS, D3DGS, and VolSegGS) for the Tangaroa, vortex, and combustion datasets. 
While VolSegGS delivers the best generation quality within the GS methods, it also produces competitive results compared to ViSNeRF for the five jets and mantle datasets.

{\bf Rendering framerate.} 
The GS methods offer real-time framerates, thanks to their rasterization-based rendering pipeline. 
Other methods evaluated in our experiments fail to meet the performance requirements for real-time exploration of dynamic scenes. 
Among the Gaussian-based approaches, 4DGS achieves the highest FPS. 
This is due to 4DGS directly utilizing fully explicit 4D Gaussians to represent dynamic scenes, making it more efficient in rendering than D3DGS and VolSegGS, which rely on deformation networks. 
However, 4DGS requires several regularization terms during optimization to ensure training stability, significantly increasing the overall training cost. 
Specifically, the regularization process involves computationally expensive k-nearest-neighbor calculations to enhance consistency among neighboring Gaussians. 
In contrast, D3DGS and VolSegGS employ deformation networks with an MLP that takes Gaussian means and time as inputs, naturally providing spatial smoothness without additional regularization.

Despite having the second-highest FPS, VolSegGS still comfortably meets real-time requirements, with a minimum of 87 FPS. 
In contrast, D3DGS achieves only about 36 FPS, making it less ideal than VolSegGS for typical 60Hz displays. 
The lower FPS of D3DGS is primarily due to its reliance on a large MLP-based deformation network for predicting the deformation of 3D Gaussians. 
VolSegGS, on the other hand, employs a hybrid deformation network consisting of an explicit feature grid combined with a lightweight MLP. 
This architectural difference allows VolSegGS to achieve faster training speeds than D3DGS, although it slightly increases the model size.

\vspace{-0.1in}
\begin{table}[htb]
\caption{Render performance comparison: DVR vs.\ VolSegGS.}
\vspace{-0.1in}
\centering
% {\scriptsize
% {
% \fontsize{5.75pt}{5.75pt}\selectfont
\resizebox{\columnwidth}{!}{
\begin{tabular}{c|ccc|ccc}
% & CPU/GPU & I/O & render & CPU/GPU & render\\ 
& \multicolumn{3}{c|}{DVR} & \multicolumn{3}{c}{VolSegGS} \\ \hline
%& ParaView & & & VolSegGS &\\ \hline
 & CPU/GPU & loading & rendering & CPU/GPU & \hot{preparation+training} & rendering \\ 
dataset & memory (GB) &  time (s) & time (ms) & memory (GB) & \hot{time (min)} & time (ms)\\ \hline
five jets & 1.23/0.91  & 0.68  & 191  & 2.46/1.53 & \hot{1.38+16.07} & 11  \\
Tangaroa & 1.72/0.90  & 1.84  & 247 & 2.51/1.54 & \hot{3.08+16.80} & 11  \\
mantle & 2.54/0.88  & 3.33  & 212  & 2.50/1.56 & \hot{2.52+15.98} & 11  \\
vortex & 3.20/1.42  & 4.55  & 342  & 2.53/1.53 & \hot{5.70+15.90} & 11  \\
combustion &  5.14/1.88  & 9.07  & 761  & 2.51/1.47 & \hot{15.96+16.92} & 11  \\
\end{tabular}
}
\label{tab:render-metrics}
\end{table}

{\bf Comparison with DVR.} 
VolSegGS achieves a rendering speedup of 17$\times$ to 69$\times$ compared with DVR using ParaView, as shown in Table~\ref{tab:render-metrics}. 
In the table, loading time refers to the average time needed to load the volume data of a single timestep into memory, and rendering time is the average time required to render a test view. 
\hot{For VolSegGS, the preparation time includes the full duration required to load data into ParaView and render all training views across the sampled timesteps listed in Table~\ref{tab:nvs-dataset}. 
At the cost of the upfront preparation and training time,} VolSegGS eliminates data loading overhead when switching timesteps during inference. 
In contrast, DVR methods incur significant data loading costs, which prevent them from reaching the real-time rendering speeds demonstrated by VolSegGS.
This makes VolSegGS particularly promising for real-time \hot{exploratory visualization} %interactive exploration 
of dynamic scenes and provides a reliable foundation for subsequent interactive segmentation and tracking tasks.

{\bf Summary.} 
VolSegGS offers the best balance of generation quality, rendering efficiency, and training cost among all the methods compared. 
\hot{While VolSegGS does not significantly outperform existing deformable Gaussian methods in rendering quality, it provides a solid foundation for subsequent segmentation and tracking in exploratory visualization.}

% \begin{table}[htb]
% \caption{Datasets for visualization generation of dynamic scenes, with training images in PNG format.}
% \vspace{-0.1in}
% \centering
% {\scriptsize
% % \resizebox{\columnwidth}{!}{
% \begin{tabular}{c|ccc}
%     & selected & \# views for  &  SAM mask  \\ 
% dataset & timestep & mask generation  & generation time (s)  \\ \hline
% five jets & 61  & 20 &  \\ 
% Tangaroa & 1  & 30 &  \\
% mantle & 1  & 10 & 36.04 \\
% vortex & 31  & 20 &  \\
% combustion & 100 & 20 &  \\
% \end{tabular}
% }
% \label{tab:seg-dataset}
% \end{table}

\vspace{-0.05in}
\subsection{3D Segmentation for Static Scenes}
% Although we are aware of a few concurrent works~\cite{Ji-arXiv24, Li-arXiv24} on 4D segmentation, we lack of complete open source codes to run the comparison in dynamic scenes.
In this section, we assess the 3D segmentation quality of VolSegGS on static scenes, comparing it against baseline 3D segmentation methods.

{\bf Datasets.}
To compare segmentation methods specifically on static scenes, we select a representative timestep from each dataset listed in Table~\ref{tab:nvs-dataset}, as indicated in Table~\ref{tab:metrics-seg}.
For each selected timestep, we render 30 views at a resolution of 800$\times$800, which serve as input images for training a 3DGS model. 
These training images are also segmented using SAM to provide 2D mask supervision for VolSegGS and baseline 3D segmentation methods. 
We use the same view sampling method described in Section~\ref{sec:exp-nvs} to generate 30 training and 181 test views. 
To quantitatively evaluate segmentation quality, we render manually segmented volumes using DVR as the GT and compute PSNR, SSIM, and LPIPS scores across 181 test views. 
Additionally, we generate corresponding 2D masks for these test views and use {\em intersection over union} (IoU) for comparison.

{\bf Baselines.}
We select two state-of-the-art 3D Gaussian segmentation methods, SAGA~\cite{Cen-arXiv23} and SAGD~\cite{Hu-arXiv24}, as baseline methods. 
To ensure a fair comparison, we pretrain a single 3DGS model per dataset and apply VolSegGS and the baseline segmentation methods to the same pretrained model. 
Each 3DGS model is trained for 30,000 iterations without limiting the number of Gaussians while keeping all other settings at their default values. 
Additionally, we include SAM 2~\cite{Ravi-SAM2-arXiv24}, given its capability for 2D video segmentation. 
Since our test images are rendered along a predefined camera path, we assemble these images into a video sequence and apply SAM 2 for video segmentation.

\begin{myitemize} 
\vspace{-0.05in} 
\item SAM 2\cite{Ravi-SAM2-arXiv24}: A foundation model that provides promptable visual segmentation for images and videos. 
It leverages a transformer architecture combined with streaming memory to enhance the performance of segment tracking.
\item SAGA\cite{Cen-arXiv23}: An efficient segmentation approach specifically designed for 3DGS. 
It employs scale-gated affinity features for each Gaussian to effectively capture inter-Gaussian relationships. 
The affinity features are optimized using contrastive learning, guided by automatically generated 2D masks obtained from SAM.
\item SAGD~\cite{Hu-arXiv24}: A training-free segmentation method tailored for 3DGS. 
It projects 2D masks produced by SAM onto 3D Gaussians from each view and determines segmentation labels through a voting strategy that aggregates binary mask labels across all views.
\vspace{-0.15in} 
\end{myitemize}

{\bf Training.}
All experiments are conducted on a machine with an NVIDIA RTX 4090 GPU with 24 GB of video memory. 
We train the affinity field network of VolSegGS for 5,000 iterations using a batch size of 8,192 pixels and the Adam optimizer with a learning rate of $1e^{-3}$. 
For the baseline methods, we use the publicly available pretrained checkpoint (\texttt{sam2.1\_hiera\_large}) for SAM 2. 
For SAGA, the affinity features are optimized following the original paper's settings, with 10,000 iterations and a batch size of 1,000 pixels. 
SAGD does not require training; therefore, we directly apply this method to the pretrained 3DGS model using SAM masks.

\begin{figure}[tb]
%\vspace{-0.1in}
\begin{center}
$\begin{array}{c@{\hspace{0.025in}}c@{\hspace{0.025in}}c@{\hspace{0.025in}}c@{\hspace{0.025in}}c}
    \includegraphics[width=0.1525\linewidth]{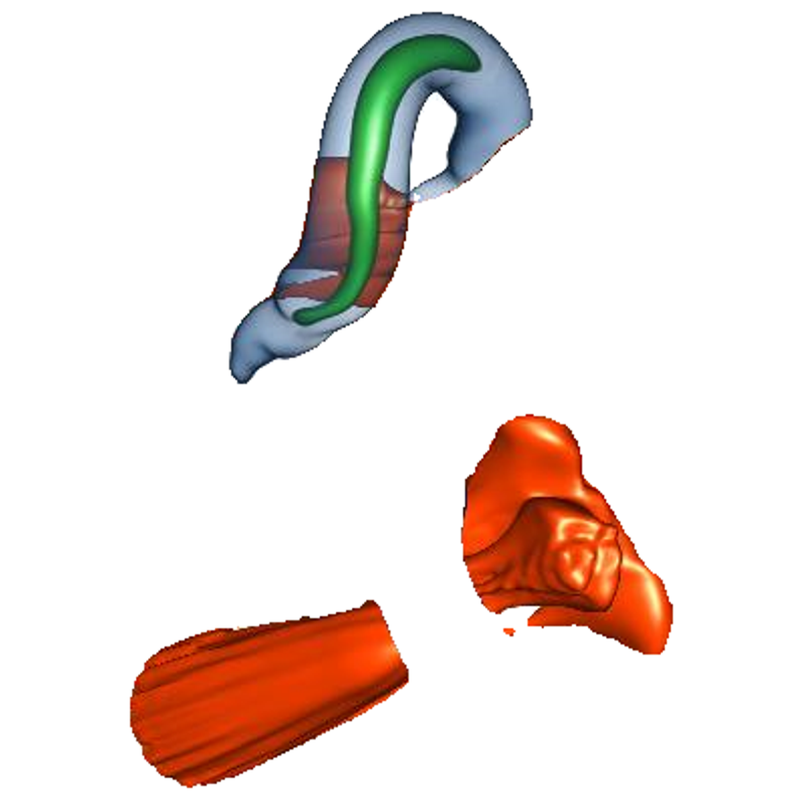}&
    \includegraphics[width=0.1525\linewidth]{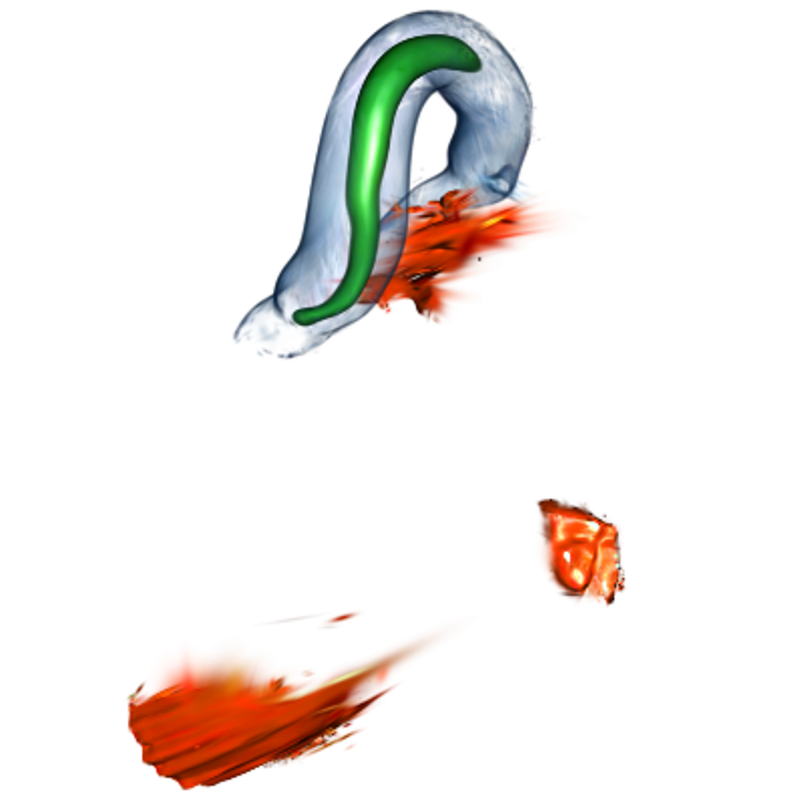}&
    \includegraphics[width=0.1525\linewidth]{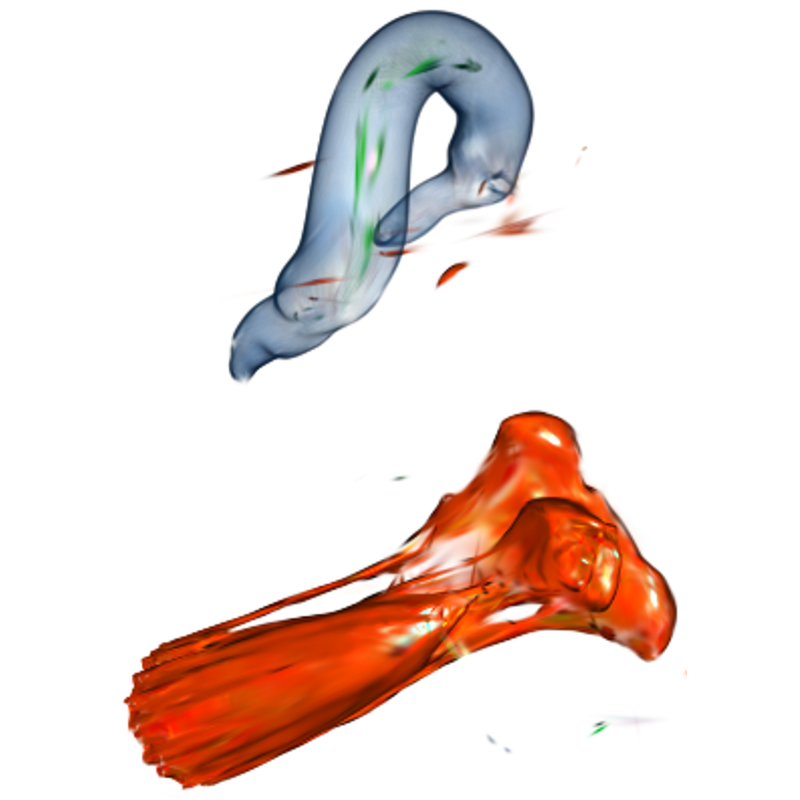}&
    \includegraphics[width=0.1525\linewidth]{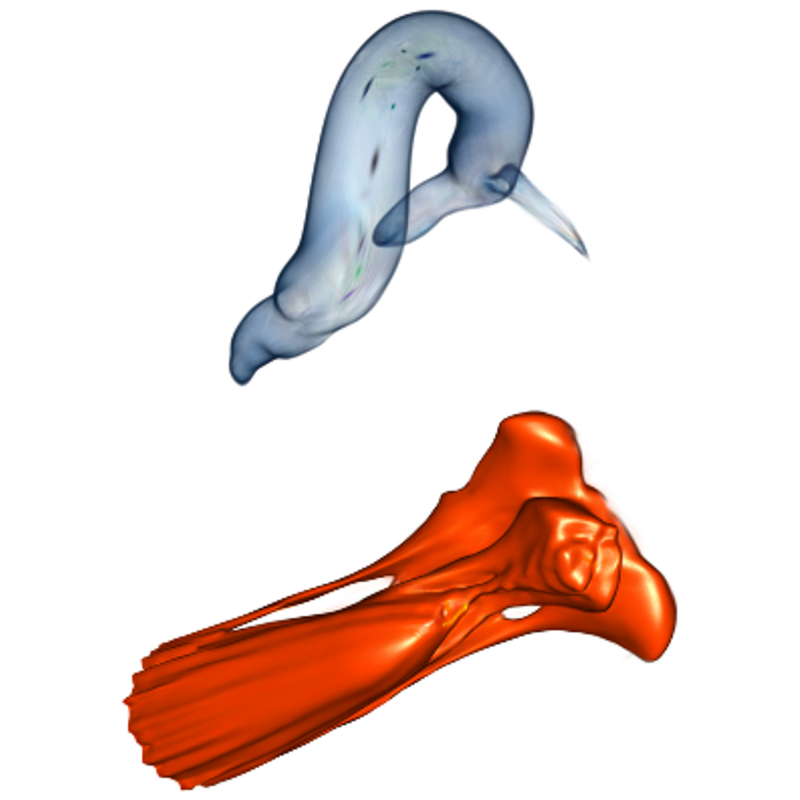}&
    \includegraphics[width=0.305\linewidth]{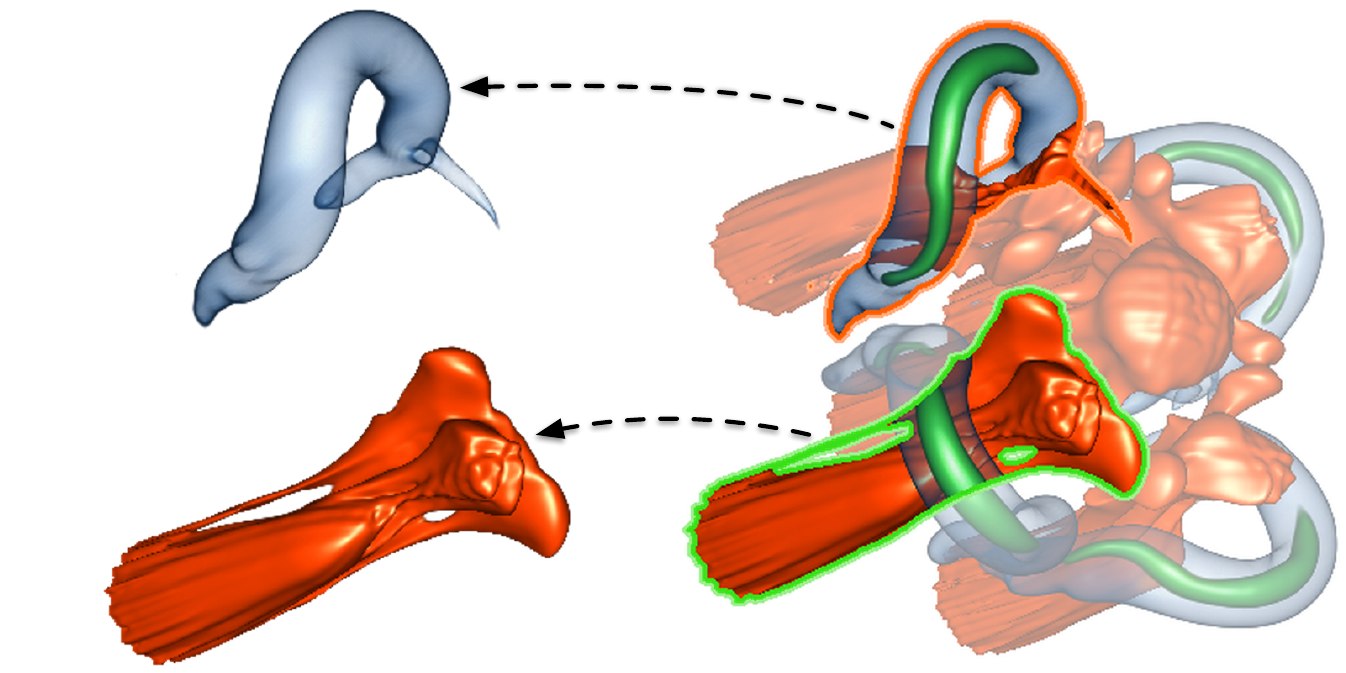}\\
    \includegraphics[width=0.1525\linewidth]{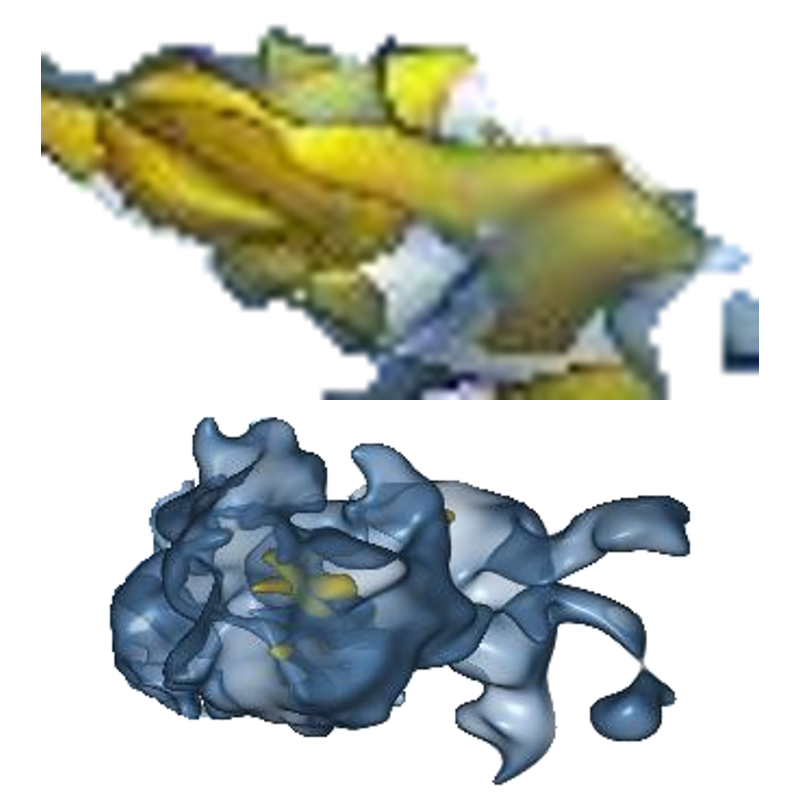}&
    \includegraphics[width=0.1525\linewidth]{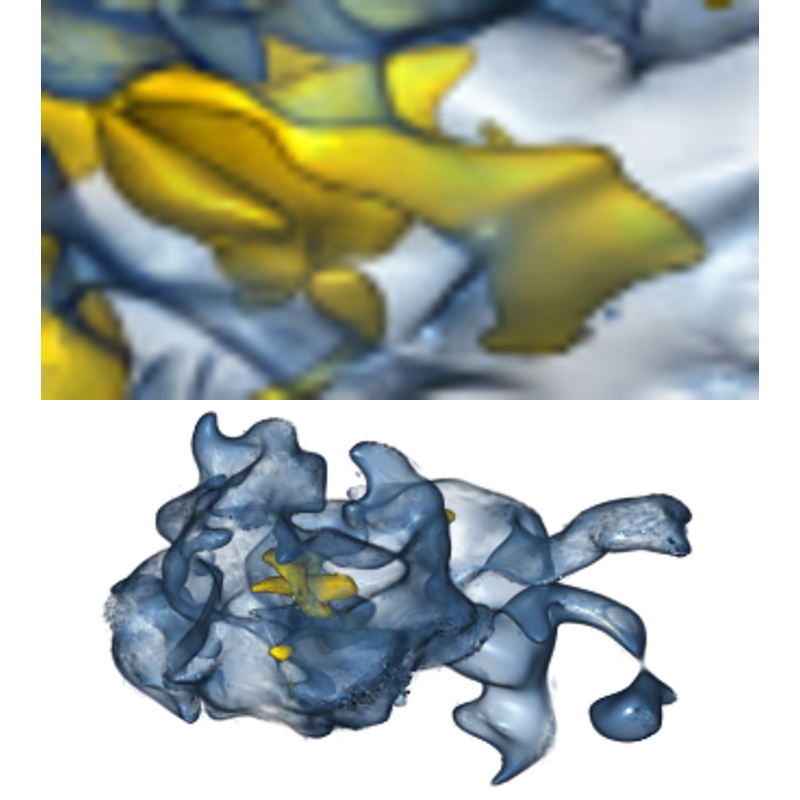}&
    \includegraphics[width=0.1525\linewidth]{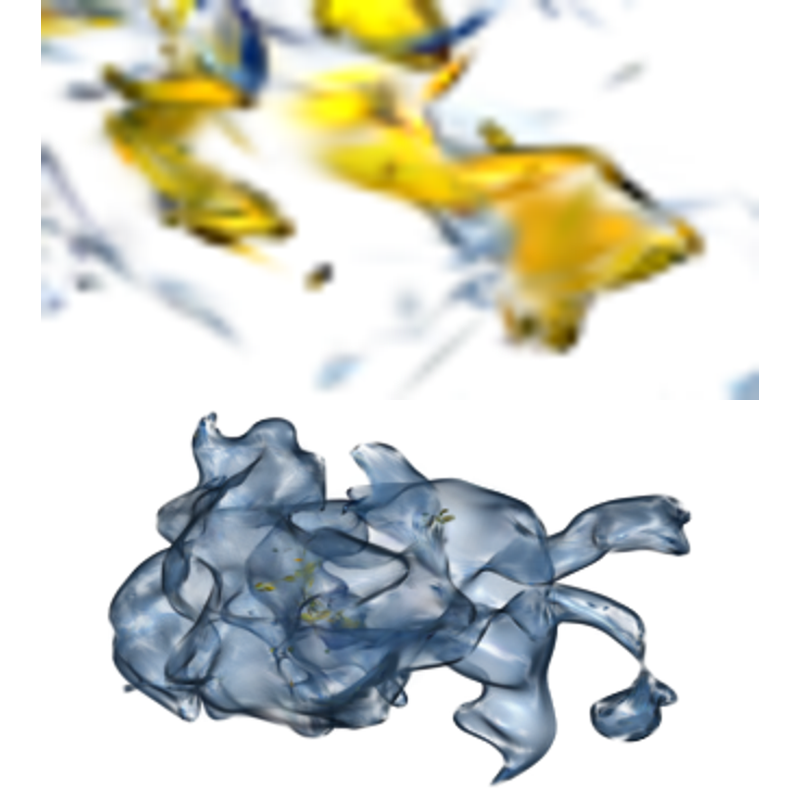}&
    \includegraphics[width=0.1525\linewidth]{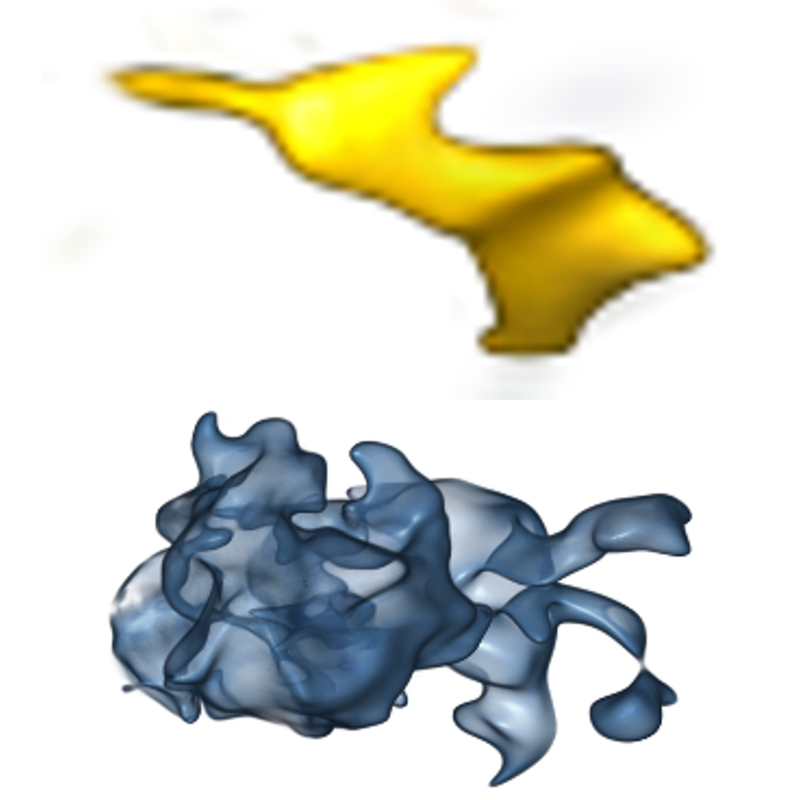}&
    \includegraphics[width=0.305\linewidth]{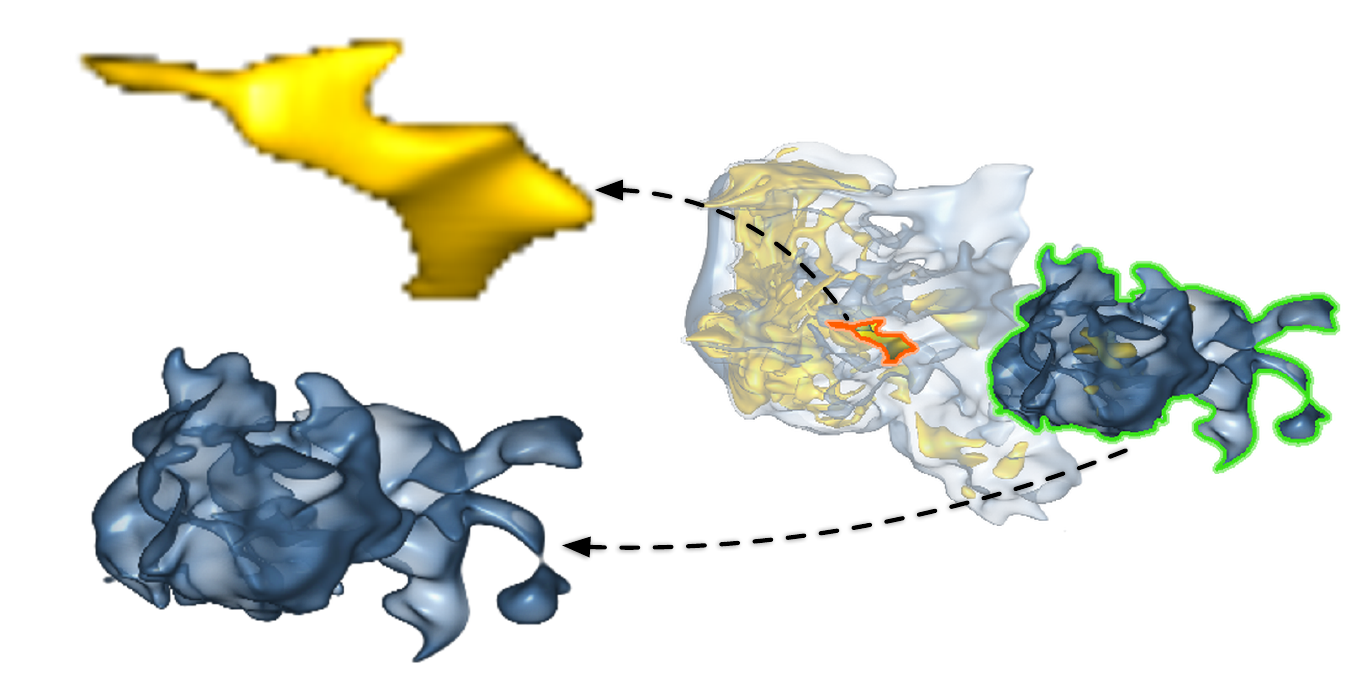}\\
    \includegraphics[width=0.1525\linewidth]{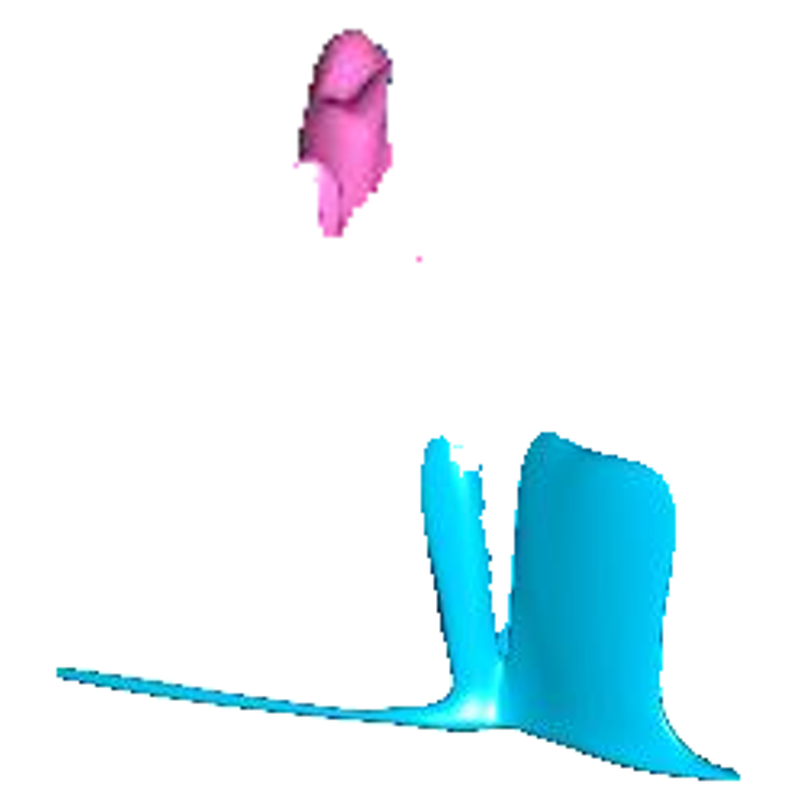}&
    \includegraphics[width=0.1525\linewidth]{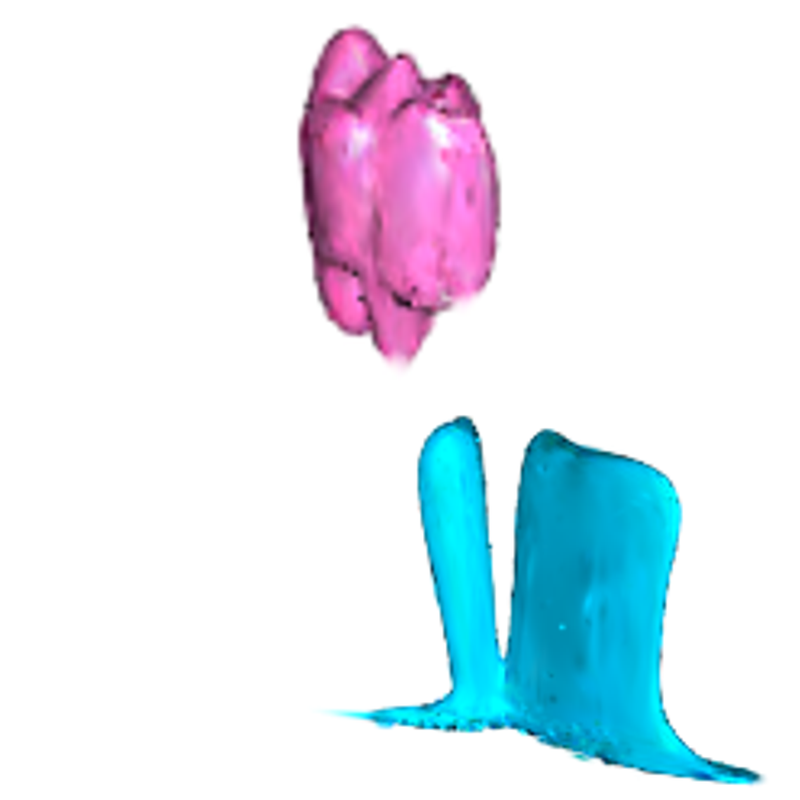}&
    \includegraphics[width=0.1525\linewidth]{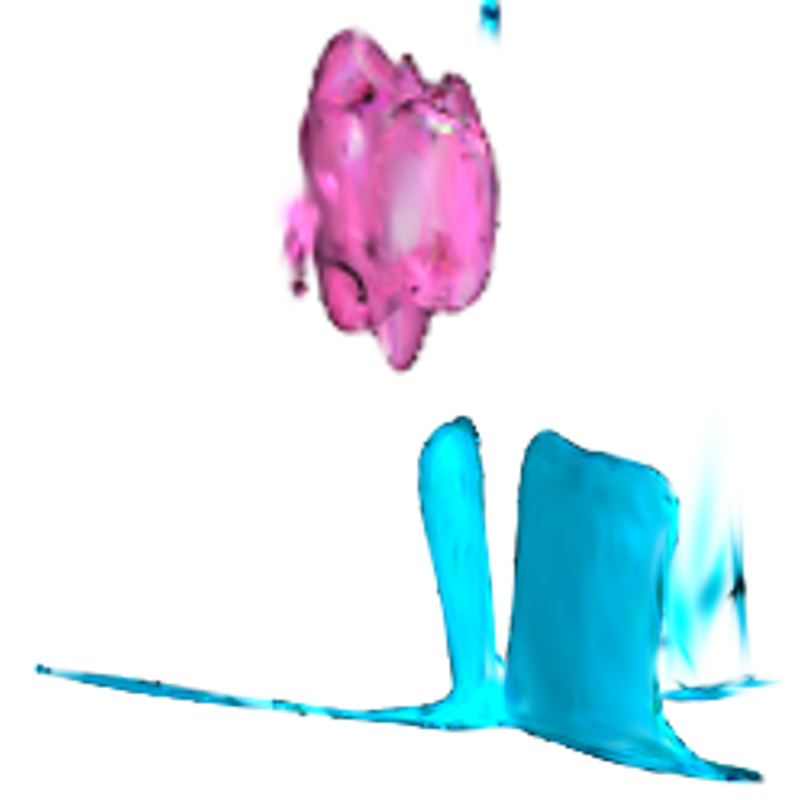}&
    \includegraphics[width=0.1525\linewidth]{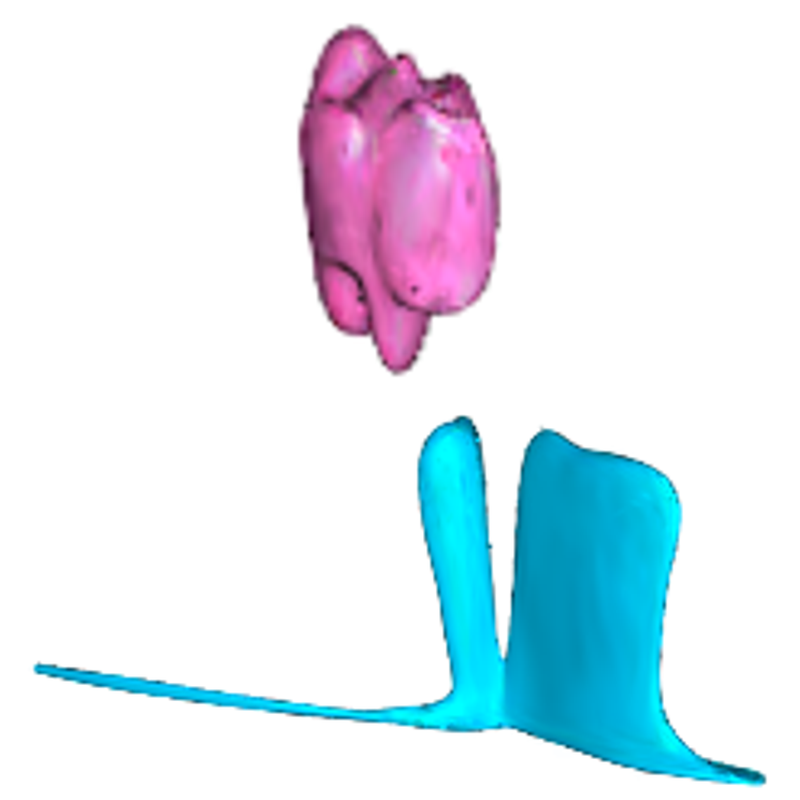}&
    \includegraphics[width=0.305\linewidth]{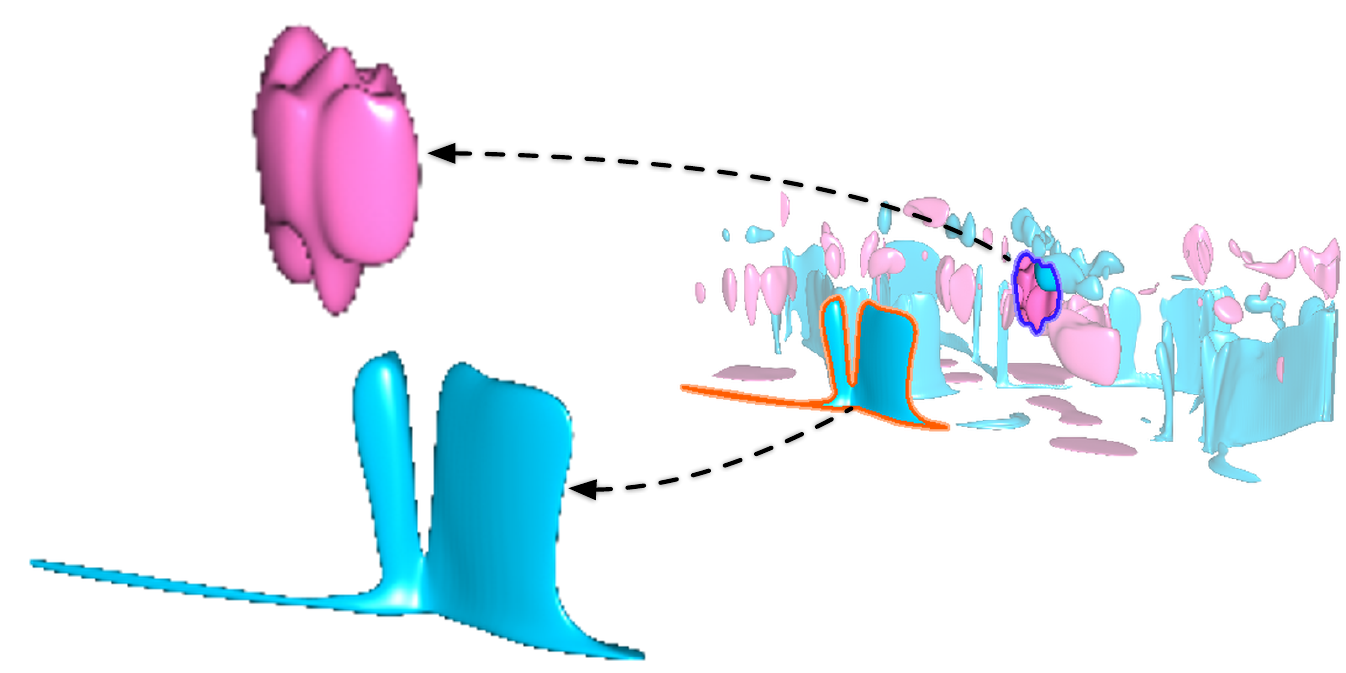}\\
    \includegraphics[width=0.1525\linewidth]{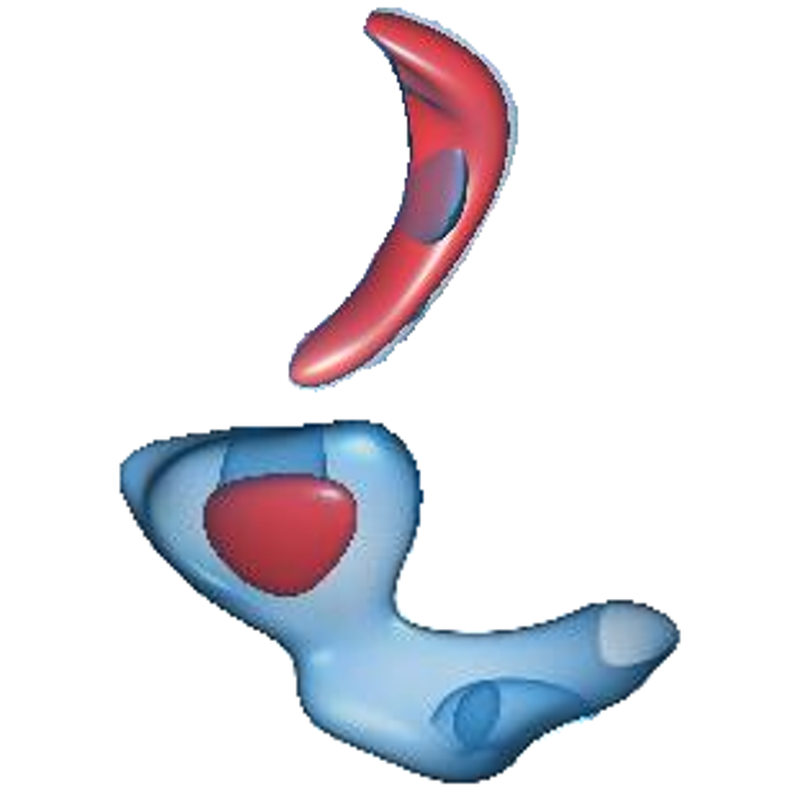}&
    \includegraphics[width=0.1525\linewidth]{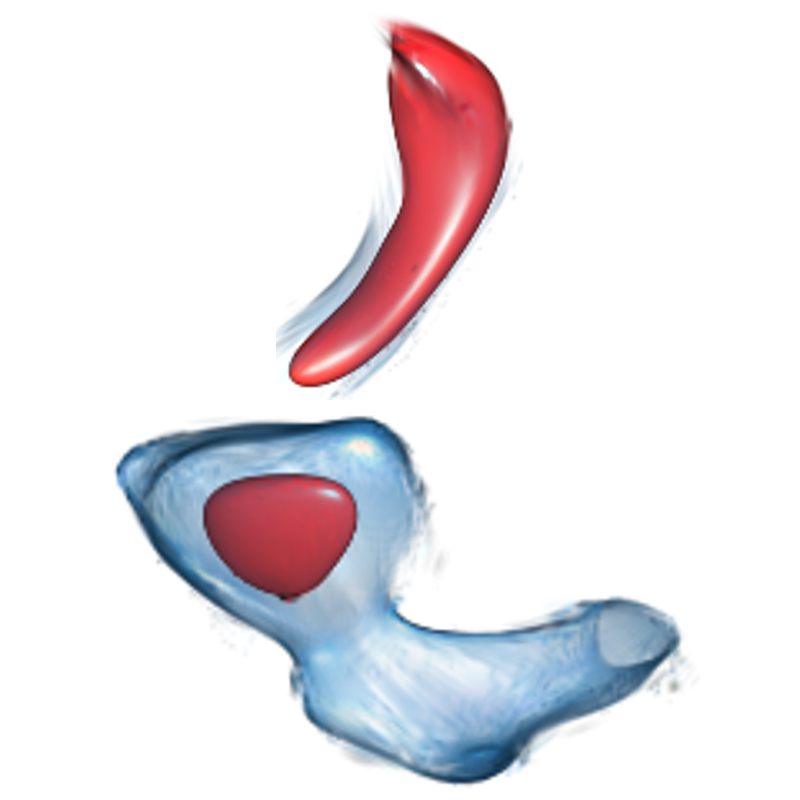}&
    \includegraphics[width=0.1525\linewidth]{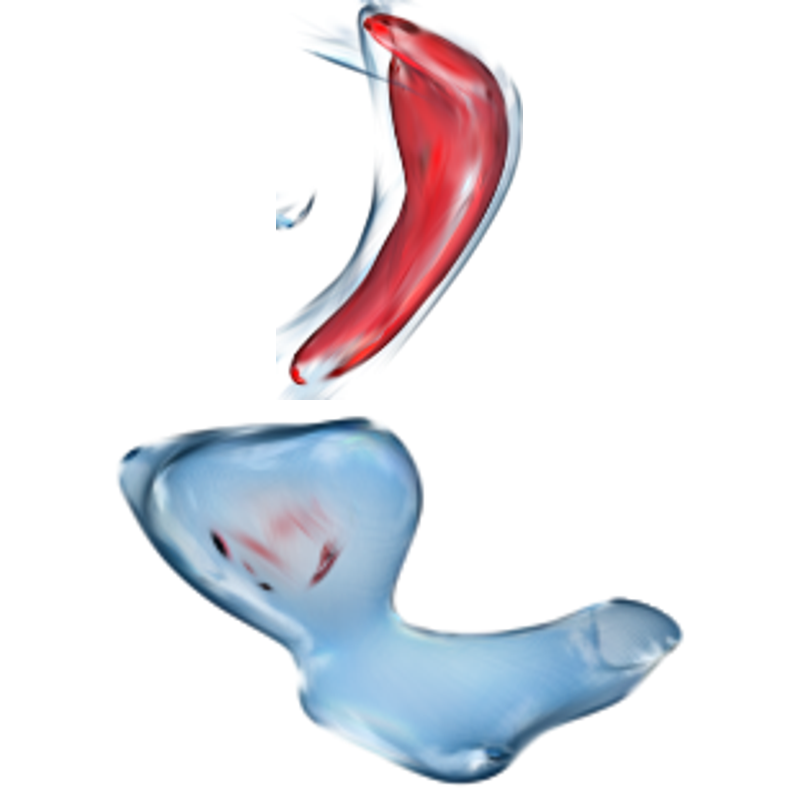}&
    \includegraphics[width=0.1525\linewidth]{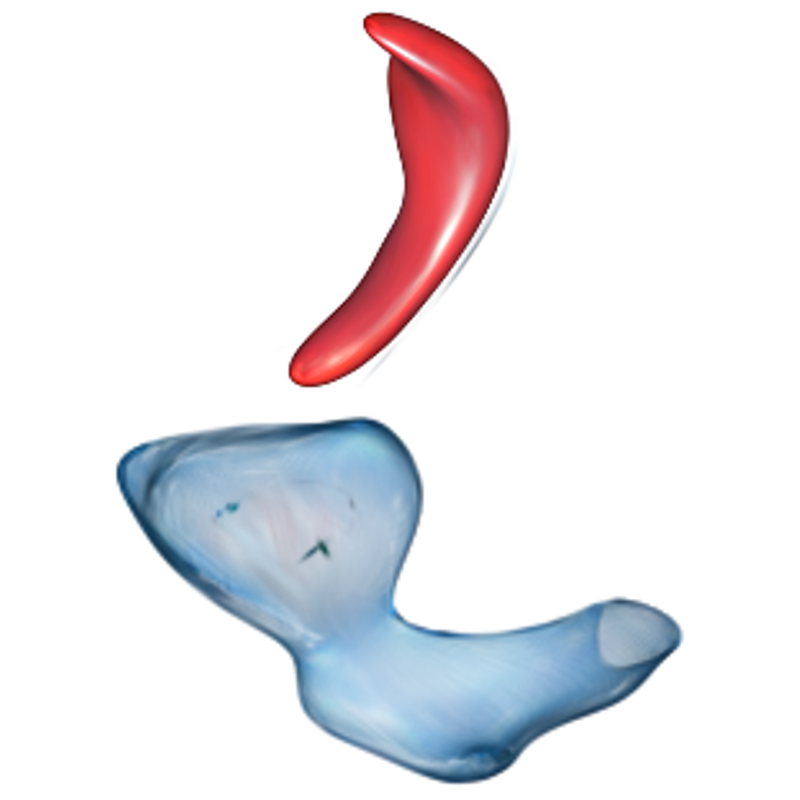}&
    \includegraphics[width=0.305\linewidth]{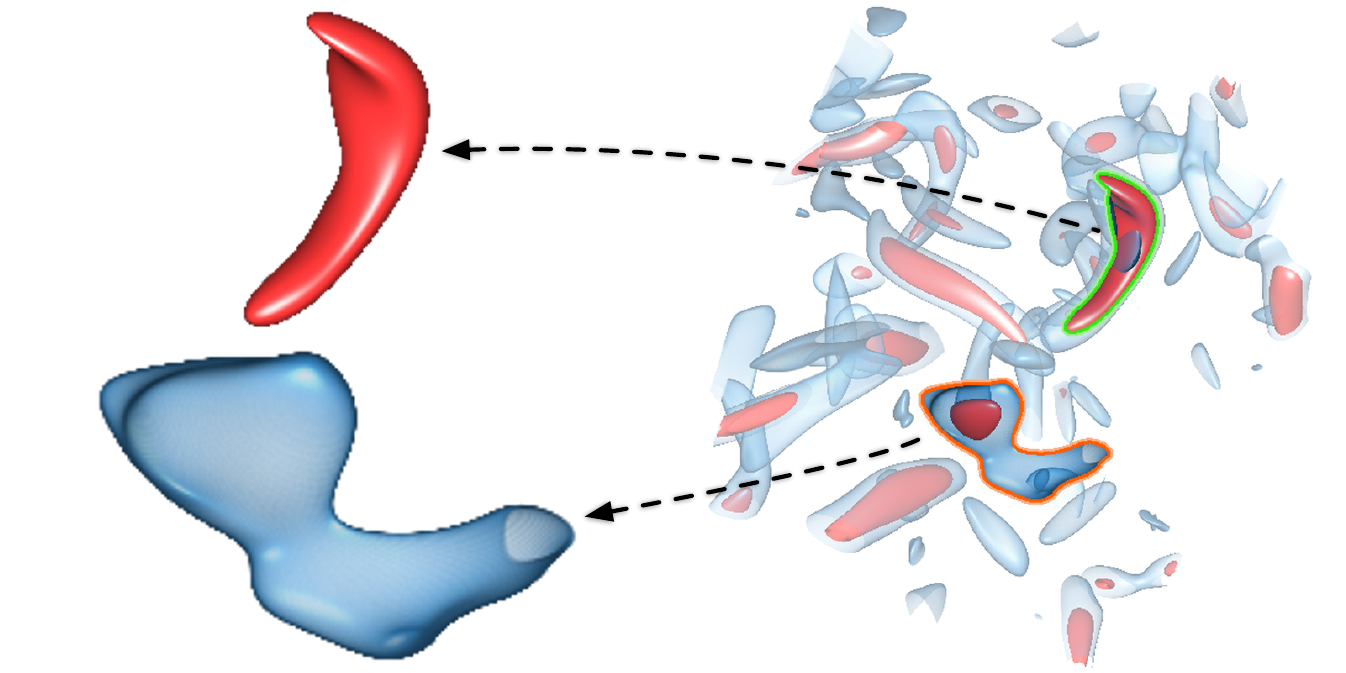}\\
    \includegraphics[width=0.1525\linewidth]{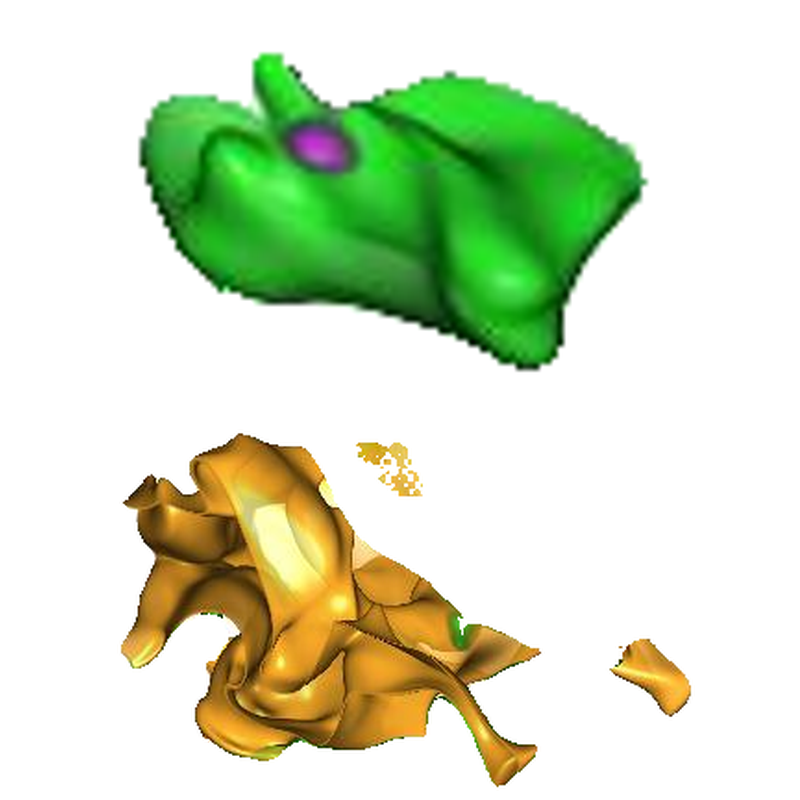}&
    \includegraphics[width=0.1525\linewidth]{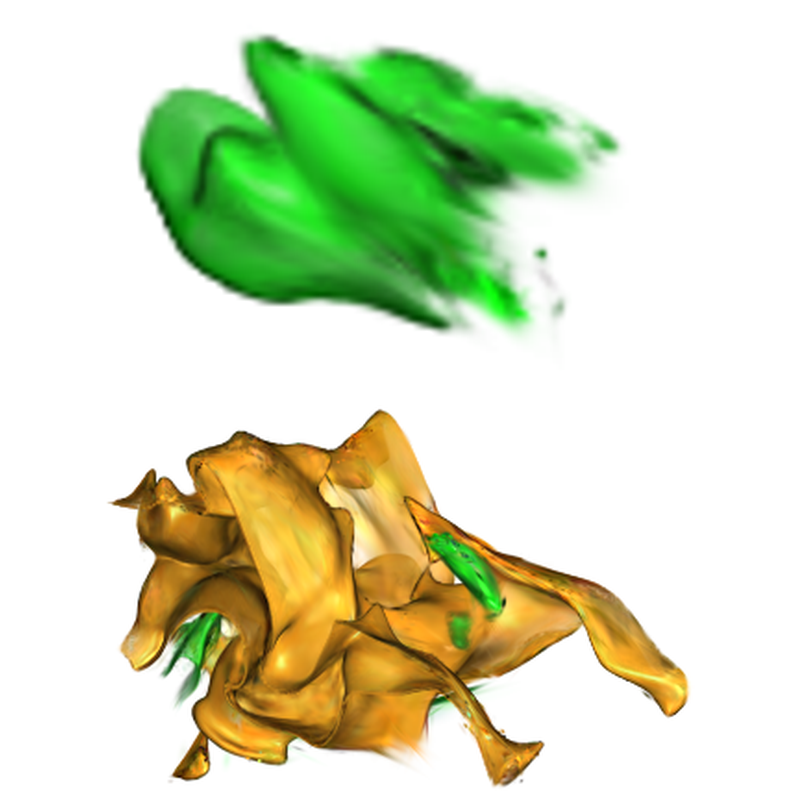}&
    \includegraphics[width=0.1525\linewidth]{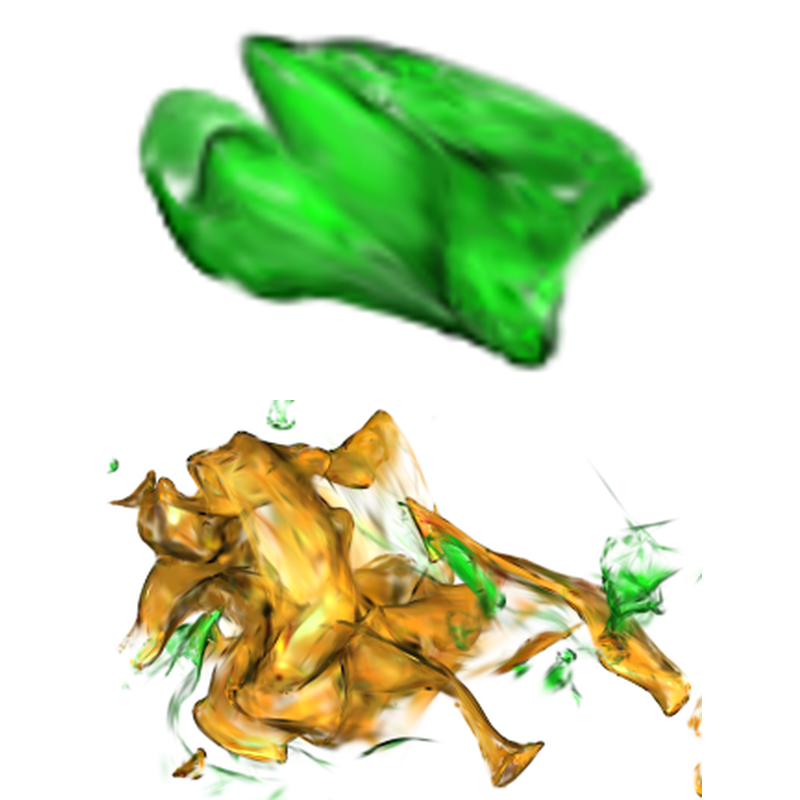}&
    \includegraphics[width=0.1525\linewidth]{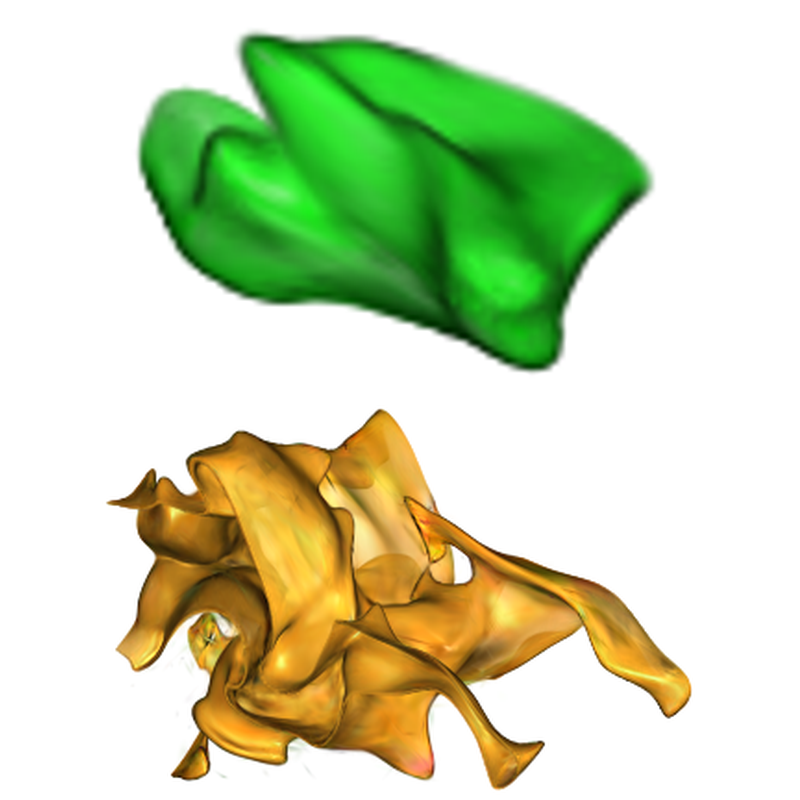}&
    \includegraphics[width=0.305\linewidth]{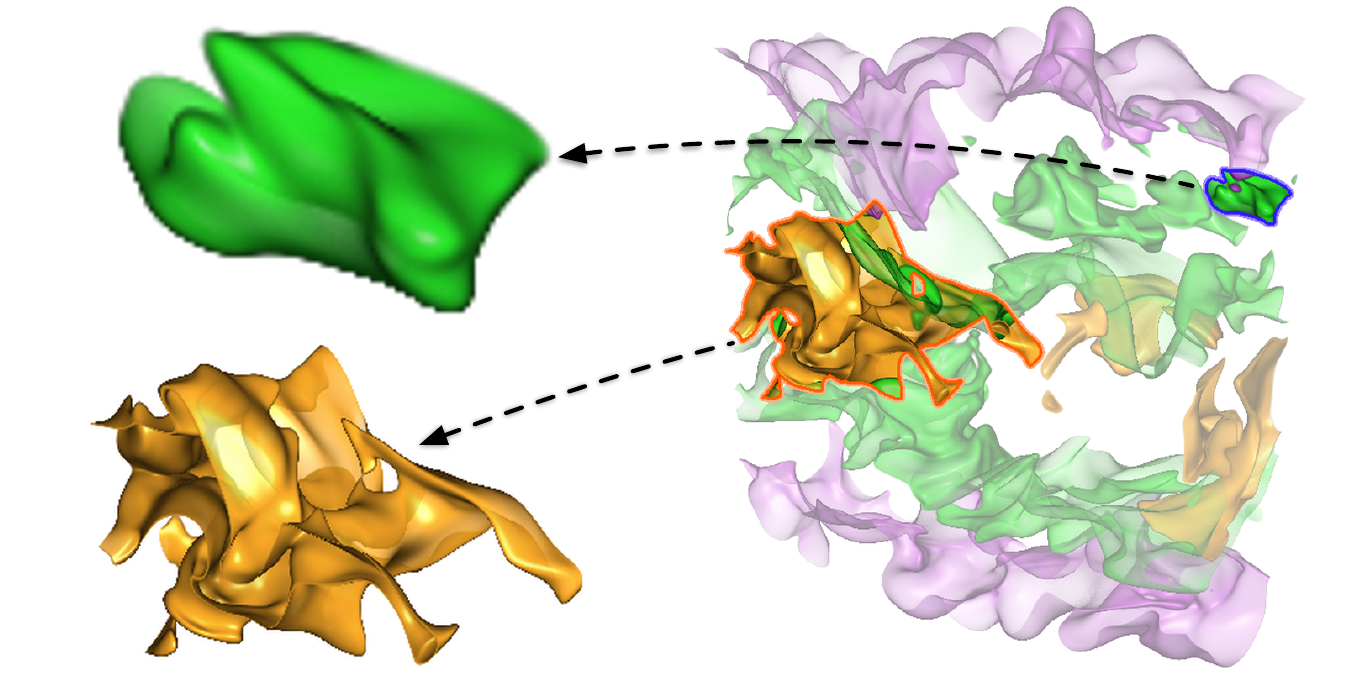}\\
    \mbox{\footnotesize (a) SAM 2} & \mbox{\footnotesize (b) SAGD} & \mbox{\footnotesize (c) SAGA} & \mbox{\footnotesize (d) VolSegGS} & \mbox{\footnotesize (e) GT}
\end{array}$
\end{center}
\vspace{-.25in} 
\caption{3D segmentation. Top to bottom: selected segmentation results of five jets, Tangaroa, mantle, vortex, and combustion.} 
\label{fig:comp-seg}
\end{figure}

% \vspace{-0.1in}
\begin{table}[htb]
    \caption{3D segmentation for static scenes: average PSNR (dB), SSIM, LPIPS, and IoU across all 181 synthesized views, training time (TT, in minutes), and segmentation time (ST, in seconds). Refer to Figure~\ref{fig:comp-seg} for the actual segments. The best ones are highlighted in bold.}
    \vspace{-0.1in}
    \centering
    %{\scriptsize
    {\fontsize{5.75pt}{5.75pt}\selectfont
    %\resizebox{\columnwidth}{!}{
    \begin{tabular}{c|c|cccccc}

        dataset         & method    & PSNR$\uparrow$ & SSIM$\uparrow$ & LPIPS$\downarrow$ & IoU$\uparrow$ & TT$\downarrow$ & ST$\downarrow$ \\ \hline
                        & SAM 2 & 26.49 & 0.983 & 0.032 & 66.99 & -- & \bf 0.05 \\
        five jets      & SAGD  & 25.70 & 0.982 & 0.032 & 74.91 & --  & 1.26 \\
        $t=61$            & SAGA   & 32.62 & 0.990 & 0.022 & 86.76 & 8.67 & 0.14 \\
        blue segment                & VolSegGS & \bf 37.49 & \bf 0.995 & \bf 0.007 & \bf 96.88 & 1.53 & 0.15 \\ \hdashline
                        & SAM 2 & 20.88 & 0.967 & 0.064 & 44.78 & -- & \bf 0.05 \\
        five jets      & SAGD  & 20.09 & 0.971 & 0.054 & 46.38 & --  & 5.13 \\
        $t=61$            & SAGA   & 29.94 & 0.983 & 0.024 & 91.95 & 8.67 & 0.14 \\
        red segment               & VolSegGS & \bf 37.06 & \bf 0.995 & \bf 0.005 & \bf 94.21 & 1.53 & 0.16 \\ \hline
                        & SAM 2 & 29.89 & 0.993 & 0.029 & 20.34 & -- & \bf 0.05 \\
        Tangaroa      & SAGD  & 18.14 & 0.946 & 0.098 & 2.50 & -- & 7.44 \\
        $t=1$            & SAGA   & 23.68 & 0.969 & 0.067 & 4.69 & 9.81 & 0.15 \\
        yellow segment                & VolSegGS & \bf 46.94 & \bf 0.999 & \bf 0.005 & \bf 62.45 & 1.85 & 0.15 \\ \hdashline
                        & SAM 2 & 29.19 & 0.988 & 0.024 & 90.91 & -- & \bf 0.05 \\
        Tangaroa      & SAGD  & 29.76 & 0.984 & 0.027 & 92.95 & -- & 11.52 \\
        $t=1$            & SAGA   & 29.16 & 0.980 & 0.020 & 97.64 & 9.81 & 0.14 \\
        blue segment                & VolSegGS & \bf 37.60 & \bf 0.996 & \bf 0.006 & \bf 97.71 & 1.85 & 0.15 \\ \hline
                    & SAM 2 & 34.28 & 0.996 & 0.019 & 36.67 & -- & \bf 0.05 \\
        mantle      & SAGD  & 42.14 & \bf 0.999 & 0.002 & 89.59 & -- & 1.05 \\
        $t=1$            & SAGA   & 38.25 & 0.996 & 0.043 & 61.27 & 9.21 & 0.14 \\
        pink segment                & VolSegGS & \bf 42.59 & \bf 0.999 & \bf 0.001 & \bf 90.39 & 1.67 & 0.15 \\ \hdashline
                        & SAM 2 & 32.93 & 0.995 & 0.016 & 72.34 & -- & \bf 0.04 \\
        mantle      & SAGD  & 33.84 & 0.995 & 0.016 & 78.54 & -- & 3.27 \\
        $t=1$            & SAGA   & 32.31 & 0.992 & 0.057 & 62.89 & 9.21 & 0.14 \\
        cyan segment                & VolSegGS & \bf 39.08 & \bf 0.998 & \bf 0.003 & \bf 87.75 & 1.67 & 0.15 \\ \hline
                        & SAM 2 & 29.50 & 0.994 & 0.033 & 40.79 & -- & \bf 0.05 \\
        vortex      & SAGD  & 35.02 & 0.996 & 0.008 & 68.60 & -- & 1.41 \\
        $t=31$            & SAGA   & 31.98 & 0.991 & 0.019 & 54.75 & 10.72 & 0.14 \\
        red segment                & VolSegGS & \bf 41.03 & \bf 0.999 & \bf 0.003 & \bf 83.56 & 1.78 & 0.15 \\ \hdashline
                        & SAM 2 & 31.74 & 0.994 & 0.012 & 84.35 & -- & \bf 0.05 \\
        vortex      & SAGD  & 33.04 & 0.994 & 0.009 & 88.67 & -- & 4.37 \\
        $t=31$            & SAGA   & 32.90 & 0.992 & 0.048 & 81.59 & 10.72 & 0.14 \\
        blue segment                & VolSegGS & \bf 39.05 & \bf 0.997 & \bf 0.005 & \bf 95.76 & 1.78 & 0.15 \\ \hline        
                        & SAM 2 & 30.92 & 0.996 & 0.017 & 34.77 & -- & \bf 0.05 \\
        combustion      & SAGD  & 33.92 & 0.998 & 0.005 & 84.45 & -- & 2.21 \\
        $t=100$            & SAGA   & 36.99 & 0.998 & 0.007 & 77.60 & 10.01 & 0.14 \\
        green segment                & VolSegGS & \bf 45.74 & \bf 0.999 & \bf 0.001 & \bf 91.30 & 1.55 & 0.15 \\ \hdashline   
                        & SAM 2 & 19.80 & 0.964 & 0.087 & 35.98 & -- & \bf 0.05 \\
        combustion      & SAGD  & 28.08 & 0.977 & 0.025 & 90.41 & -- & 7.55 \\
        $t=100$            & SAGA   & 23.45 & 0.942 & 0.098 & 65.33 & 10.01 & 0.15 \\
        yellow segment                & VolSegGS & \bf 32.02 & \bf 0.984 & \bf 0.015 & \bf 92.65 & 1.55 & 0.15 \\
  \end{tabular}
    }
    \label{tab:metrics-seg}
\end{table}

{\bf Qualitative results.}
Figure~\ref{fig:comp-seg} highlights VolSegGS's superior segmentation performance compared to SAM 2, SAGD, and SAGA. 
% \hot{We primarily discuss the segmentation results presented in the top sub-row of each dataset.}
Due to minimal occlusions, all methods perform well for the least challenging mantle dataset. 
However, VolSegGS excels in preserving the most accurate shape. 
Specifically, VolSegGS retains most of the original 3D Gaussians for the mantle dataset's cyan segment, whereas SAGA and SAGD either omit necessary Gaussians or introduce extraneous ones. 
This advantage stems from VolSegGS's MLP-based deformation field network, which enables smooth and precise spatial segmentation. 
In contrast, methods relying entirely on explicit features of 3D Gaussians tend to introduce significant noise. 
A similar trend is observed in both segments of the five jets dataset: although SAGA performs relatively well, it still misses some Gaussians and introduces unnecessary ones, ultimately degrading the quality.

All baseline methods struggle to separate segments beneath overlying layers for the yellow segment of the Tangaroa dataset. 
VolSegGS, however, successfully splits segments beneath thin, semi-transparent layers by leveraging its two-level segmentation strategy. 
Despite verifying the correctness of input prompts using SAM masks, SAM 2 and SAGD still fail to produce accurate segmentation. 
SAM 2 fails due to inherent ambiguity in segment tracking, while SAGD's voting strategy proves ineffective in resolving segmentation under such conditions. 
SAGA tries to segment occluded parts but is significantly impacted by noisy affinity features on Gaussians, leading to suboptimal results.

The two-level segmentation strategy of VolSegGS also effectively removes inner parts of different colors within semi-transparent segments in other datasets.
SAM 2 and SAGD struggle to remove the red part for the blue segment of the vortex dataset because they rely on 2D input prompts. 
If users manually exclude the red part, the resulting mask remains incomplete and fails to fully capture the blue segment, leading to suboptimal segmentation. 
SAGA, on the other hand, leverages affinity features that effectively distinguish the blue part from the red. 
However, the red part is not entirely removed due to noise, leaving behind residual artifacts.
Similar phenomena can also be observed from the blue segments of the five jets and the Tangaroa datasets.
For the yellow segment of the combustion dataset, the close proximity of the green and yellow parts causes segmentation errors when using 2D masks for supervision. 
During projection onto 3D Gaussians, this overlap results in unintended blending, hindering SAGD and SAGA from extracting the yellow segment cleanly without interference from the green part.

{\bf Quantitative results.}
Table~\ref{tab:metrics-seg} presents the quantitative evaluation of 3D segmentation quality across all datasets. 
VolSegGS consistently outperforms baseline methods across all datasets in terms of PSNR, SSIM, LPIPS, and IoU, aligning with the qualitative results. 
From an efficiency standpoint, SAGA and VolSegGS optimize affinity features for segmenting 3D Gaussians. 
However, VolSegGS achieves faster training times than SAGA due to its affinity field network, which ensures smoothness and consistency among neighboring Gaussians. 
In contrast, SAGA explicitly optimizes affinity features for individual Gaussians, requiring more iterations to achieve the same level of consistency.
Although SAGD does not require training, it is less efficient than VolSegGS in segmentation speed. 
Each time users provide a point prompt, SAGD introduces noticeable delays compared to SAGA and VolSegGS, making real-time interaction less responsive.

{\bf Summary.}
\hot{Across all static visualization scenes, VolSegGS consistently outperforms baseline segmentation methods in terms of segmentation quality. 
It offers a reliable solution for 3D segmentation in static scenes and establishes a strong foundation for enabling interactive segmentation and tracking in exploratory visualization.}

\vspace{-0.05in}
\subsection{Segment Tracking for Dynamic Scenes}
In this section, we present the tracking results of VolSegGS on dynamic scenes.
While we acknowledge a few unpublished concurrent works on 4D segmentation with Gaussians~\cite{Ji-arXiv24, Li-arXiv24}, we cannot directly compare them due to their lack of open-source code.
Instead, we present three use-case scenarios to illustrate the accuracy and robustness of VolSegGS in segment-tracking tasks.

\begin{figure}[htb]
\vspace{-0.1in}
\begin{center}
$\begin{array}{c@{\hspace{0.01in}}c@{\hspace{0.01in}}c@{\hspace{0.01in}}c@{\hspace{0.01in}}c}
    \includegraphics[width=0.19\linewidth]{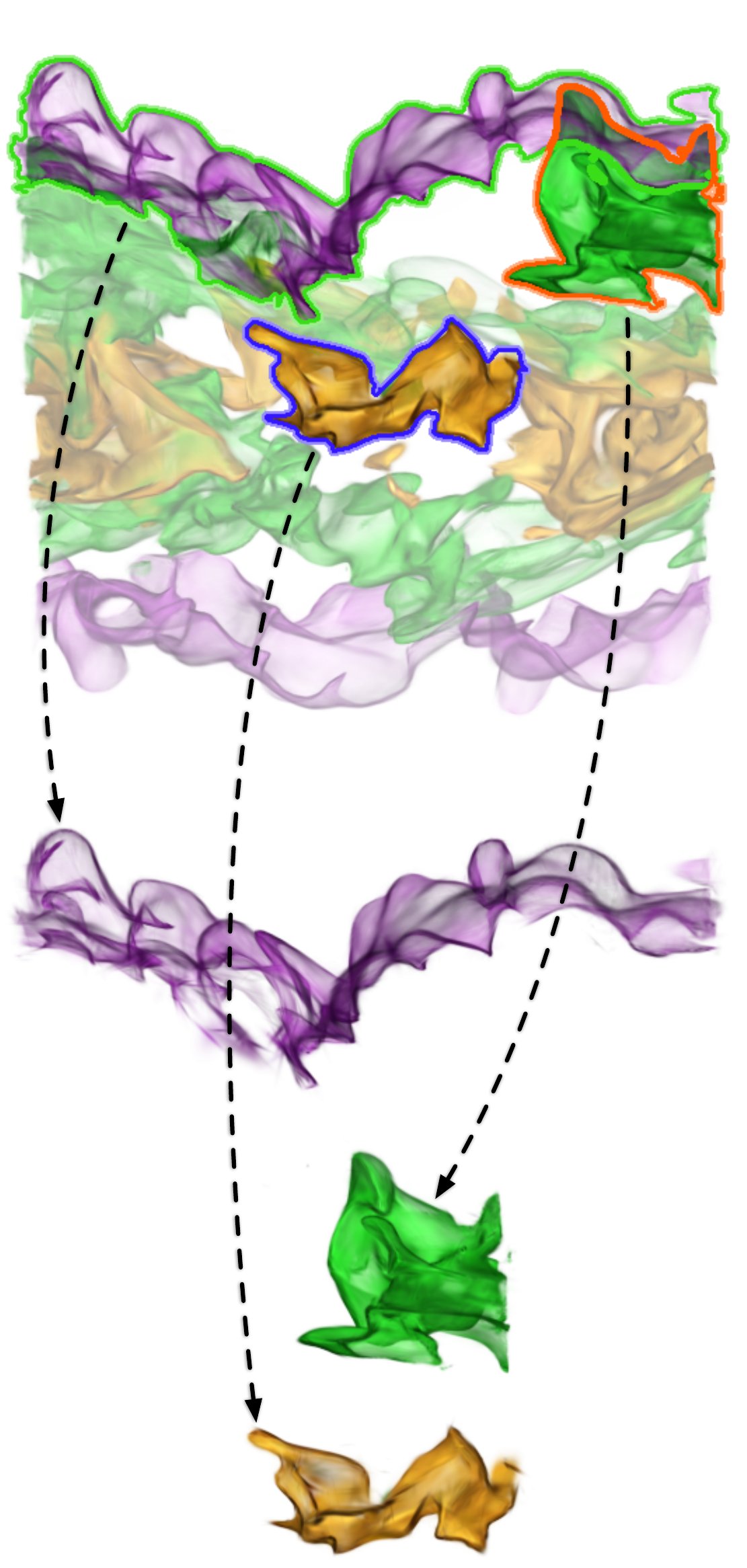}&
    \includegraphics[width=0.19\linewidth]{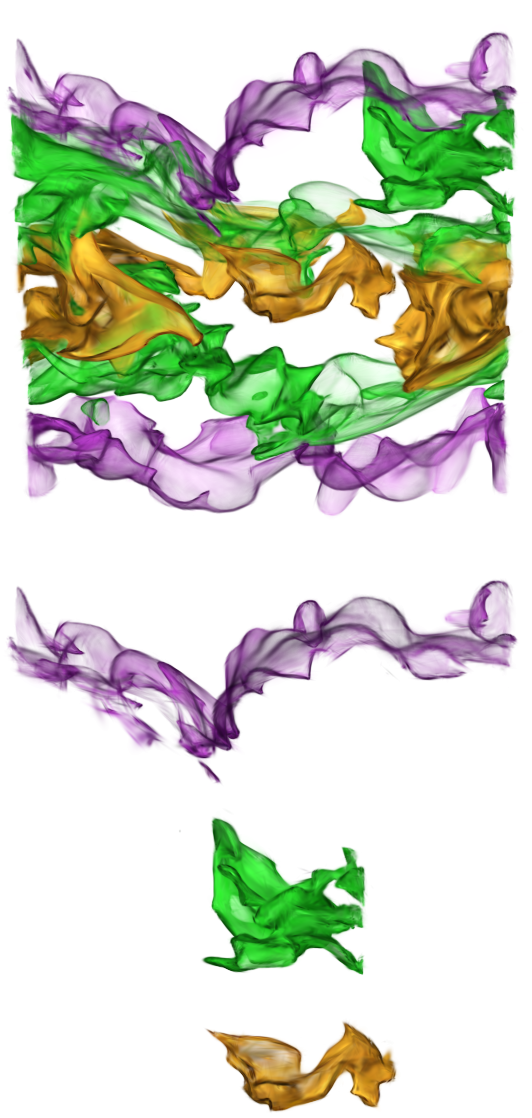}&
    \includegraphics[width=0.19\linewidth]{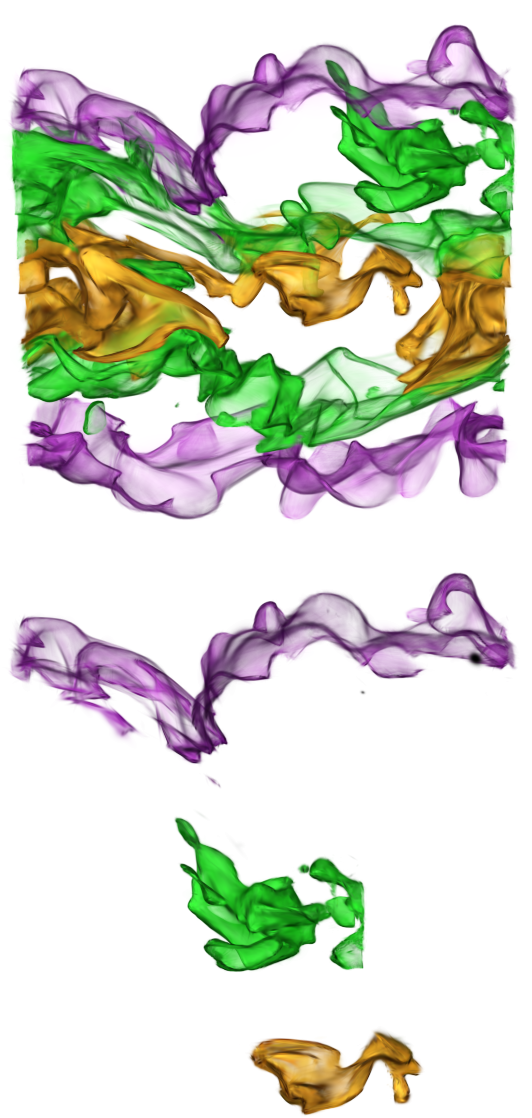}&
    \includegraphics[width=0.19\linewidth]{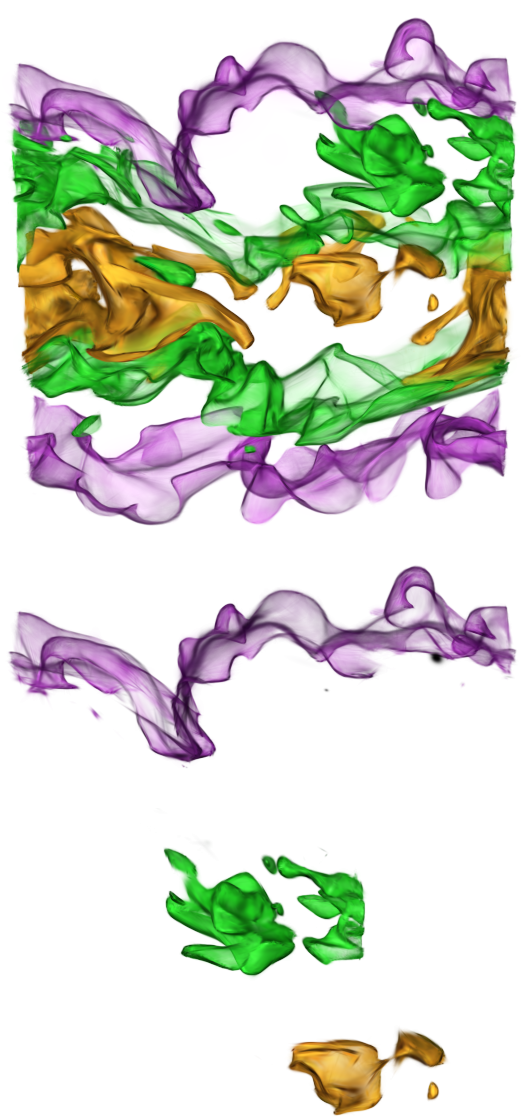}&
    \includegraphics[width=0.19\linewidth]{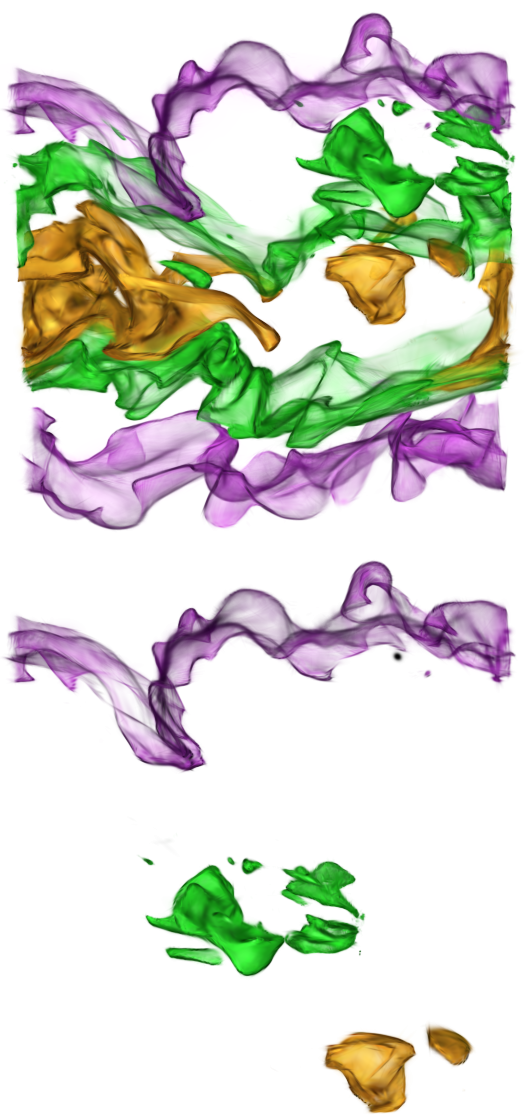}\\
    \mbox{\footnotesize (a) $t=20$} & \mbox{\footnotesize (b) $t=40$} & \mbox{\footnotesize (c) $t=60$} & \mbox{\footnotesize (d) $t=80$} & \mbox{\footnotesize (e) $t=100$}
\end{array}$
\end{center}
\vspace{-.25in} 
\caption{Single-segment tracking with combustion. Top row: the full scene. Rest rows: tracking results of individual segments.} 
\label{fig:tracking-combustion}
\end{figure}

{\bf Single-segment tracking.}
In this scenario, we select three segments at timestep 20 from the combustion dataset and track each segment individually until timestep 100.
Figure~\ref{fig:tracking-combustion} presents the masked results of the full scene and individually tracked segments across five selected timesteps.
\hot{The results in Table~\ref{tab:tracking-combustion} indicate robust tracking performance, with all individual segments maintaining an IoU above 80 throughout the sequence.}
By focusing on single-segment tracking, users can analyze the evolution of specific segments without distractions from the surrounding scene.

\vspace{-0.2in}
\begin{table}[htb]
    \caption{\hot{Single-segment tracking with combustion: average PSNR (dB), SSIM, LPIPS, and IoU across 181 synthesized views.}}
    \vspace{-0.1in}
    \centering
    %{\scriptsize
    \resizebox{2.25in}{!}{
    % \resizebox{\columnwidth}{!}{
    \begin{tabular}{c|c|cccc}
        \hot{segment} & \hot{timestep} & \hot{PSNR$\uparrow$} & \hot{SSIM$\uparrow$} & \hot{LPIPS$\downarrow$} & \hot{IoU$\uparrow$} \\ \hline
        & \hot{20} & \hot{29.46} & \hot{0.963} & \hot{0.045} & \hot{86.86} \\
        & \hot{40} & \hot{29.23} & \hot{0.960} & \hot{0.049} & \hot{85.77} \\
        \hot{purple} & \hot{60} & \hot{29.07} & \hot{0.964} & \hot{0.045} & \hot{88.08} \\
        & \hot{80} & \hot{28.42} & \hot{0.960} & \hot{0.042} & \hot{88.46}\\
        & \hot{100} & \hot{29.04} & \hot{0.959} & \hot{0.048} & \hot{86.96}\\ \hline
        & \hot{20} & \hot{33.13} & \hot{0.989} & \hot{0.015} & \hot{88.01} \\
        & \hot{40} & \hot{34.55} & \hot{0.992} & \hot{0.011} & \hot{87.55} \\
        \hot{yellow} & \hot{60} & \hot{35.05} & \hot{0.992} & \hot{0.011} & \hot{88.46} \\
        & \hot{80} & \hot{33.60} & \hot{0.992} & \hot{0.011} & \hot{83.59}\\
        & \hot{100} & \hot{36.61} & \hot{0.995} & \hot{0.011} & \hot{86.63}\\ \hline
        & \hot{20} & \hot{33.81} & \hot{0.986} & \hot{0.014} & \hot{91.51} \\
        & \hot{40} & \hot{32.86} & \hot{0.986} & \hot{0.013} & \hot{88.87} \\
        \hot{green} & \hot{60} & \hot{32.34} & \hot{0.986} & \hot{0.013} & \hot{87.53} \\
        & \hot{80} & \hot{30.33} & \hot{0.982} & \hot{0.012} & \hot{86.08}\\
        & \hot{100} & \hot{33.53} & \hot{0.989} & \hot{0.011} & \hot{83.77}\\
        \end{tabular}
    }
    \label{tab:tracking-combustion}
\end{table}

\begin{figure}[htb]
%\vspace{-0.1in}
\begin{center}
\setlength{\tabcolsep}{0cm}
\begin{tabular}{ccccc}
\multirow{4}{7em}{\vspace{-0.35in} \includegraphics[width=\linewidth]{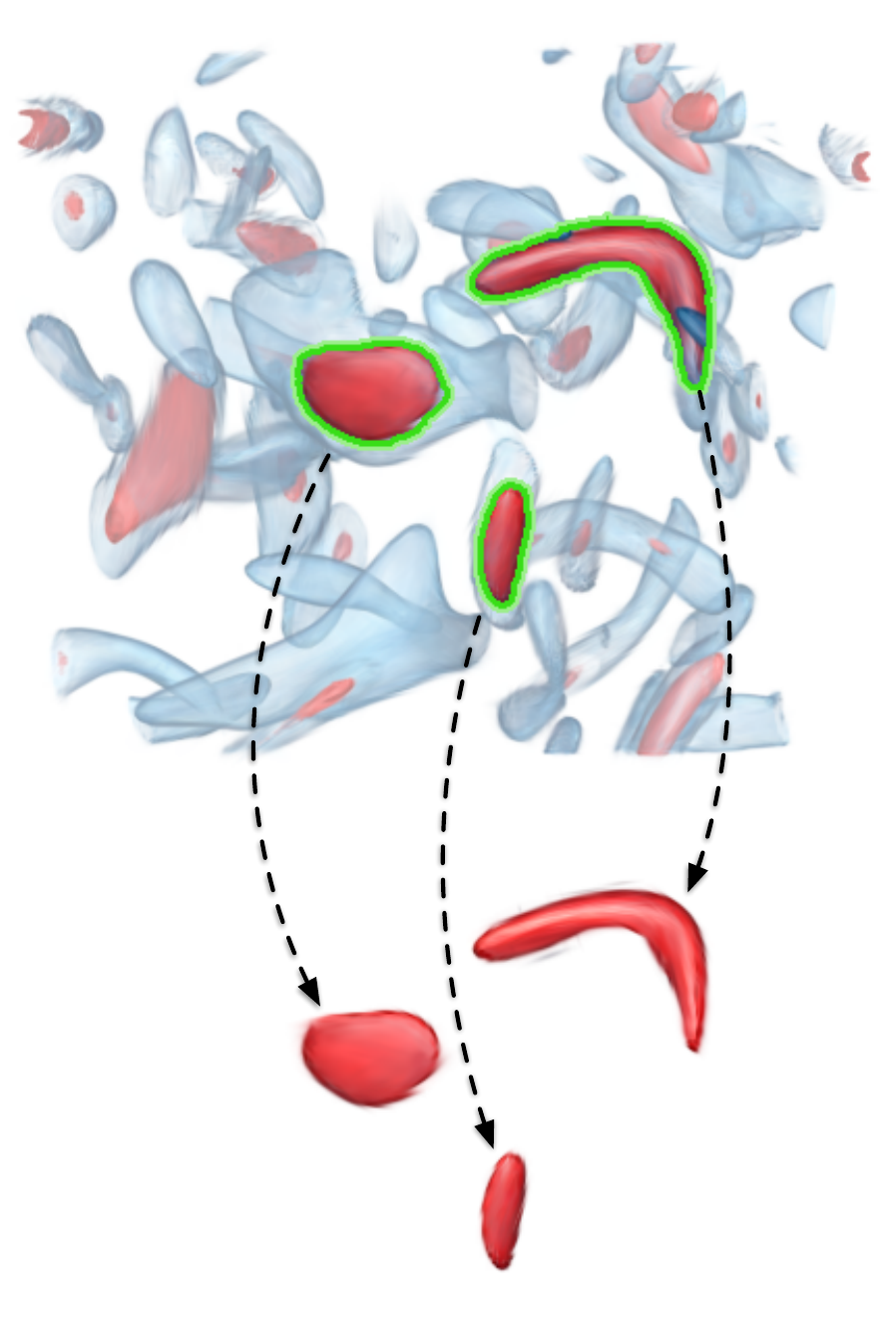}} &
    \includegraphics[width=0.1825\linewidth]{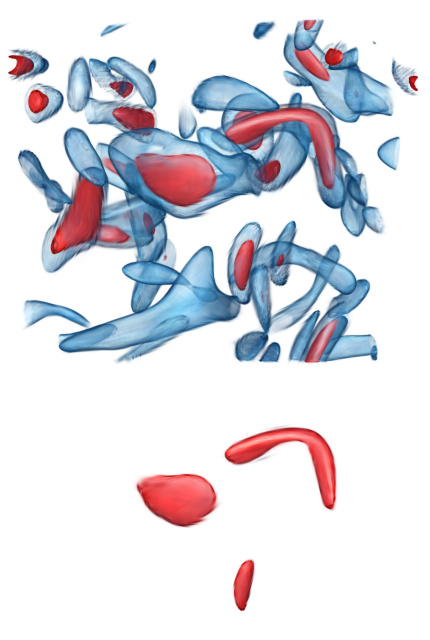} &
    \includegraphics[width=0.1825\linewidth]{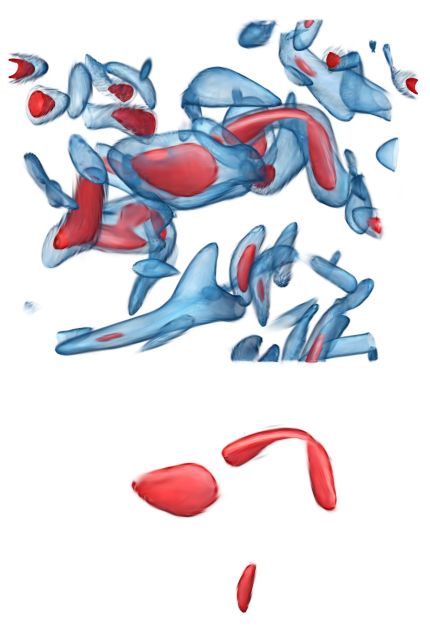} &
    \includegraphics[width=0.1825\linewidth]{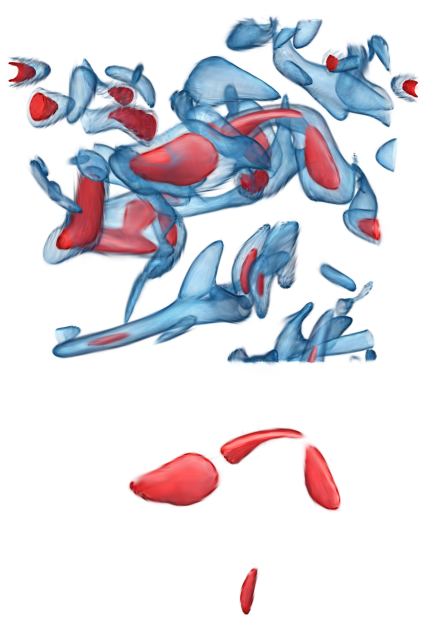} &
    \includegraphics[width=0.1825\linewidth]{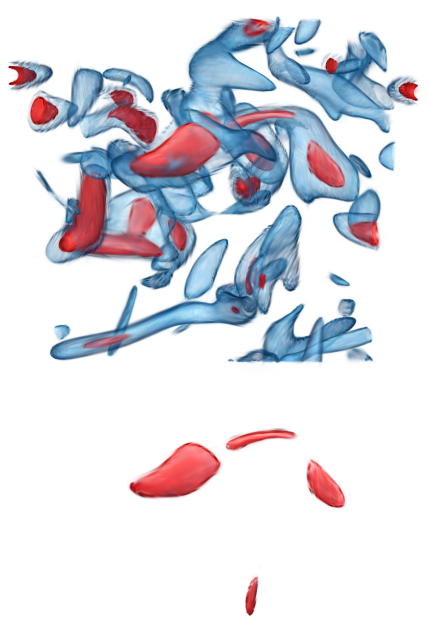} \\
    & \mbox{\footnotesize (b) $t=60$} & \mbox{\footnotesize (c) $t=70$} & \mbox{\footnotesize (d) $t=80$} & \mbox{\footnotesize (e) $t=90$} \\
    \mbox{\footnotesize (a) $t=50$} & \includegraphics[width=0.1825\linewidth]{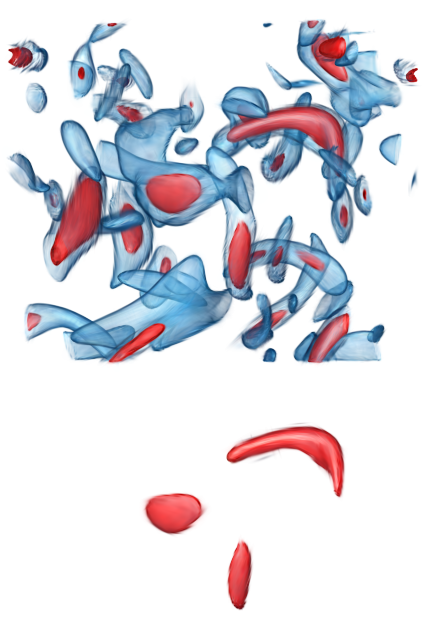} & 
\includegraphics[width=0.1825\linewidth]{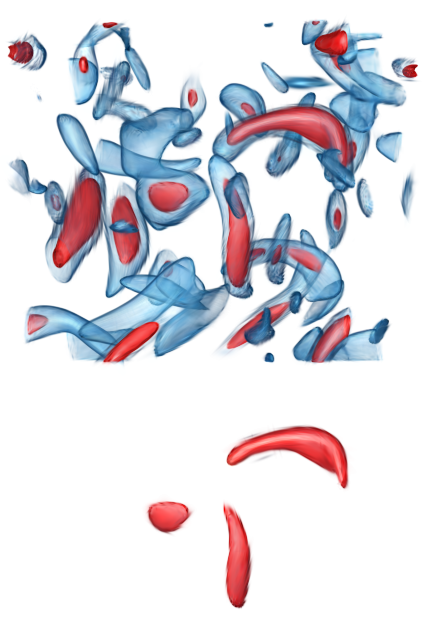} &
\includegraphics[width=0.1825\linewidth]{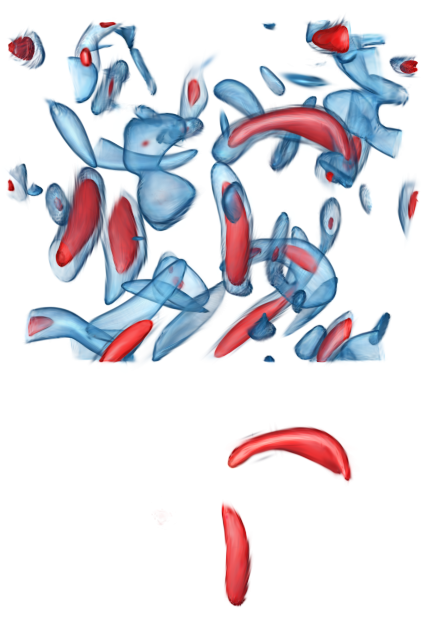} &
\includegraphics[width=0.1825\linewidth]{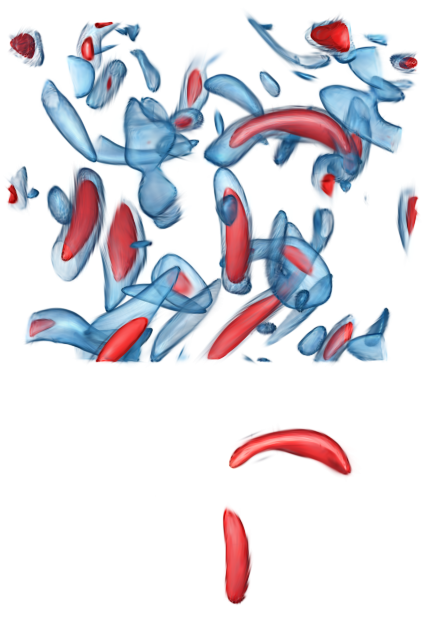} \\ 
& \mbox{\footnotesize (f) $t=40$} & \mbox{\footnotesize (g) $t=30$} & \mbox{\footnotesize (h) $t=20$} & \mbox{\footnotesize (i) $t=10$} \\
\vspace{-.25in}
\end{tabular}
\caption{Grouped segment tracking with vortex. (a) shows segmentation processed at timestep 50. (b) to (e) are forward tracking results, and (f) to (i) are backward tracking results.}
\label{fig:tracking-vortex}
\end{center}
\vspace{-.25in}
\end{figure}

{\bf Grouped segment tracking.}
In this scenario, we select multiple segments at timestep 50 from the vortex dataset and group them for tracking as a whole over time.
As shown in the top row of Figure~\ref{fig:tracking-vortex}, for the forward tracking, we observe the segment on the top-right corner splitting into two segments at timestep 80.
In fact, if we start with four segments at timestep 90 and do backward tracking, the results will be the same as shown in (d) to (a), and the two segments on the right join into one segment at timestep 70.
For the backward tracking, the segment in the middle disappears at timestep 20.
VolSegGS maintains the group as a whole, ensuring that the segmentation is not disrupted by individual segments' split, join, or disappearance. 
Changes in the rest of the scene also do not affect the tracked group.
\hot{Table~\ref{tab:tracking-vortex} shows that the grouped segment consistently achieves an IoU above 80 in both directions, reflecting reliable and robust tracking performance.}
With grouped segment tracking, users can perform a comparative analysis of multiple segments and study their collective behaviors over time.

\begin{table}[htb]
    \caption{\hot{Grouped segment tracking with vortex: average PSNR (dB), SSIM, LPIPS, and IoU across 181 synthesized views.}}
    \vspace{-0.1in}
    \centering
    %{\scriptsize
    \resizebox{2in}{!}{
    % \resizebox{\columnwidth}{!}{
    \begin{tabular}{c|cccc}
        \hot{timestep} & \hot{PSNR$\uparrow$} & \hot{SSIM$\uparrow$} & \hot{LPIPS$\downarrow$} & \hot{IoU$\uparrow$} \\ \hline
        \hot{10} & \hot{37.04} & \hot{0.996} & \hot{0.008} & \hot{82.07} \\
        \hot{20} & \hot{36.28} & \hot{0.996} & \hot{0.007} & \hot{84.01} \\
        \hot{30} & \hot{34.12} & \hot{0.994} & \hot{0.009} & \hot{82.15} \\
        \hot{40} & \hot{33.87} & \hot{0.994} & \hot{0.009} & \hot{82.42} \\
        \hot{50} & \hot{33.99} & \hot{0.994} & \hot{0.009} & \hot{82.09} \\
        \hot{60} & \hot{35.28} & \hot{0.995} & \hot{0.009} & \hot{83.09} \\
        \hot{70} & \hot{34.93} & \hot{0.995} & \hot{0.009} & \hot{82.28} \\
        \hot{80} & \hot{34.81} & \hot{0.995} & \hot{0.009} & \hot{81.65} \\
        \hot{90} & \hot{34.96} & \hot{0.995} & \hot{0.009} & \hot{80.35} \\
        \end{tabular}
    }
    \label{tab:tracking-vortex}
\end{table}

\begin{figure}[htb]
\vspace{-0.1in}
\begin{center}
$\begin{array}{c@{\hspace{0.01in}}c@{\hspace{0.01in}}c@{\hspace{0.01in}}c}
    \includegraphics[width=0.235\linewidth]{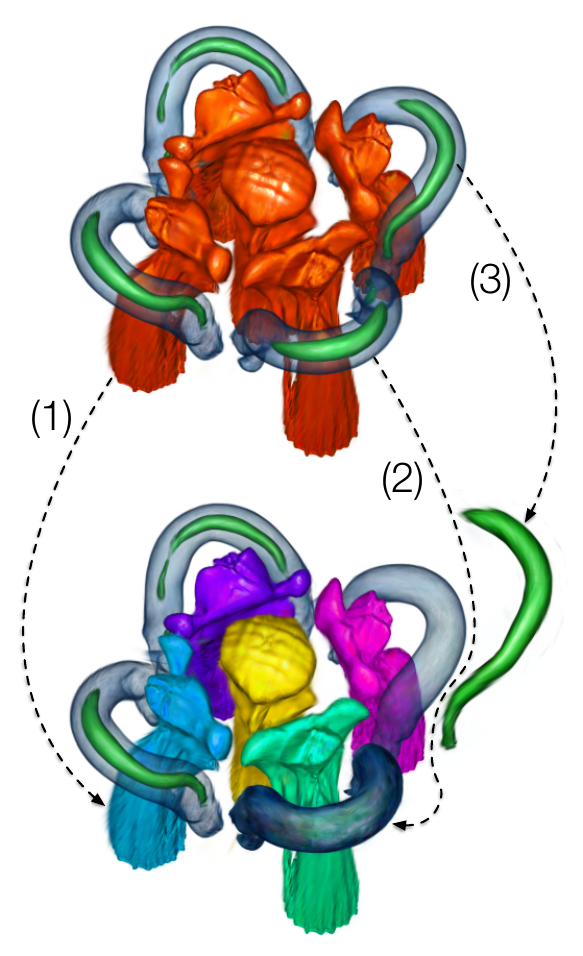}&
    \includegraphics[width=0.235\linewidth]{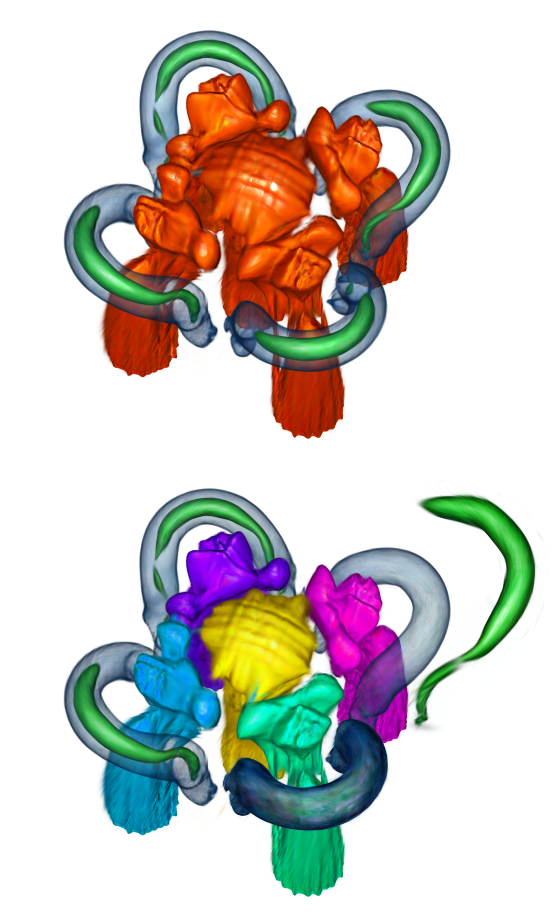}&
    \includegraphics[width=0.235\linewidth]{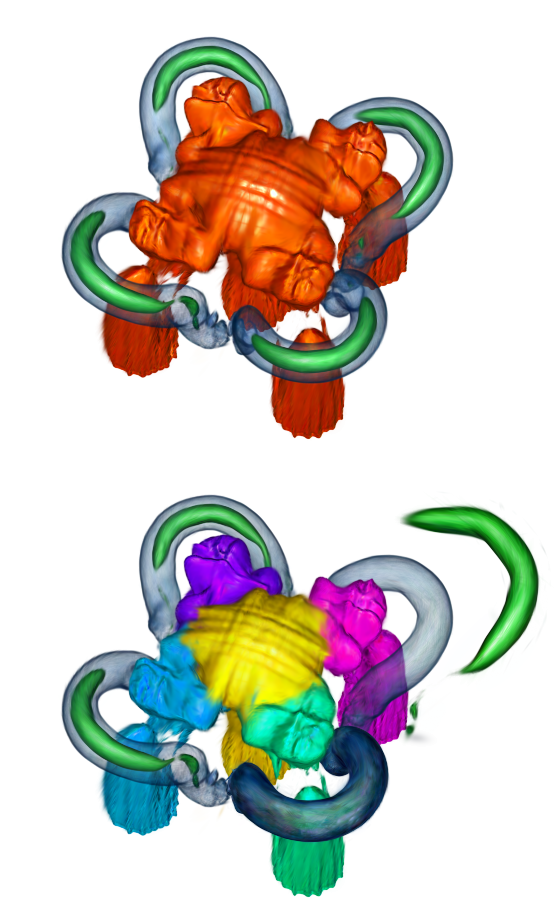}&
    \includegraphics[width=0.235\linewidth]{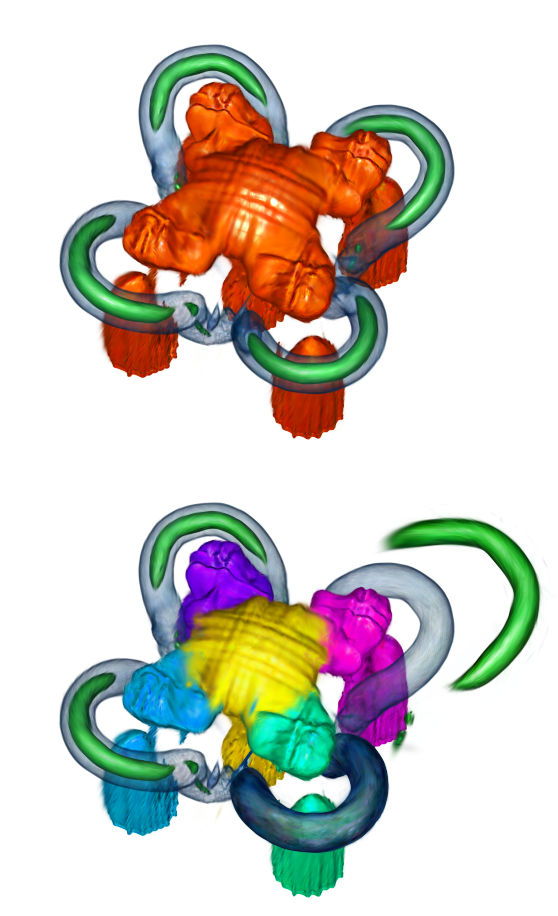}\\
    \mbox{\footnotesize (a) $t=60$} & \mbox{\footnotesize (b) $t=40$} & \mbox{\footnotesize (c) $t=20$} & \mbox{\footnotesize (d) $t=1$}
\end{array}$
\end{center}
\vspace{-.25in} 
\caption{Edited segment tracking with five jets. 
%Top row: the original scene. Bottom row: the modified scene with edited segments being tracked. 
Each edited segment is modified with a different (1) color, (2) opacity, or (3) transformation.} 
\label{fig:tracking-fivejets}
\end{figure}

{\bf Edited segment tracking.} 
In this scenario, we demonstrate that VolSegGS enables versatile editing of segments while effectively tracking the edited segments across multiple timesteps. 
To illustrate this capability, we select several segments from the five jets dataset at timestep 60. 
As shown in Figure~\ref{fig:tracking-fivejets}, we first (1) divide the red region into five segments, assigning each a unique color. 
Next, we (2) increase the opacity of a specific segment within the blue region. 
Finally, we (3) transform a segment from the green region, repositioning it away from the center and scaling it up. 
The backward tracking results from timestep 60 reveal that these edited segments remain consistent throughout the dynamic scene. 
For instance, the recolored segments remain visually distinguishable even when merging at the top and splitting vertically at timestep 20. 
The segment with increased opacity clearly maintains its opacity over time. 
Additionally, the transformed segment is consistently visible at an enlarged scale and remains separate from the main structure. 
This demonstrates that VolSegGS allows users to conveniently edit segments of interest at any timestep, with modifications seamlessly propagated across all other timesteps.

{\bf Summary.}
We show the tracking capability of VolSegGS through three use-case scenarios: single, grouped, and edited segment tracking.
VolSegGS effectively tracks individual, group, and edited segments over time, establishing a reliable foundation for the interactive exploration of dynamic scenes.
Furthermore, VolSegGS enables real-time segment tracking throughout the dynamic scene by following the deformation of 3D Gaussians. 
This real-time tracking capability greatly enhances the user experience, facilitating segment evolution analysis in dynamic scenes and enabling comparative studies across multiple segments.
% This real-time tracking capability is particularly valuable for users conducting segment evolution analysis in dynamic scenes and performing comparative studies across multiple segments.

% \vspace{-0.1in}
% \begin{table}[htb]
% \caption{Segmentation methods comparison: traditional vs.\ VolSegGS.}
% \vspace{-0.1in}
% \centering
% % {\scriptsize
% %{\fontsize{6pt}{7pt}\selectfont
% \resizebox{\columnwidth}{!}{
% \begin{tabular}{c|ccccc}
%     & volume data & segmentation & segmentation & segmentation & segment \\ 
% method & requirement &  mechanism & granularity & (training) time & inference \\ \hline
% \cite{Huang-RGVis-PG03}               & yes  & region-grow  & coarse  & fast  & --  \\
% \cite{Tzeng-HiDimCla-TVCG05}       & yes  & sparse coding  & medium  & slow  & slow \\
% \cite{Ip-HistSeg-TVCG12}          & yes  & 2D histogram  & medium  & moderate  & fast  \\
% \cite{Soundararajan-LPTF-CGF15}          & yes  & probability  & coarse  & fast  & fast  \\
% \cite{Ma-FeatCla-TVCG18}          & yes  & isosurface  & coarse  & fast  & slow \\
% \cite{Quan-H3DCSC-TVCG18}       & yes  & sparse coding  & medium  & slow  & moderate \\
% VolSegGS                                    & no  & visual  & fine & moderate & fast  \\
% \end{tabular}
% }
% \label{tab:diss-seg-comp}
% \end{table}

\vspace{-0.05in}
\subsection{Limitations}

%{\bf Limitations.}
% VolSegGS is primarily limited by its reliance on DVR images. 
% While this approach enables real-time dynamic visualization scene exploration and segmentation, it is restricted to visible regions only. 
% Additionally, the overall effectiveness of segmentation and tracking is heavily influenced by the TFs used to render the images. 
\hot{While VolSegGS learns dynamic visualization scenes from DVR images and enables real-time scene exploration and segmentation, its reliance on DVR images rather than volume data introduces three key limitations. 
First, the surrogate-level segmentation labels generated by VolSegGS are derived from rendered visualizations and are not directly transferable to the original volumetric data.
While these labels are sufficient to support exploratory visualization tasks, such as interactive segmentation, object tracking, and downstream editing, for broader applicability, future work could investigate robust methods for mapping these labels back to volume data.
Second, the DVR images used to train VolSegGS are rendered using a single TF.
As a result, the overall effectiveness of segmentation and tracking is highly dependent on the chosen TF.
This may limit the flexibility of VolSegGS in handling complex scenes with intricate geometries and heavy occlusions, where multiple TFs are often required to comprehensively capture structural details.
A potential solution is to train multiple sets of deformable Gaussians on scenes rendered with different TFs, providing users with greater flexibility for segmentation and tracking across varying TFs.
This is feasible because the Gaussians are relatively independent of one another, making merging them relatively straightforward.
We view multi-TF support as a natural extension and future improvement. %, but it has not been implemented at this stage due to the focus of work.
Third, as shown in Tables~\ref{tab:nvs-dataset}~and~\ref{tab:render-metrics}, both preparation time and storage requirement for generating training images tend to increase with the increasing volume size.
Few-shot learning should offer a potential future direction to reduce the required training data.}

Another limitation of VolSegGS is its potential difficulty with long-term tracking. 
For extended sequences, the deformation field network may struggle to capture the changes in the entire scene over time, especially if there are large intervals between frames or significant overall deformations. 
This issue could be mitigated by increasing the number of 3D Gaussians and concatenating multiple deformation field networks to handle longer sequences. 
However, this would increase the model size and training cost, which could reduce its overall usability.

\vspace{-0.05in}
\section{Conclusions and Future Work}

We introduced VolSegGS, a novel dynamic volumetric scene segmentation framework that enables real-time rendering using deformable Gaussian representations trained on volume rendering images. 
Our method employs a robust two-level segmentation strategy:
(1) coarse-level segmentation partitions the scene based on estimated view-independent colors of Gaussians, and 
(2) fine-level segmentation refines these results through an affinity field network optimized with automatically generated 2D mask supervision. 
Furthermore, we utilize Gaussian deformation to track segments over time while maintaining real-time performance. 
Our evaluations on multiple time-varying datasets show that VolSegGS outperforms state-of-the-art methods in dynamic scene representation and 3D segmentation. 
VolSegGS provides a practical solution for 
\hot{exploratory visualization of large-scale volumetric datasets by enabling interactive segmentation, tracking, and editing}.
%large-scale volumetric data exploration by enabling real-time interactive visualization with minimal hardware requirements.

% Our future work will extend VolSegGS in the following ways.
\hot{Our future work will focus on addressing the current limitations of VolSegGS and further enhancing its capabilities for dynamic volume visualization and analysis.
First, we plan to investigate reliable methods for mapping image-based segmentation labels back to the original volumetric data, enabling more fundamental analysis tasks.
We also aim to improve the flexibility of VolSegGS by introducing multi-TF support, allowing users to seamlessly switch between different TFs.
To reduce data preparation time and storage costs, we will explore few-shot learning techniques to minimize the number of required training images.
We also intend to develop efficient methods for concatenating multiple dynamic scene clips, enabling VolSegGS to handle longer and more complex temporal sequences.
Additionally, given the reliance on SAM masks, a promising direction is to fine-tune the SAM model on volumetric datasets and evaluate its impact on segmentation accuracy and robustness. 
Finally, we envision VolSegGS as a platform for broader innovation in scientific visualization. 
Future research may incorporate language-guided interaction~\cite{Ai-VIS25} or enable complex operations such as relighting, inpainting, and spatiotemporal enhancement.}

% First, VolSegGS provides a strong foundation for integrating language features into dynamic Gaussian representations, allowing for more intuitive, language-based queries.
% Second, the fine-grained segmentation capabilities can improve analysis by targeting finer structural details.
% Third, while the current version supports basic adjustments (e.g., color, opacity, and transformations), we plan to integrate advanced operations such as relighting, stylization, inpainting, temporal extrapolation, and spatial super-resolution. 
% Given the real-time tracking capabilities, these operations can be applied efficiently to dynamic volumetric datasets.

% if specified like this the section will be committed in review mode
\vspace{-0.05in}
\acknowledgments{This research was supported in part by the U.S.\ National Science Foundation through grants IIS-1955395, IIS-2101696, OAC-2104158, and IIS-2401144, and the U.S.\ Department of Energy through grant DE-SC0023145. The authors would like to thank the anonymous reviewers for their insightful comments.}

%\newpage
%\clearpage
\vspace{-0.05in}
\section*{Appendix}

\setcounter{section}{0}
\setcounter{figure}{0}
\setcounter{table}{0}
%\setcounter{page}{1}

%\vspace{-0.05in}
\section{Ablation Study}

% We study three aspects that impact the performance of VolSegGS: the loss function, initialization of canonical 3D Gaussians, and Gaussian opacity deformation.
\hot{
We conduct an ablation study on two key components of VolSegGS: dynamic scene representation learning and segmentation.
First, to train deformable 3D Gaussians, we analyze the impact of several factors, including the choice of loss function, the initialization of canonical 3D Gaussians, Gaussian opacity deformation, and the structure of the deformation field network.
Next, we examine how the proposed two-level segmentation strategy contributes to overall segmentation quality improvement.
}

%\vspace{-0.1in}
\begin{table}[htb]
    \caption{Comparison of training VolSegGS on different loss combinations using the mantle dataset: average PSNR (dB), SSIM, and LPIPS across all 181 synthesized views. Training time (TT, in minutes) is also reported. The best ones are highlighted in bold.}
    \vspace{-0.1in}
    \centering
    {\scriptsize
    % \resizebox{\columnwidth}{!}{
    \begin{tabular}{c|cccc}
        loss    & PSNR$\uparrow$ & SSIM$\uparrow$ & LPIPS$\downarrow$ & TT$\downarrow$\\ \hline
        L1 & 27.94 & 0.974 & 0.036 & 15.70 \\
        L2 & \bf 29.18 & 0.974 & 0.032 & \bf 15.52 \\
        L2+SSIM & 29.02 & \bf 0.979 & \bf 0.030  & 15.98 \\
        \end{tabular}
    }
    \label{tab:ablation-loss}
\end{table}

\begin{figure}[htb]
%\vspace{-0.1in}
\begin{center}
$\begin{array}{c@{\hspace{0.05in}}c}
    \includegraphics[width=0.4\linewidth]{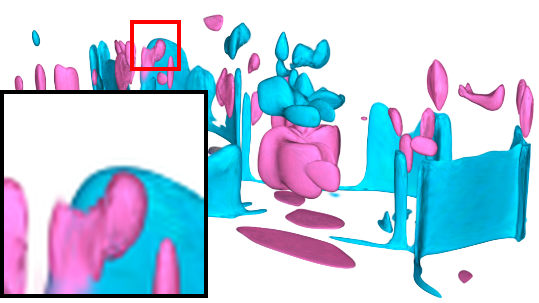}&
    \includegraphics[width=0.4\linewidth]{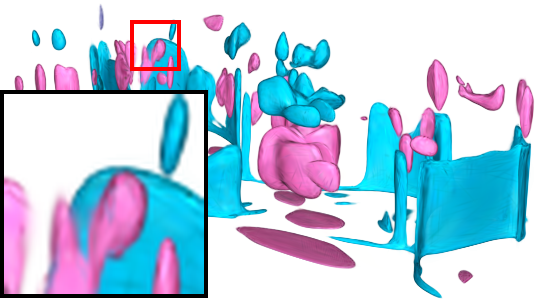}\\
    \mbox{\footnotesize (a) L1} & \mbox{\footnotesize (b) L2} \\
    \includegraphics[width=0.4\linewidth]{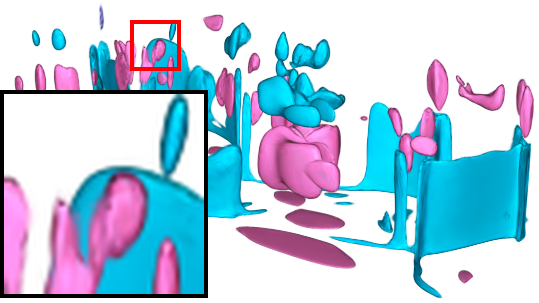}&
    \includegraphics[width=0.4\linewidth]{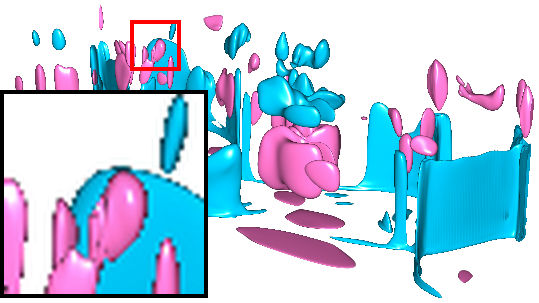}\\
    \mbox{\footnotesize (c) L2 + DSSIM} & \mbox{\footnotesize (d) GT}
\end{array}$
\end{center}
\vspace{-.25in} 
\caption{Comparison of training VolSegGS on different loss combinations using the mantle dataset.} 
\label{fig:ablation-loss}
\end{figure}

\begin{table}[htb]
    \caption{Comparison of VolSegGS on the TV loss using the combustion dataset: average PSNR (dB), SSIM, and LPIPS across all 181 synthesized views. Training time (TT, in minutes) is also reported. The best ones are highlighted in bold.}
    \vspace{-0.1in}
    \centering
    {\scriptsize
    % \resizebox{\columnwidth}{!}{
    \begin{tabular}{c|cccc}
        TV loss   & PSNR$\uparrow$ & SSIM$\uparrow$ & LPIPS$\downarrow$ & TT$\downarrow$\\ \hline
        without & 18.90 & 0.759 & 0.296 &  \bf 16.87 \\
        with & \bf 25.76 & \bf 0.897 & \bf 0.092 &  16.92 \\
        \end{tabular}
    }
    \label{tab:ablation-tv}
\end{table}

\begin{figure}[htb]
%\vspace{-0.1in}
\begin{center}
$\begin{array}{c@{\hspace{0.05in}}c@{\hspace{0.05in}}c}
    \includegraphics[width=0.3\linewidth]{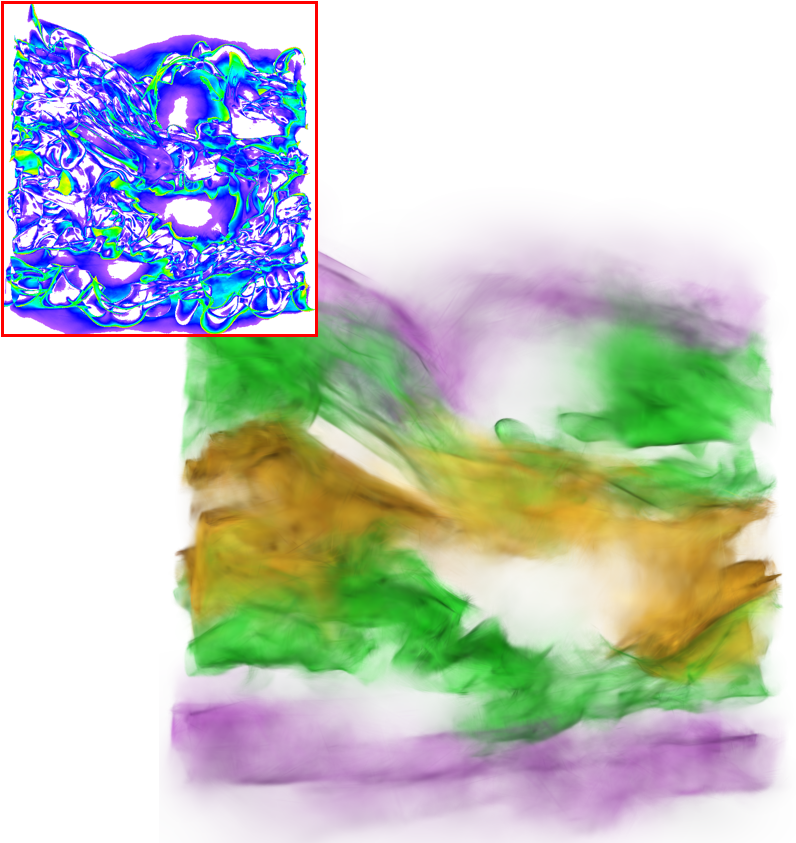}&
    \includegraphics[width=0.3\linewidth]{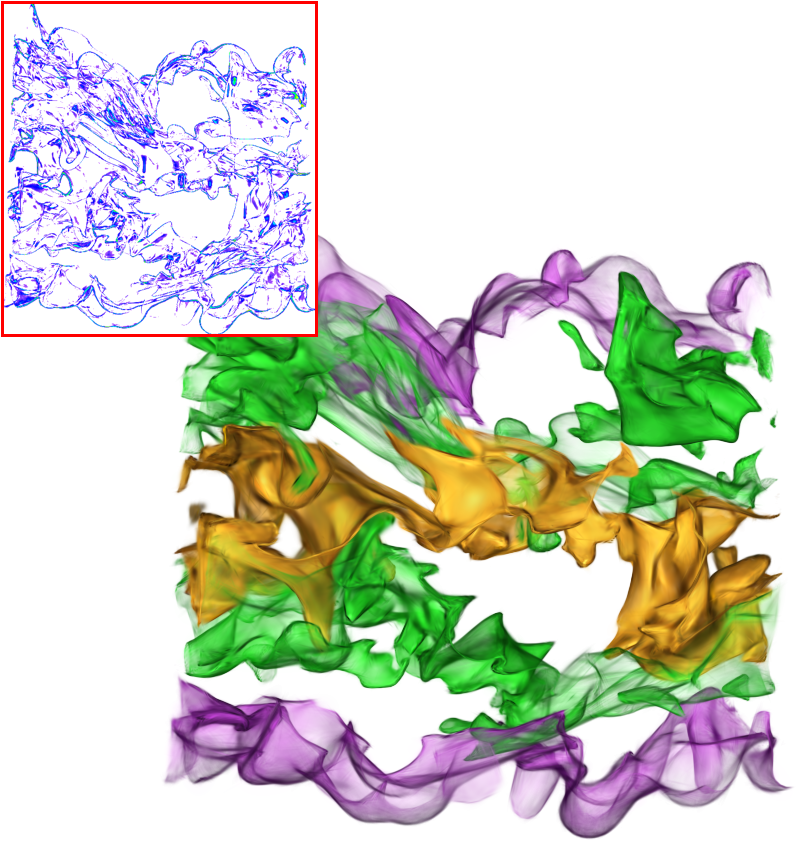}&
    \includegraphics[width=0.3\linewidth]{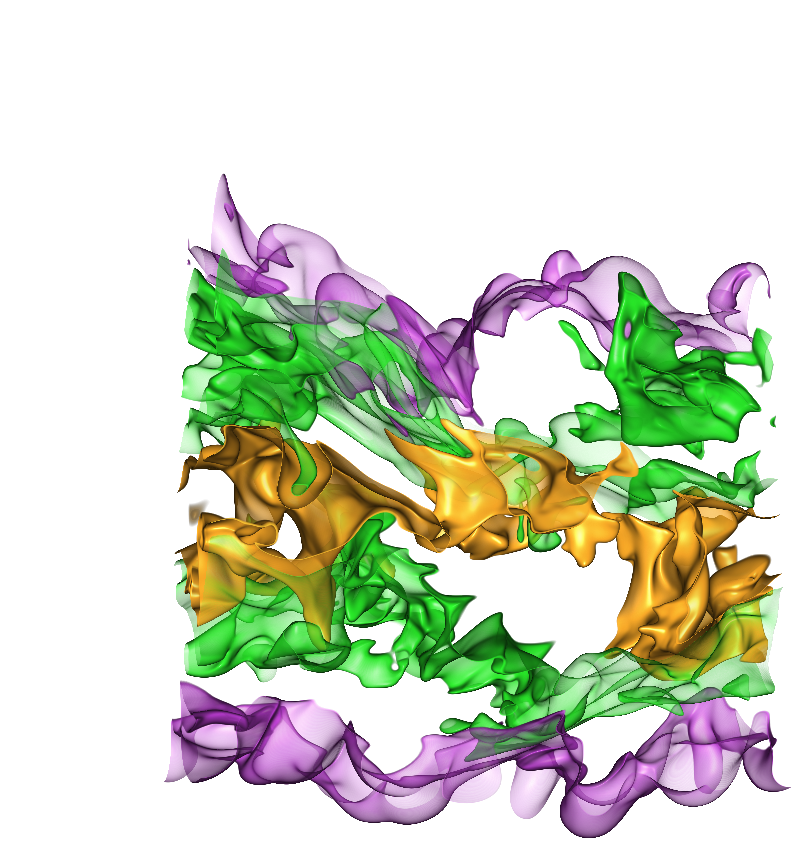}\\
    \mbox{\footnotesize (a) w/o TV loss} & \mbox{\footnotesize (b) w/ TV loss} & \mbox{\footnotesize (c) GT}
\end{array}$
\end{center}
\vspace{-.25in} 
\caption{Comparison of VolSegGS on the TV loss using the combustion dataset.} 
\label{fig:ablation-tv}
\end{figure}

\begin{table}[htb]
    \caption{Comparison of VolSegGS on initializing the canonical 3D Gaussians using the five jets dataset: average PSNR (dB), SSIM, LPIPS, and rendering framerate (FPS) across all 181 synthesized views. Training time (TT, in minutes) is also reported. The best ones are highlighted in bold.}
    \vspace{-0.1in}
    \centering
    {\scriptsize
    % \resizebox{\columnwidth}{!}{
    \begin{tabular}{c|ccccc}
        initialization   & PSNR$\uparrow$ & SSIM$\uparrow$ & LPIPS$\downarrow$ & FPS$\uparrow$ & TT$\downarrow$\\ \hline
        without & 26.17 & 0.957 & 0.039 & \bf 89.36 & \bf 15.73 \\
        with & \bf 27.16 & \bf 0.965 & \bf 0.030 & 88.31 & 16.07 \\
        \end{tabular}
    }
    \label{tab:ablation-init}
\end{table}

\begin{figure}[htb]
%\vspace{-0.1in}
\begin{center}
$\begin{array}{c@{\hspace{0.05in}}c@{\hspace{0.05in}}c}
    \includegraphics[width=0.3\linewidth]{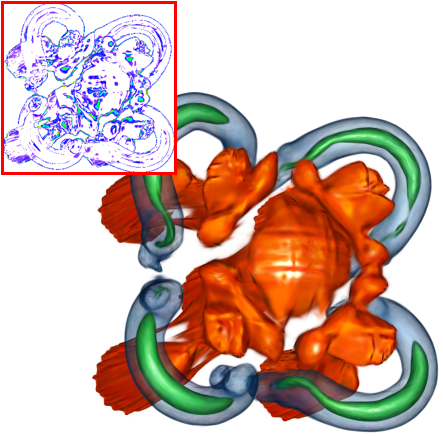}&
    \includegraphics[width=0.3\linewidth]{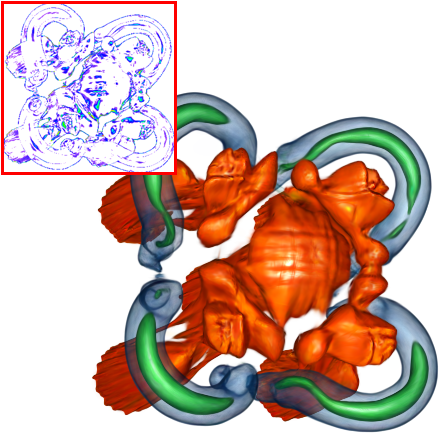}&
    \includegraphics[width=0.3\linewidth]{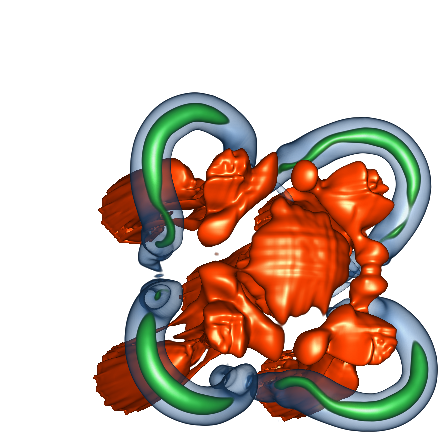}\\
    \mbox{\footnotesize (a) w/o initialization} & \mbox{\footnotesize (b) w/ initialization} & \mbox{\footnotesize (c) GT}
\end{array}$
\end{center}
\vspace{-.25in} 
\caption{Comparison of VolSegGS on initializing the canonical 3D Gaussians using the five jets dataset.} 
\label{fig:ablation-init}
\end{figure}

\begin{table}[htb]
    \caption{Comparison of VolSegGS on the Gaussian opacity deformation using the vortex dataset: average PSNR (dB), SSIM, and LPIPS across all 181 synthesized views. Training time (TT, in minutes) is also reported. The best ones are highlighted in bold.}
    \vspace{-0.1in}
    \centering
    {\scriptsize
    % \resizebox{\columnwidth}{!}{
    \begin{tabular}{c|cccc}
        opacity    & PSNR$\uparrow$ & SSIM$\uparrow$ & LPIPS$\downarrow$ & TT$\downarrow$\\ \hline
        fixed & 26.02 & 0.937 & 0.094 & 15.95 \\
        deformable & \bf 26.37 & \bf 0.947 & \bf 0.074 & \bf 15.90 \\
        \end{tabular}
    }
    \label{tab:ablation-do}
\end{table}

\begin{figure}[htb]
%\vspace{-0.1in}
\begin{center}
$\begin{array}{c@{\hspace{0.05in}}c@{\hspace{0.05in}}c}
    \includegraphics[width=0.3\linewidth]{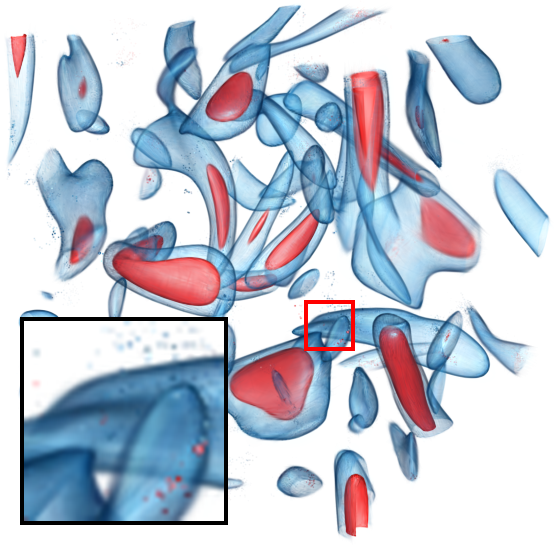}&
    \includegraphics[width=0.3\linewidth]{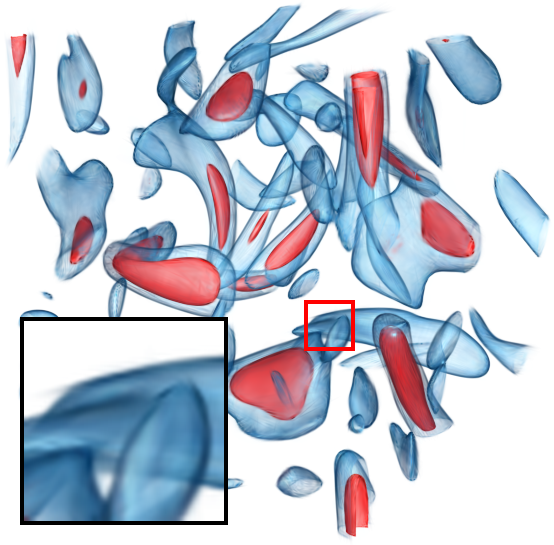}&
    \includegraphics[width=0.3\linewidth]{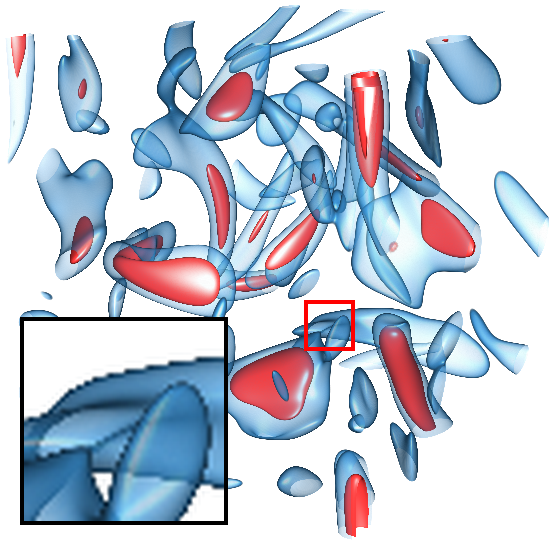}\\
    \mbox{\footnotesize (a) fixed opacity} & \mbox{\footnotesize (b) deformable opacity} & \mbox{\footnotesize (c) GT}
\end{array}$
\end{center}
\vspace{-.25in} 
\caption{Comparison of VolSegGS on the Gaussian opacity deformation using the vortex dataset.} 
\label{fig:ablation-do}
\end{figure}

\begin{table}[htb]
    \caption{Comparison of VolSegGS on different structures of deformation field network using the Tangaroa dataset: average PSNR (dB), SSIM, LPIPS, and rendering framerate (FPS) across all 181 synthesized views, training time (TT, in minutes), and model size (MS, in MB). The best ones are highlighted in bold.}
    \vspace{-0.1in}
    \centering
    {\scriptsize
    % \resizebox{\columnwidth}{!}{
    \begin{tabular}{c|cccccc}

        structure    & PSNR$\uparrow$ & SSIM$\uparrow$ & LPIPS$\downarrow$ & FPS$\uparrow$ & TT$\downarrow$ & MS$\downarrow$ \\ \hline
        implicit & 26.32 & 0.931 & 0.067 & 37.58 & 23.48 & \bf 37.36 \\
        hybrid & \bf 28.15 & \bf 0.953 & \bf 0.042 & \bf 88.65 & \bf 16.80 & 43.48 \\
    \end{tabular}
    }
    \label{tab:ablation-dnet}
\end{table}

\begin{figure}[htb]
%\vspace{-0.1in}
\begin{center}
$\begin{array}{c@{\hspace{0.05in}}c@{\hspace{0.05in}}c}
    \includegraphics[width=0.3\linewidth]{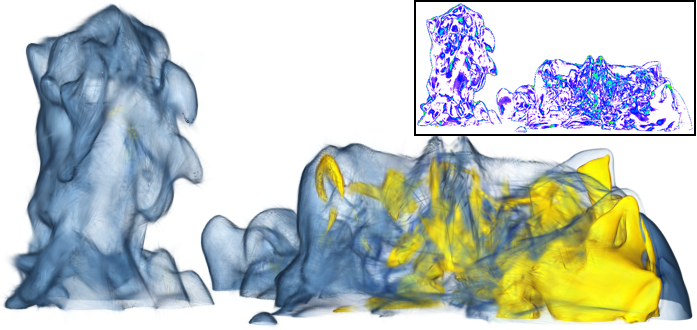}&
    \includegraphics[width=0.3\linewidth]{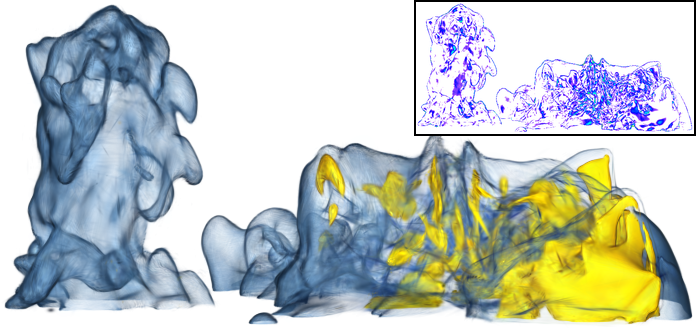}&
    \includegraphics[width=0.3\linewidth]{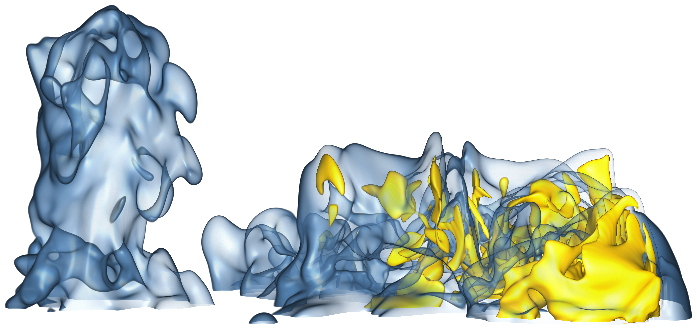}\\
    \mbox{\footnotesize (a) implicit} & \mbox{\footnotesize (b) hybrid} & \mbox{\footnotesize (c) GT}
\end{array}$
\end{center}
\vspace{-.25in} 
\caption{Comparison of VolSegGS on different structures of deformation field network using the Tangaroa dataset.} 
\label{fig:ablation-dnet}
\end{figure}

\begin{table}[htb]
    \caption{\hot{Comparison of VolSegGS on different segmentation methods using the vortex dataset: average PSNR (dB), SSIM, LPIPS, and IoU across all 181 synthesized views. The best ones are highlighted in bold.}}
    \vspace{-0.1in}
    \centering
    {\scriptsize
    % \resizebox{\columnwidth}{!}{
    \begin{tabular}{c|c|cccc}
        \hot{segment} & \hot{method} & \hot{PSNR$\uparrow$} & \hot{SSIM$\uparrow$} & \hot{LPIPS$\downarrow$} & \hot{IoU$\uparrow$} \\ \hline
        & \hot{coarse only} & \hot{18.77} & \hot{0.954} & \hot{0.223} & \hot{12.56} \\
        \hot{red} & \hot{fine only} & \hot{31.44} & \hot{0.993} & \hot{0.017} & \hot{45.84} \\
        & \hot{coarse+fine} & \hot{\bf 42.01} & \hot{\bf 0.999} & \hot{\bf 0.003} & \hot{\bf 84.22}\\ \hline
        & \hot{coarse only} & \hot{16.32} & \hot{0.841} & \hot{0.339} & \hot{7.62} \\
        \hot{blue} & \hot{fine only} & \hot{37.78} & \hot{0.997} & \hot{0.007} & \hot{\bf 95.26} \\
        & \hot{coarse+fine} & \hot{\bf 40.51} & \hot{\bf 0.998} & \hot{\bf 0.005} & \hot{95.21}\\
        \end{tabular}
    }
    \label{tab:ablation-seg}
\end{table}

\begin{figure}[htb]
%\vspace{-0.1in}
\begin{center}
$\begin{array}{c@{\hspace{0.1in}}c@{\hspace{0.1in}}c@{\hspace{0.1in}}c}
    \includegraphics[width=0.1525\linewidth]{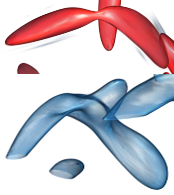}&
    \includegraphics[width=0.1525\linewidth]{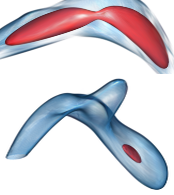}&
    \includegraphics[width=0.1525\linewidth]{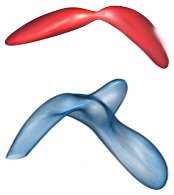}&
    \includegraphics[width=0.315\linewidth]{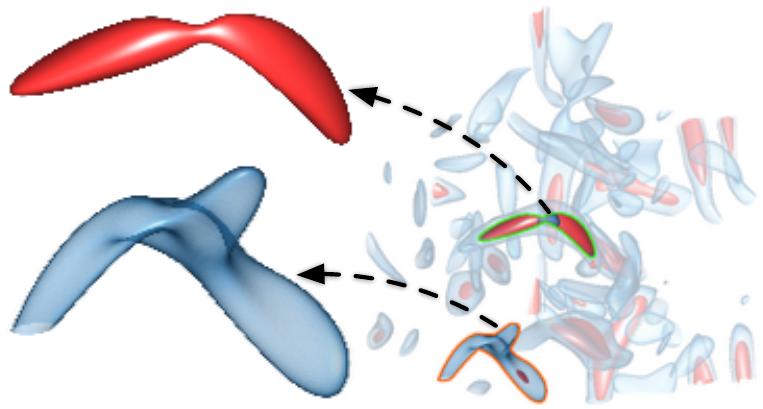}\\
    \mbox{\footnotesize (a) coarse only} & \mbox{\footnotesize (b) fine only} & \mbox{\footnotesize (c) coarse+fine} & \mbox{\footnotesize (d) GT}
\end{array}$
\end{center}
\vspace{-.25in} 
\caption{\hot{Comparison of VolSegGS on different segmentation methods using the vortex dataset.}} 
\label{fig:ablation-seg}
\end{figure}

\begin{table}[htb]
    \caption{Comparison of training VolSegGS on different numbers of sampled timesteps using the combustion dataset: average PSNR (dB), SSIM, and LPIPS across all 181 synthesized views. The best ones are highlighted in bold.}
    \vspace{-0.1in}
    \centering
    {\scriptsize
    % \resizebox{\columnwidth}{!}{
    \begin{tabular}{c|ccc}
        \# timesteps    & PSNR$\uparrow$ & SSIM$\uparrow$ & LPIPS$\downarrow$ \\ \hline
        10 & 23.91 & 0.857 & 0.114 \\
        20 & 25.07 & 0.881 & 0.100 \\
        30 & 25.76 & 0.897 & 0.092 \\
        40 & \bf 25.98 & \bf 0.901 & \bf 0.090 \\
        \end{tabular}
    }
    \label{tab:hyper-ntime}
\end{table}

\begin{figure}[htb]
%\vspace{-0.1in}
\begin{center}
$\begin{array}{c@{\hspace{0.05in}}c@{\hspace{0.05in}}c@{\hspace{0.05in}}c@{\hspace{0.05in}}c}
    \includegraphics[width=0.18\linewidth]{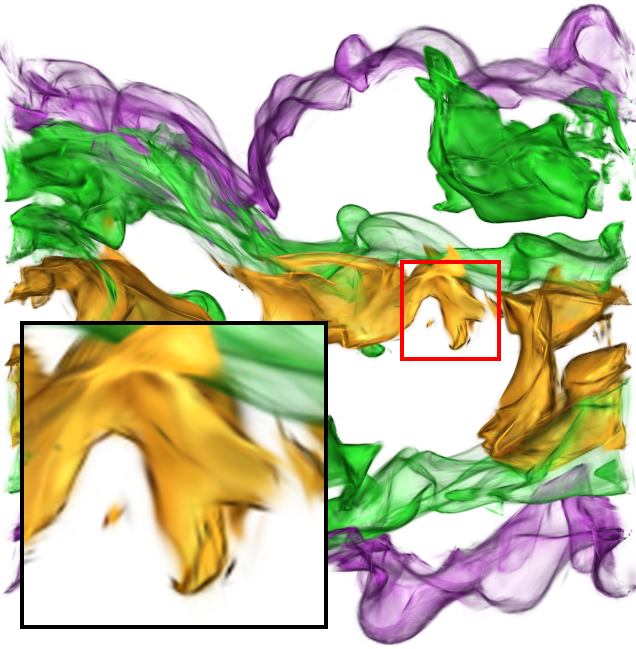}&
    \includegraphics[width=0.18\linewidth]{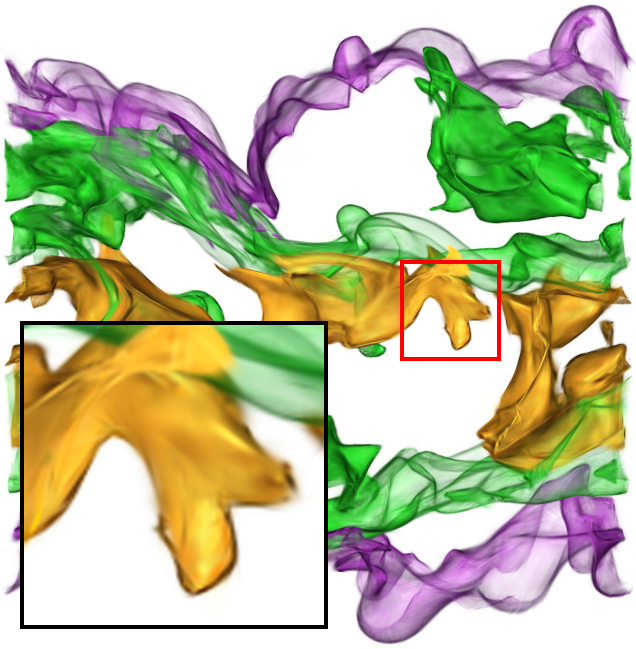}&
    \includegraphics[width=0.18\linewidth]{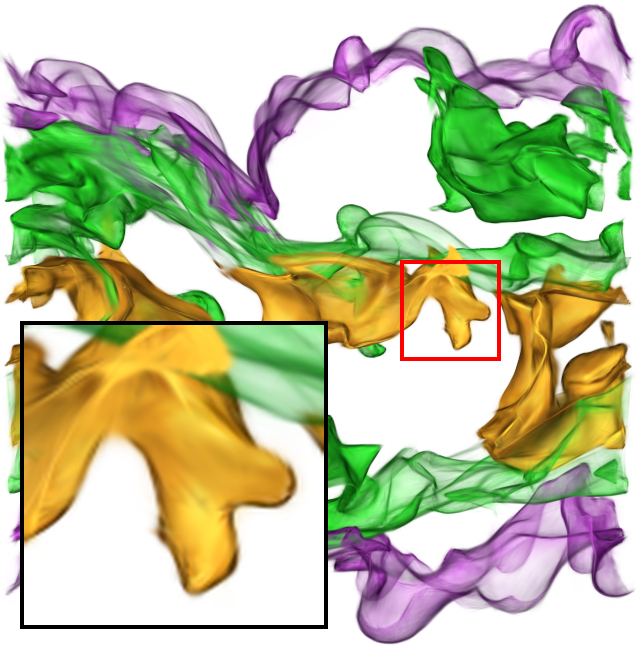}&
    \includegraphics[width=0.18\linewidth]{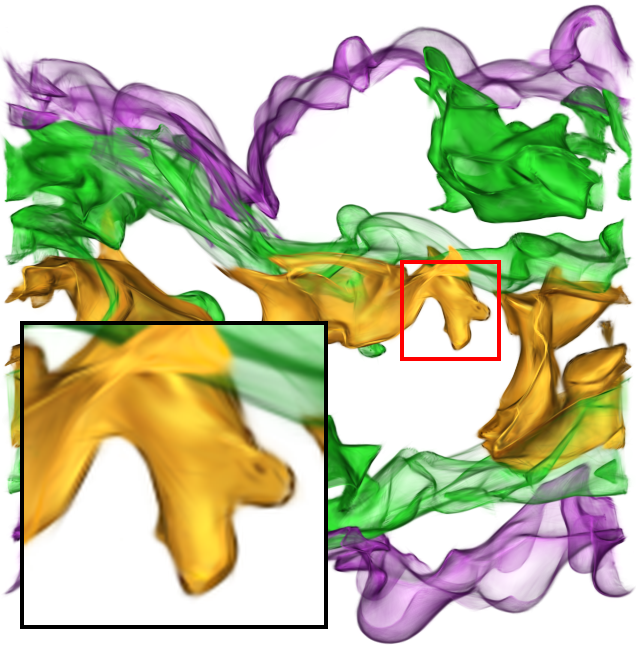}&
    \includegraphics[width=0.18\linewidth]{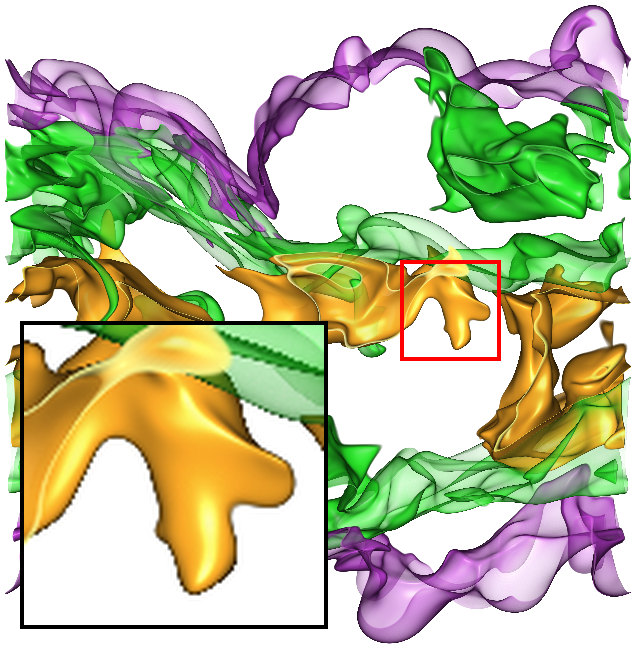}\\
    \mbox{\footnotesize (a) 10} & \mbox{\footnotesize (b) 20} & \mbox{\footnotesize (c) 30} & \mbox{\footnotesize (d) 40} & \mbox{\footnotesize (e) GT}
\end{array}$
\end{center}
\vspace{-.25in} 
\caption{Comparison of training VolSegGS on different numbers of sampled timesteps using the combustion dataset.} 
\label{fig:hyper-ntime}
\end{figure}

\begin{table}[htb]
    \caption{Comparison of training VolSegGS on different numbers of sampled views per timestep using the Tangaroa dataset: average PSNR (dB), SSIM, and LPIPS across all 181 synthesized views. The best ones are highlighted in bold.}
    \vspace{-0.1in}
    \centering
    {\scriptsize
    % \resizebox{\columnwidth}{!}{
    \begin{tabular}{c|ccc}
        \# timesteps    & PSNR$\uparrow$ & SSIM$\uparrow$ & LPIPS$\downarrow$ \\ \hline
        10 & 25.27 & 0.918 & 0.058 \\
        20 & 27.43 & 0.945 & 0.045 \\
        30 & 28.15 & 0.953 & 0.042 \\
        40 & \bf 28.37 & \bf 0.955 & \bf 0.041 \\
        \end{tabular}
    }
    \label{tab:hyper-nview}
\end{table}

\begin{figure}[htb]
%\vspace{-0.1in}
\begin{center}
$\begin{array}{c@{\hspace{0.05in}}c@{\hspace{0.05in}}c}
    \includegraphics[width=0.3\linewidth]{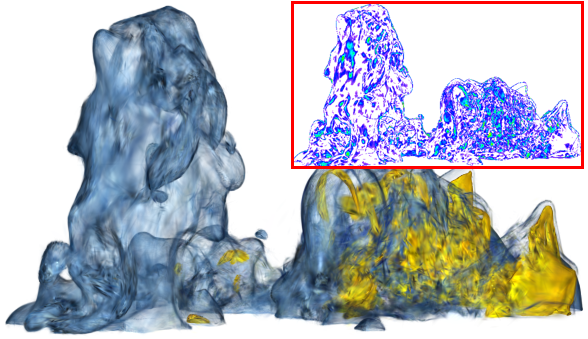}&
    \includegraphics[width=0.3\linewidth]{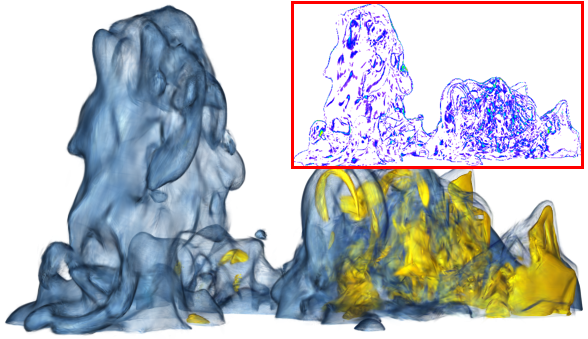}&
    \includegraphics[width=0.3\linewidth]{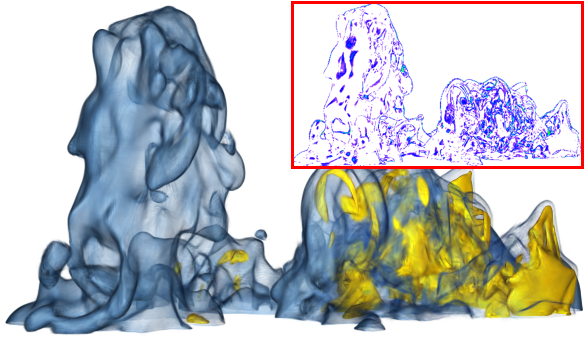}\\
    \mbox{\footnotesize (a) 10} & \mbox{\footnotesize (b) 20} & \mbox{\footnotesize (c) 30} \\
    \includegraphics[width=0.3\linewidth]{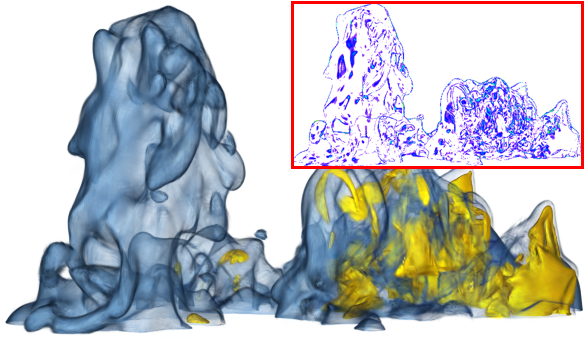}&
    \includegraphics[width=0.3\linewidth]{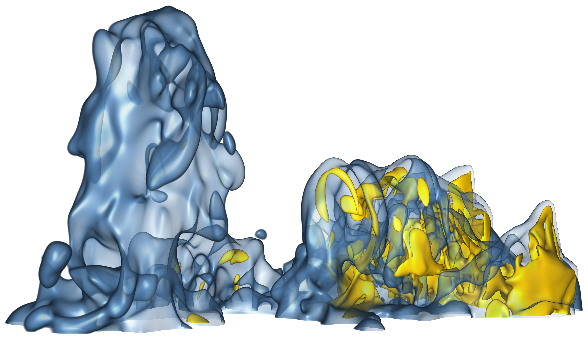}&\\
    \mbox{\footnotesize (d) 40} & \mbox{\footnotesize (e) GT} & 
\end{array}$
\end{center}
\vspace{-.25in} 
\caption{Comparison of training VolSegGS on different numbers of sampled views per timestep using the Tangaroa dataset.} 
\label{fig:hyper-nview}
\end{figure}

\begin{table}[htb]
    \caption{Comparison of training VolSegGS on different numbers of iterations using the five jets dataset: average PSNR (dB), SSIM, and LPIPS across all 181 synthesized views. Training time (TT, in minutes) is also reported. The best ones are highlighted in bold.}
    \vspace{-0.1in}
    \centering
    {\scriptsize
    % \resizebox{\columnwidth}{!}{
    \begin{tabular}{c|cccc}
        \# iterations    & PSNR$\uparrow$ & SSIM$\uparrow$ & LPIPS$\downarrow$ & TT$\downarrow$ \\ \hline
        5,000 & 23.63 & 0.926 & 0.068 & \bf 3.85 \\
        10,000 & 25.42 & 0.952 & 0.038 & 7.78 \\
        15,000 & 26.64 & 0.962 & 0.034 & 11.85 \\
        20,000 & 27.16 & 0.965 & 0.030 & 16.07 \\
        25,000 & 27.20 & 0.965 & \bf 0.029 & 19.80 \\
        30,000 & \bf 27.40 & \bf 0.966 & \bf 0.029 & 24.02 \\
        \end{tabular}
    }
    \label{tab:hyper-iter}
\end{table}

\begin{figure}[htb]
%\vspace{-0.1in}
\begin{center}
$\begin{array}{c@{\hspace{0.05in}}c@{\hspace{0.05in}}c@{\hspace{0.05in}}c}
    \includegraphics[width=0.23\linewidth]{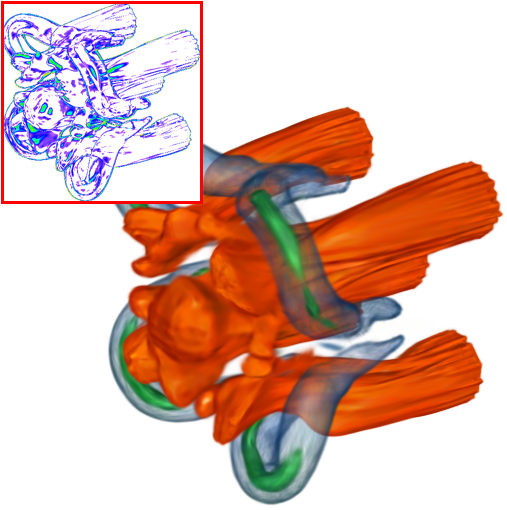}&
    \includegraphics[width=0.23\linewidth]{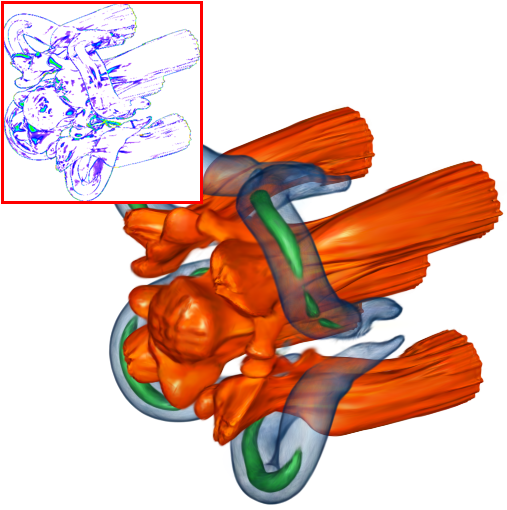}&
    \includegraphics[width=0.23\linewidth]{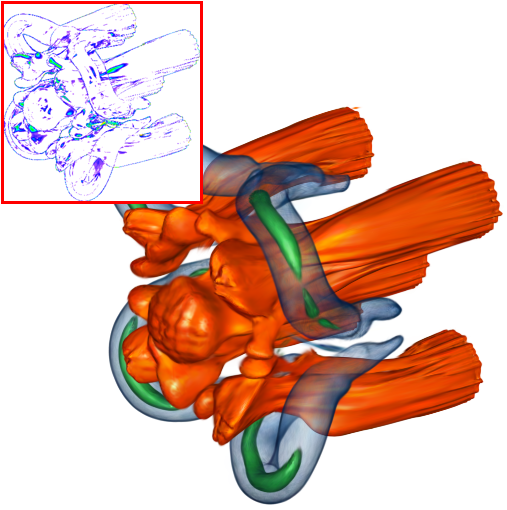}&
    \includegraphics[width=0.23\linewidth]{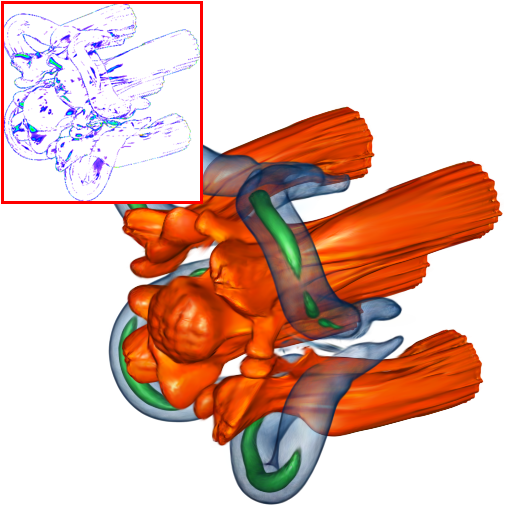}\\
    \mbox{\footnotesize (a) 5,000} & \mbox{\footnotesize (b) 10,000} & \mbox{\footnotesize (c) 15,000} & \mbox{\footnotesize (d) 20,000}\\
    \includegraphics[width=0.23\linewidth]{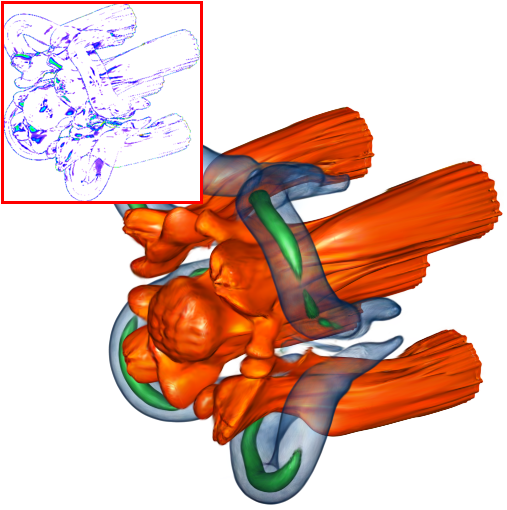}&
    \includegraphics[width=0.23\linewidth]{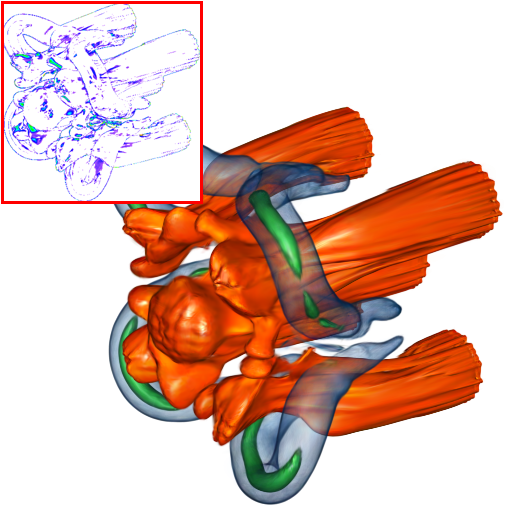}&
    \includegraphics[width=0.23\linewidth]{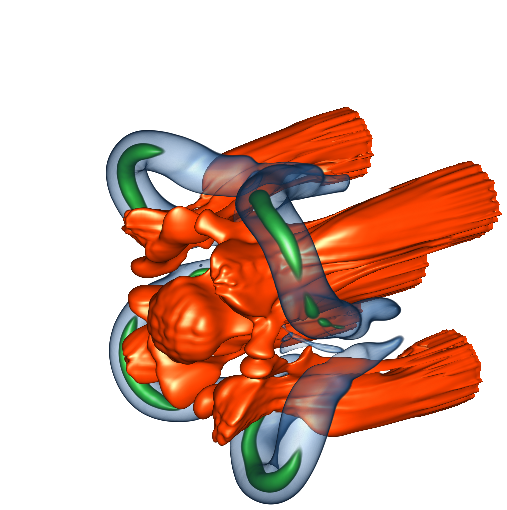}&\\
    \mbox{\footnotesize (e) 25,000} & \mbox{\footnotesize (f) 30,000} & \mbox{\footnotesize (g) GT}
\end{array}$
\end{center}
\vspace{-.25in} 
\caption{Comparison of training VolSegGS on different numbers of iterations using the five jets dataset.} 
\label{fig:hyper-iter}
\end{figure}

\begin{table}[htb]
    \caption{\hot{Comparison of VolSegGS on different numbers and spatial distributions of views using the combustion dataset: average PSNR (dB), SSIM, LPIPS, and IoU across all 181 synthesized views. The best ones are highlighted in bold.}}
    \vspace{-0.1in}
    \centering
    {\scriptsize
    % \resizebox{\columnwidth}{!}{
    \begin{tabular}{c|c|cccc}
        \hot{segment} & \hot{\# views} & \hot{PSNR$\uparrow$} & \hot{SSIM$\uparrow$} & \hot{LPIPS$\downarrow$} & \hot{IoU$\uparrow$} \\ \hline
        & \hot{30 (evenly)} & \hot{\bf 38.03} & \hot{\bf 0.996} & \hot{\bf 0.005} & \hot{\bf 94.14} \\
        & \hot{10 (evenly)} & \hot{32.44} & \hot{0.993} & \hot{0.016} & \hot{89.65} \\
        \hot{green} & \hot{10 ($x$-axis)} & \hot{37.04} & \hot{0.996} & \hot{0.008} & \hot{90.91}\\ 
        & \hot{10 ($y$-axis)} & \hot{37.11} & \hot{0.995} & \hot{0.006} & \hot{88.53}\\ 
        & \hot{10 ($z$-axis)} & \hot{37.34} & \hot{0.995} & \hot{0.006} & \hot{91.71}\\ \hline
        & \hot{30 (evenly)} & \hot{43.48} & \hot{\bf 0.999} & \hot{0.004} & \hot{93.48} \\
        & \hot{10 (evenly)} & \hot{42.04} & \hot{\bf 0.999} & \hot{0.010} & \hot{91.88} \\
        \hot{yellow} & \hot{10 ($x$-axis)} & \hot{43.48} & \hot{\bf 0.999} & \hot{\bf  0.003} & \hot{93.48}\\ 
        & \hot{10 ($y$-axis)} & \hot{\bf 43.49} & \hot{\bf 0.999} & \hot{\bf 0.003} & \hot{\bf 93.49}\\ 
        & \hot{10 ($z$-axis)} & \hot{43.48} & \hot{\bf 0.999} & \hot{\bf 0.003} & \hot{93.48}\\ 
        \end{tabular}
    }
    \label{tab:ablation-sam}
\end{table}

\begin{figure}[htb]
%\vspace{-0.1in}
\begin{center}
$\begin{array}{c@{\hspace{0.1in}}c@{\hspace{0.1in}}c@{\hspace{0.1in}}c@{\hspace{0.1in}}c@{\hspace{0.1in}}c}
    \includegraphics[width=0.1\linewidth]{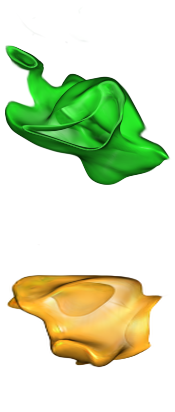}&
    \includegraphics[width=0.1\linewidth]{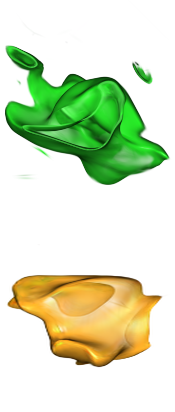}&
    \includegraphics[width=0.1\linewidth]{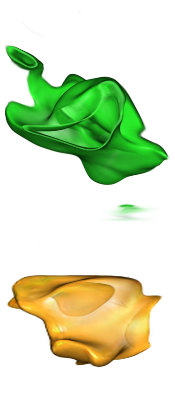}&
    \includegraphics[width=0.1\linewidth]{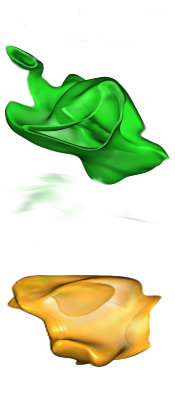}&
    \includegraphics[width=0.1\linewidth]{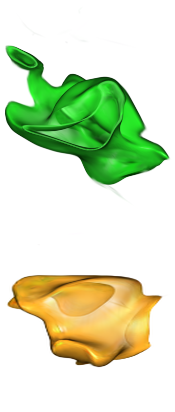}&
    \includegraphics[width=0.33\linewidth]{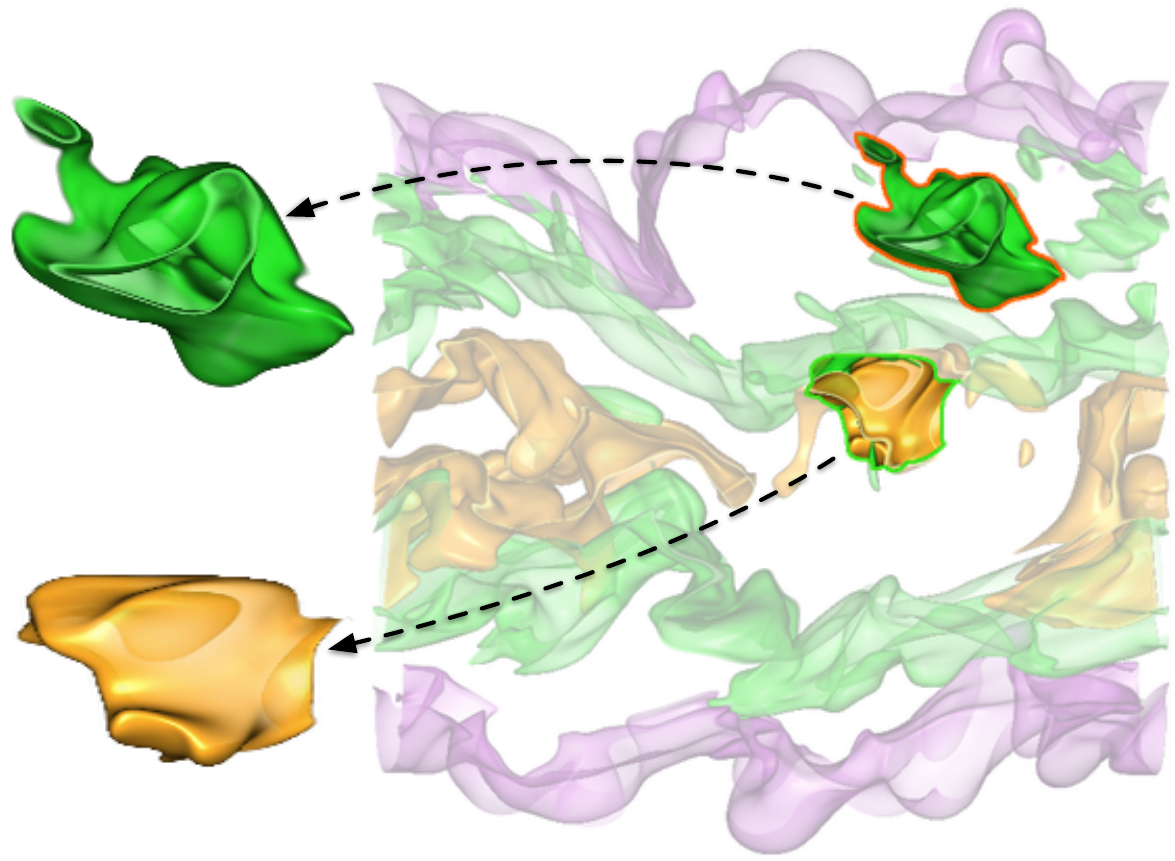}\\
    \mbox{\footnotesize (a) 30} & \mbox{\footnotesize (b) 10} & \mbox{\footnotesize (c) $x$} & \mbox{\footnotesize (d) $y$} & \mbox{\footnotesize (e) $z$} & \mbox{\footnotesize (f) GT}
\end{array}$
\end{center}
\vspace{-.25in} 
\caption{\hot{Comparison of VolSegGS on different numbers and spatial distributions of views using the combustion dataset. 30 and 10 refer to segmentation results using SAM masks generated from 30 and 10 evenly distributed views, respectively. $x$, $y$, and $z$ refer to segmentation results using SAM masks generated from 10 views biased along the $x$-, $y$-, and $z$-axis, respectively.}} 
\label{fig:ablation-sam}
\end{figure}

\begin{figure}[!h]
%\vspace{-0.1in}
\begin{center}
$\begin{array}{c@{\hspace{0.05in}}c@{\hspace{0.05in}}c}
    \includegraphics[height=0.6in]{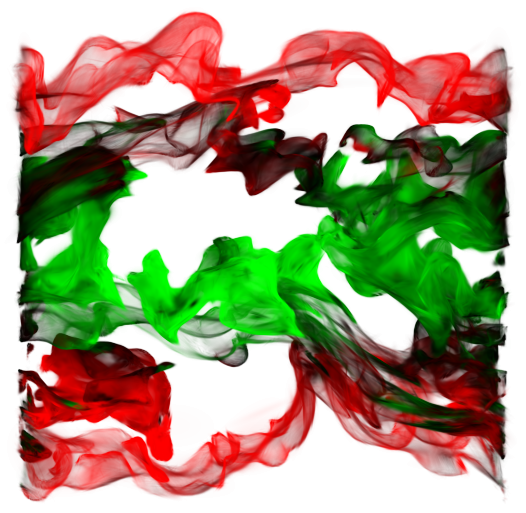}&
    \includegraphics[height=0.6in]{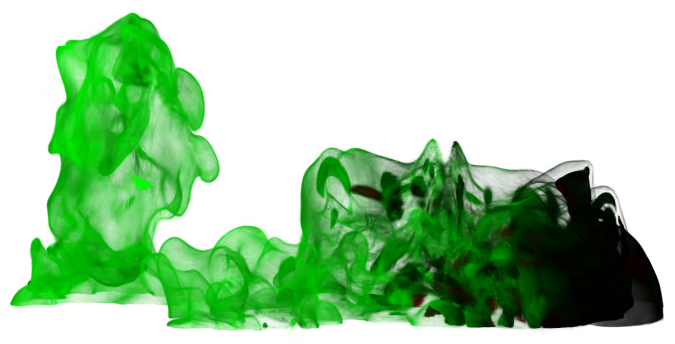}&
    \includegraphics[height=0.6in]{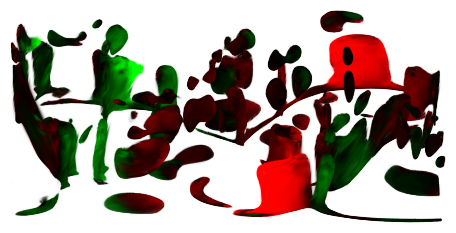}\\
%    \mbox{\footnotesize (a)} & \mbox{\footnotesize (b)} & \mbox{\footnotesize (c)}
\end{array}$
\end{center}
\vspace{-.25in} 
\caption{Visualization of the $x$-axis deformation velocity: red indicates positive values, green indicates negative values, and brightness represents magnitude. Left to right: combustion, Tangaroa, and mantle.} 
\label{fig:diss-deform}
\end{figure}

{\bf Loss function.}
From Table~\ref{tab:ablation-loss}, we observe that the L2 loss function achieves the highest PSNR, while the combination of L2 and SSIM losses yields the best performance in terms of SSIM and LPIPS.
As illustrated in Figure~\ref{fig:ablation-loss}, the L1 loss results in smooth outputs with missing fine structures, whereas the L2 loss better preserves subtle details but introduces noticeable artifacts. 
The L2+SSIM loss produces the most visually appealing results, retaining fine details while effectively reducing artifacts.
Moreover, we find that TV loss plays a critical role in the convergence of the deformation field network, as evidenced by the results in Table~\ref{tab:ablation-tv} and Figure~\ref{fig:ablation-tv}.
Without TV loss, the model fails to learn a coherent deformation field, likely due to the lack of spatiotemporal neighborhood consistency.

{\bf Initialization of Canonical 3D Gaussians.}
As shown in Table~\ref{tab:ablation-init}, initializing the canonical 3D Gaussians for 3,000 iterations provides performance improvements with minimal increase in training time.
The gains in PSNR, SSIM, and LPIPS exceed those achieved by an additional 5,000 iterations of joint training, as shown in Table~\ref{tab:hyper-iter}.
Figure~\ref{fig:ablation-init} further illustrates that initialization leads to visibly enhanced detail reconstruction.

{\bf Gaussian opacity deformation.}
From Table~\ref{tab:ablation-do}, we observe that incorporating deformable opacity enables VolSegGS to achieve higher performance across PSNR, SSIM, and LPIPS.
As shown in Figure~\ref{fig:ablation-do}, using fixed opacity leads to visible artifacts caused by small floating Gaussians, whereas deformable opacity more accurately models the disappearance, resulting in cleaner renderings.

{\bf Structure of deformation field network.}
Table~\ref{tab:ablation-dnet} shows that the hybrid design delivers superior performance in PSNR, SSIM, and LPIPS compared to the fully implicit design. 
When both are trained for 30,000 iterations jointly with the warmed-up canonical 3D Gaussians, the hybrid design converges faster, requiring less training time.
Additionally, it achieves a higher rendering framerate, despite having a slightly larger model size.
Figure~\ref{fig:ablation-dnet} further highlights that the fully implicit design leads to blurred reconstructions, whereas the hybrid design enables more accurate recovery of details.

\hot{\bf Two-level segmentation.}
\hot{
As shown in Table~\ref{tab:ablation-seg} and Figure~\ref{fig:ablation-seg}, the coarse-level segmentation primarily relies on color, making it difficult to distinguish individual components that share similar colors.
Fine-level segmentation captures structure but ignores color, which can make it difficult to separate inner and outer parts with different appearances.
Our two-level approach successfully combines both, enabling a clear separation of regions based on color and structure. 
This validates its effectiveness in segmenting volume visualization scenes.
}

\vspace{-0.05in}
\section{Hyperparameter Analysis}

For hyperparameter analysis, we investigate three aspects that impact the \hot{rendering quality using deformable 3D Gaussians in VolSegGS}: the number of sampled timesteps for training, the number of sampled views per timestep for training, and the number of joint training iterations.
\hot{Additionally, we evaluate the effect of the number and diversity of SAM masks from different views on the performance of the affinity field network.
}

{\bf Number of sampled timesteps for training.}
Table~\ref{tab:hyper-ntime} shows that, with an insufficient number of sampled timesteps for training, VolSegGS may have difficulty reconstructing the scene accurately. 
Figure~\ref{fig:hyper-ntime}, allocating 30 timesteps allows the model to recover most of the details in the scene of the combustion dataset.

{\bf Number of sampled views per timestep for training.}
According to Table~\ref{tab:hyper-nview}, training VolSegGS with a limited number of views per timestep reduces reconstruction quality.
Figure~\ref{fig:hyper-nview} illustrates that with 30 views per timestep, the model could recover most details in the Tangaroa scene.

{\bf Number of joint training iterations.}
Table~\ref{tab:hyper-iter} suggests that 20,000 iterations are sufficient for jointly training the canonical 3D Gaussians and the deformation field network.
As shown in Figure~\ref{fig:hyper-iter}, VolSegGS can reconstruct most of the fine details in the five jets dataset after being trained for 20,000 iterations.

\hot{\bf Number and view distribution of SAM masks.}
\hot{
In this analysis, we investigate the impact of different numbers and spatial distributions of views on the segmentation performance of VolSegGS.
As shown in Table~\ref{tab:ablation-sam} and Figure~\ref{fig:ablation-sam}, our default setting generates SAM masks from 30 views, corresponding to the number of training views per timestep.
We then evaluate reduced configurations using only 10 views, either evenly distributed or biased in the viewing direction along the $x$-, $y$-, or $z$-axis, respectively.

The results show that the affinity field network trained with SAM masks remains largely robust even when the number of views is reduced from 30 to 10.
The performance drop is minimal, indicating that the network can still effectively leverage limited 2D segmentation input.
However, both the number and spatial distribution of views do influence segmentation quality, as noise and ambiguity in SAM masks can lead to localized errors.
Interestingly, we observe consistent improvements when the views are biased toward a specific direction.
In these cases, clustering views spatially enhances the consistency of SAM masks, and for less occluded regions, such as the yellow segment, this strategy can even outperform the evenly distributed setting with more views.
In contrast, more heavily occluded regions, such as the green segment, require a greater number of diverse viewpoints to achieve satisfactory segmentation results.

Note that, in the paper, we use evenly distributed views to ensure fair, consistent, and standardized experimental conditions.
}

\vspace{-0.05in}
\section{Method Comparison and Additional Discussion}

{\bf Comparison with segmentation methods.}
Existing volume segmentation methods~\cite{Huang-RGVis-PG03, Tzeng-HiDimCla-TVCG05, Ip-HistSeg-TVCG12, Soundararajan-LPTF-CGF15, Ma-FeatCla-TVCG18, Quan-H3DCSC-TVCG18, Sharma-CGF20, Kim-ACCESS21, He-GCNFCV-JV22} primarily rely on TFs to classify voxels. 
Earlier methods~\cite{Huang-RGVis-PG03, Tzeng-HiDimCla-TVCG05, Ip-HistSeg-TVCG12, Soundararajan-LPTF-CGF15, Ma-FeatCla-TVCG18, Quan-H3DCSC-TVCG18} improved segmentation quality by incorporating higher-dimensional features and multi-dimensional TFs. 
However, they often suffer from increased computational overhead and the complexity of designing multi-dimensional TFs. 
More recent methods~\cite{Sharma-CGF20, Kim-ACCESS21, He-GCNFCV-JV22} have shifted toward leveraging deep learning to assist in TF design, yet this significantly increases segmentation time. 

In contrast, VolSegGS introduces a visual segmentation approach that achieves 3D segmentation by reconstructing visualizations from rendered images. 
% It complements all previously mentioned segmentation methods, as it can directly utilize rendered images produced using their TFs. 
Specifically, VolSegGS employs a color-based coarse segmentation strategy that aligns with TF-based colorization. 
Additionally, it offers a flexible, multi-scale fine segmentation capability, enabling further subdivision of coarse segments based on visual cues. 
While fine-level segmentation requires an initial preparation time of several minutes, it supports immediate inference. 
By leveraging an efficient scene representation based on 3D Gaussians instead of raw volumetric data, VolSegGS enables real-time rendering and segmentation for large-scale datasets.

\hot{
It is important to note that, unlike the previously mentioned methods, VolSegGS does not support direct segmentation on raw volume data. 
This limitation may restrict its applicability in certain use cases and hinder direct performance comparisons with volume-based approaches. 
Rather than serving as a replacement, VolSegGS can complement existing methods by leveraging their TFs for coarse-level segmentation of 3D scenes.
}

{\bf Comparison with feature-tracking methods.}
Existing feature-tracking methods~\cite{Silver-TVCG97, Ji-VIS03, Muelder-PVIS09, Widanagamaachchi-LDAV12, Dutta-TVCG16, Saikia-CGF17, Schnorr-TVCG20} for time-varying scalar field data primarily rely on deterministic algorithms. 
Most prior works~\cite{Silver-TVCG97, Ji-VIS03, Muelder-PVIS09, Dutta-TVCG16, Schnorr-TVCG20} track individual features by comparing voxel values or isosurfaces across adjacent timesteps.
Meanwhile, a separate line of research~\cite{Widanagamaachchi-LDAV12, Saikia-CGF17} enables global feature tracking by computing and comparing merge trees.

In contrast, VolSegGS introduces a novel feature-tracking approach by learning a deformation field from DVR images of time-varying data. 
Unlike prior methods, VolSegGS tracks global features without relying on predefined critical points, isosurfaces, or merge trees. 
Instead, it offers greater flexibility by enabling users to track arbitrary segments without requiring additional recomputation. 
The time required to train the 3D Gaussians with the deformation field network is comparable to the time needed to compute merge trees. 
However, once trained, VolSegGS enables real-time tracking and rendering of any arbitrary segment, even for large-scale datasets. 
Moreover, as illustrated in Figure~\ref{fig:diss-deform}, VolSegGS can visualize the global deformation velocity of the entire scene, providing a comprehensive understanding of the volumetric scene's evolution.

\hot{Although VolSegGS lacks the capability to directly track features in raw volume data, it is primarily designed as a visualization tool, emphasizing real-time, exploratory interaction with dynamic visualization scenes.}

{\bf SAM masks for segmentation.}
\hot{Relying on SAM masks for segmentation may present challenges as well, as SAM has not been fine-tuned on scientific datasets. 
When all SAM masks from multiple views fail to accurately capture a segment, VolSegGS could lead to incomplete segmentation or mistakenly encompass adjacent regions.
To mitigate this issue, our affinity feature network helps smooth segmentation results in the implicit space, while the multi-scale fine-level segmentation allows users to select smaller parts to assemble a complete segment.
However, this approach may be suboptimal in certain cases and is intended only as a workaround.
It would be valuable for future work to investigate fine-tuning SAM on visualization datasets for segmentation quality improvement.}

\vspace{-0.05in}
\bibliographystyle{abbrv-doi-hyperref}
\bibliography{template}

\end{document}